\documentclass{ws-ijmpa}

\usepackage{latexsym}

\def\Dslash{\hspace{2pt}\raisebox{0.7pt}{$\slash$} \hspace{-7pt} D}
\def\Dslashs{\hspace{1pt}\raisebox{0.75pt}{$\scriptstyle{\slash}$} \hspace{-6pt} D}
\def\bea{\begin{eqnarray}}
\def\eea{\end{eqnarray}}
\def\be{\begin{equation}}
\def\ee{\end{equation}}
\def\nn{\nonumber}

\def\a{&}
\def\Z{{\bf Z}}

\let\trimmarks

\makeatletter
\def\leaderfill{\leaders\hbox to 1em{\hss.\hss}\hfill}
\newcommand\@pnumwidth{2.55em}
\newcommand\@tocrmarg{2.55em}
\newcommand\@dotsep{4.5}
\newcommand\tableofcontents{
{\global
\@topnum\z@
\@afterindentfalse
\if@twocolumn
\@restonecoltrue
\onecolumn
\else
\@restonecolfalse
\fi
\vspace*{10pt}
\noindent
{\bf Contents}\par
\vskip1em
\nobreak}
{\small
\@starttoc{toc}
}\if@restonecol
\twocolumn
\fi}
\newcommand*\l@section[2]{
\ifnum \c@tocdepth >\z@
\addpenalty\@secpenalty
\setlength\@tempdima{1.4em}
\begingroup
\parindent \z@
\rightskip
\@pnumwidth \parfillskip -\@pnumwidth
\leavevmode
\advance\leftskip\@tempdima
\hskip -\leftskip
#1\nobreak\leaderfill\nobreak
\hb@xt@\@pnumwidth{\hss #2}\par
\endgroup
\fi}
\newcommand*\l@subsection{\@dottedtocline{2}{1.4em}{2.2em}}
\newcommand*\l@subsubsection{\@dottedtocline{2}{3.6em}{3em}}
\newcommand*\l@appendix{\@dottedtocline{2}{0em}{6.2em}}
\def\numberline#1{\hb@xt@\@tempdima{#1.\hfil}}
\makeatother

\begin{document}

% elimina la scritta appendix nel titolo delle appendici %
\renewcommand\appendixname{\hspace{-3pt}}

% fa alte anche le citazioni nelle frasi %
%\renewcommand\refcite{$\!\!$\cite}

\markboth{C.~A.~Scrucca and M.~Serone}
{Anomalies in field theories with extra dimensions}

% Aggiunge la singla della rivista all'inizio %
%\catchline{}{}{}

\title{\vspace{-65pt}\flushright{\tiny \em CERN-PH-TH/2004-041 \\[-2mm]
SISSA-17/2004/EP}\\ \vspace{40pt}
Anomalies in field theories with extra dimensions}

\author{CLAUDIO A. SCRUCCA}

\address{Physics Department, Theory Division, CERN, CH-1211 Geneva 23, Switzerland}

\author{MARCO SERONE}

\address{ISAS-SISSA and INFN, Via Beirut 2-4, I-34013 Trieste, Italy}

\vspace{-50pt}

\maketitle

% aggiunge le date di pubblicazione %
%\pub{Received (Day Month Year)}{Revised (Day Month Year)}

\begin{abstract}
We give an overview of the issue of anomalies in field theories with extra dimensions.
We start by reviewing in a pedagogical way the computation of the standard perturbative
gauge and gravitational anomalies on non-compact spaces, using Fujikawa's approach and
functional integral methods, and discuss the available mechanisms for their cancellation.
We then generalize these analyses to the case of orbifold field theories with compact
internal dimensions, emphasizing the new aspects related to the presence of orbifold
singularities and discrete Wilson lines, and the new cancellation mechanisms that are
becoming available. We conclude with a very brief discussion on global and parity anomalies.
\end{abstract}

\tableofcontents

\section{Introduction}

A symmetry of a classical description of a physical system is said to be anomalous if
it cannot be promoted to a symmetry of the full quantum description of the same system.
More precisely, anomalies can arise in quantum field theories whenever certain divergent
amplitudes --- typically involving massless chiral fermions in the internal lines --- do not
admit a regulator that is manifestly compatible with all the symmetries of the external
currents. There are several important symmetries that can be affected by anomalies.
When these are global symmetries of the system, as for example the axial symmetry in
QCD, nothing dramatic happens, although the anomaly has in general important physical
consequences. On the other hand, if the symmetry in question is local, as for example
a gauge symmetry, the anomaly makes in general the theory inconsistent. Indeed, local
symmetries do not merely represent an invariance under a group of symmetry transformations,
but rather a redundancy of the theory, redundancy that is generally necessary to decouple
unphysical states from the theory. If a local symmetry is anomalous, such a decoupling is
not possible and as a consequence the theory is, in general, inconsistent at the quantum
level. Because of the exchange of unphysical states in internal loop amplitudes, for
instance, perturbative unitarity is lost. Anomalies in local symmetries must therefore
be avoided.

Although anomalies technically originate from the bad UV behavior of certain
parity-violating amplitudes, they are well-defined IR effects that are always calculable
and finite. This is clear from the fact that, by definition, they cannot be canceled by
adding local counterterms to the effective action. In renormalizable field theories, the
only way to avoid anomalies in local symmetries is to require that the various contributions
they receive from the elementary fields of the theory should cancel against each other.
This severely constrains the spectrum of chiral fermions.
In non-renormalizable effective field theories, on the contrary, there exists the possibility
of canceling a non-vanishing anomaly by adding non-renormalizable, gauge-variant, operators.
These are typically Wess--Zumino counterterms that parametrize the effect of the massive
degrees of freedom that have been integrated out. Furthermore, by adding to the theory
antisymmetric tensor fields, one can cancel by means of the Green--Schwarz mechanism certain
kinds of anomalies. The degree of the restrictions imposed on a
theory by anomalies therefore depends significantly on whether this theory is renormalizable
and can be valid at all the energy scales, such as gauge theories in $4$ dimensions, or
non-renormalizable and valid as an effective description of a more fundamental theory
only below a certain maximal energy, such as gauge theories in more than $4$ dimensions.

Higher-dimensional field theories have received renewed interest during the last few years.
It has become clear that many peculiar physical phenomena associated to extra dimensions
can be efficiently described with a higher-dimensional effective field theory. This has
led to the concrete implementation, in phenomenological model building, of many interesting
concepts involving extra dimensions taken from string theory. The most relevant example
of these is without doubt the orbifold construction\cite{orb}, which provides an extremely
useful way of exploiting the interesting features of curved spaces in a context that remains
tractable and allows for concrete computations. Another very interesting example is the
idea of symmetry and supersymmetry breaking through twisted boundary conditions\cite{SS},
or equivalently the idea of Wilson lines\cite{hos}. It is mainly in the context of field
theories with orbifold and Wilson-line projections that most of the recent developments in
the phenomenological applications of extra dimensions have emerged. Let us mention, most
notably, the construction of grand unified theories with very economical symmetry breaking
and natural doublet--triplet splitting\cite{kaw,gut6}, ultra-softly broken and very constrained
supersymmetric models\cite{ultra,bhn}, and theories with gauge--Higgs unification and protected
effective potential\cite{hil,gh6}; see {\em e.g.} Ref.~\refcite{reviewextra} for an overview
of these developments and a complete list of references. This kind of models must be understood
as low-energy effective field theories and need in principle a consistent UV completion.
However, since anomalies are an IR effect, the issue of their cancellation must be addressed
already at the level of this effective description.

In the context of orbifold models, where a higher-dimensional gauge symmetry group is broken
down to a $4$-dimensional gauge symmetry subgroup, it will be important to distinguish
between exact and spontaneously broken symmetries. Indeed, an anomaly in an exact local
symmetry induces a breakdown of unitarity at all energy scales, whereas an anomaly in a
spontaneously broken local symmetry induces a breakdown of unitarity only above the
symmetry-breaking scale. The former case is always unacceptable, as it makes the theory
inconsistent. The latter case might instead be naturally tolerated in a low-energy effective
theory whose validity range and perturbative unitarity are anyhow bounded by a physical cut-off.
Indeed, as mentioned, it is always possible in such a situation to add a local Wess--Zumino
counterterm that
cancels the anomaly and influences physics only above the symmetry-breaking scale. In the
light of this general discussion, it might seem irrelevant to analyze anomaly cancellation
in the higher-dimensional theory, and unnecessary to require that all the local symmetries,
both those that are broken and those that are unbroken from the $4$-dimensional point of
view, be anomaly free. More precisely, a higher-dimensional anomaly that integrates to zero
from the $4$-dimensional point of view would correspond, after the standard Kaluza--Klein
decomposition, to an anomaly in the infinitely many local symmetries that are spontaneously
broken in the compactification process, and as such it can always be cured with a suitable
Wess--Zumino counterterm. However, it is important to remark in this respect that
higher-dimensional field theories do generally present extra global or local symmetries
involving the internal dimensions, and it is therefore a non-trivial issue to understand
whether the Wess--Zumino counterterms, needed to cancel the anomalies, respect or not such
symmetries. A convenient way to tackle this problem is to compute the anomalies directly in
the higher-dimensional theory and with a regulator that manifestly respects the internal
symmetries of the compact space. In this way it becomes clear that the Wess--Zumino
counterterms that cancel the anomalies always arise from a microscopic Lagrangian
involving a generalized Green--Schwarz mechanism. The latter, however, can generally
give rise also to massless scalar particles, with axion-like couplings; it is thus
not totally equivalent to adding Wess--Zumino counterterms, even at low energies.

The aim of this work is to give a comprehensive and pedagogical review of the computation
of gauge and gravitational anomalies, and the corresponding anomaly cancellation mechanisms,
in higher-dimensional field theories, focusing in particular on orbifold field theories. We
will follow Fujikawa's approach\cite{fuj}, in which anomalies are due to the non-invariance
of the integration measure in the functional integral defining the quantum effective action.
The variation of the measure encoding the anomaly is properly defined through a heat kernel
regularization, and is then computed using path-integral techniques as the high-temperature
limit of the partition function of a suitable supersymmetric quantum mechanical system, as
in Ref.~\refcite{AGW}. We mostly follow Ref.~\refcite{AGG} for the standard
characterization of anomalies, and in particular for the discussion of the Wess--Zumino
consistency conditions\cite{WZ} and their general solution through the Stora--Zumino descent
relations\cite{SZ}. We will not comment much on the many interesting mathematical aspects of
anomalies and their relations with index theorems\cite{Atiyah:tf}, but rather refer the
interested reader to other works, {\em e.g.} Ref.~\refcite{AGerice}, for an introduction to
these more formal aspects of anomalies.

The paper is organized as follows. In section 2 we review the standard computation of
perturbative gauge and gravitational anomalies in even-dimensional non-compact spaces.
We start with a general discussion of the Wess--Zumino consistency conditions and the
Stora--Zumino descent relations. We then compute the chiral anomaly for a Dirac spinor
in even dimensions, and finally deduce from it the gauge and gravitational anomalies of
chiral fermions by means of the Stora--Zumino descent procedure. In section 3 we review
the known anomaly cancellation mechanisms in any even number of dimensions. We discuss
the Green--Schwarz mechanism for the cancellation of reducible anomalies, as well as the
Wess--Zumino counterterms for the cancellation of anomalies in spontaneously broken gauge
symmetries. In section 4 we consider orbifold field theories and compute the gauge and
gravitational anomalies for the simplest examples of such spaces using the same formalism
as employed in section 2 for non-compact spaces, which manifestly respects the internal
symmetries of the compact space. Particular attention is devoted to the study of how
discrete Wilson lines modify the form of gauge and gravitational anomalies on orbifolds.
In section 5 we analyze the generalization to orbifolds
of the Green--Schwarz anomaly cancellation mechanism discussed in section 3, emphasizing
the new possibilities arising as a consequence of the compactness of the internal space
and the localization of anomalies at orbifold singularities. In section 6 we briefly review
global and parity anomalies and discuss their relevance to the consistency of orbifold
field theories. In appendix A we describe in some detail the Stora--Zumino descent procedure
that connects gauge and gravitational anomalies to chiral anomalies. In Appendix B we
end by giving some details about the derivation of the supersymmetric quantum mechanical
system that is used throughout all the review to compute anomalies.

\section{Perturbative anomalies}

The presence of perturbative gauge anomalies in $2n>4$ space-time dimensions was first
established in Ref.~\refcite{anohigher} by computing the $(n+1)$-point function at $1$
loop, which generalizes the classic triangular graph in $4$ space-time dimensions\cite{ABJ}.
Subsequently, it was pointed out in Ref.~\refcite{AGW} that theories in an even number of
dimensions, with Weyl fermions of spin $1/2$ or $3/2$, are generically afflicted by gauge
and/or gravitational anomalies.\footnote{In $4k+2$ dimensions, neutral self-dual and
antiself-dual antisymmetric tensor fields can also contribute to purely gravitational
anomalies, but we will not discuss this possibility.} Since then, a lot of progress has been
done in understanding the structure and geometrical properties of perturbative gauge and
gravitational anomalies in an arbitrary even number of dimensions, and a very elegant and
general mathematical characterization is now available; see for instance Ref.~\refcite{AGG}
for a review on the subject. In particular, it has become evident that anomalies are best
described in Euclidean space, where they affect only the imaginary part of the effective
action $\Gamma$ induced by chiral fermions\cite{AGW}. Throughout this review, we will
thus consider Euclidean spaces and only briefly comment on the continuation to Minkowski
spaces at the end of this section.

The structure of the anomalies that can occur in local symmetries is
strongly constrained by the group structure of these symmetry transformations. In
particular, two successive transformations $\delta_{\epsilon_1}$ and $\delta_{\epsilon_2}$
with parameters $\epsilon_1$ and $\epsilon_2$ must satisfy the basic property:
$\big[\delta_{\epsilon_1},\delta_{\epsilon_2}\big] = \delta_{[\epsilon_1,\epsilon_2]}$.
This implies that any anomaly ${\cal I}(\epsilon) = \delta_\epsilon \Gamma$ arising
from a non-trivial variation of the effective action $\Gamma$ under a local symmetry
transformation with parameter $\epsilon$, must satisfy the so-called Wess--Zumino
consistency condition\cite{WZ}:
\be
\delta_{\epsilon_1} {\cal I}(\epsilon_2) - \delta_{\epsilon_2} {\cal I}(\epsilon_1)
= {\cal I}([\epsilon_1,\epsilon_2]) \,.
\label{WZ2}
\ee
The general solution of this consistency condition can be characterized in
an elegant way in terms of a $(2n+2)$-form with the help of the so-called
Stora--Zumino descent relations\cite{SZ}. For any local symmetry
with transformation parameter $\epsilon$ (a $0$-form), connection $a$ (a $1$-form)
and curvature $f$ (a $2$-form), these are defined as follows. Starting from
a generic closed and invariant $(2n+2)$-form $\Omega_{2n+2}(f)$, one can define
an equivalence class of Chern--Simons $(2n+1)$-forms $\Omega_{2n+1}^{(0)}(a,f)$
through the local decomposition $\Omega_{2n+2} = d \Omega_{2n+1}^{(0)}$. This specifies
$\Omega_{2n+1}^{(0)}$ only modulo exact $2n$-forms, implementing the redundancy
associated to the local symmetry under consideration. One can then define yet
another equivalence class of $2n$-forms $\Omega_{2n}^{(1)}(\epsilon,a,f)$, modulo
exact $(2n-1)$-forms, through the transformation properties of the Chern--Simons
form under a local symmetry transformation: $\delta_\epsilon \Omega_{2n+1}^{(0)}
= d \Omega_{2n}^{(1)}$. It is the unique integral of this class of $2n$-forms
$\Omega_{2n}^{(1)}$ that gives the relevant general solution of (\ref{WZ2}):
\be
{\cal I}(\epsilon) = 2 \pi i \int_{M_{2n}} \!\!\!\! \Omega_{2n}^{(1)}(\epsilon) \;.
\label{G1}
\ee
To understand this, it is convenient to introduce the analog of the above descent
relations for manifolds. Starting from the $2n$-dimensional space-time manifold $M_{2n}$,
we define the equivalence class of $(2n+1)$-dimensional manifolds $M_{2n+1}^\prime$ whose
boundary is $M_{2n}$: $M_{2n} = \partial M_{2n+1}^\prime$, and similarly the equivalence
class of $(2n+2)$-dimensional manifolds $M_{2n+2}^{\prime\prime}$ such that
$M_{2n+1}^\prime = \partial M_{2n+2}^{\prime\prime}$. Both $M_{2n+1}^\prime$ and
$M_{2n+2}^{\prime\prime}$ are defined modulo components with no boundary, on
which exact forms integrate to zero. Since $\int_{M_{2n}} \Omega_{2n}^{(1)}(\epsilon)
= \delta_\epsilon \int_{M_{2n+1}^\prime} \!\!\! \Omega_{2n+1}^{(0)}
= \delta_\epsilon \int_{M_{2n+2}^{\prime\prime}} \!\!\! \Omega_{2n+2}$,
eq.~(\ref{G1}) can be rewritten as
\be
{\cal I}(\epsilon)
= 2 \pi i\, \delta_\epsilon \int_{M_{2n+1}^{\prime}} \hspace{-11pt} \Omega_{2n+1}^{(0)}
= 2 \pi i\, \delta_\epsilon \int_{M_{2n+2}^{\prime\prime}} \hspace{-11pt} \Omega_{2n+2} \;.
\label{G2}
\ee
It is clear that (\ref{G2}) provides a solution of (\ref{WZ2}), since it is manifestly
in the form of the variation of some functional under the symmetry transformation.
In fact, the descent construction also insures that (\ref{G2}) actually characterizes
the most general non-trivial solution of (\ref{WZ2}), modulo possible local
counterterms. This can be understood more precisely within the BRST formulation of
the Stora--Zumino descent relations, which is described in Appendix A. In that language,
the Wess--Zumino consistency condition implies that anomalies must be associated to
BRST-closed objects, whereas trivial anomalies corresponding to the variation of local
counterterms are associated to BRST-exact objects. The non-trivial solutions of the
consistency conditions are therefore characterized by the BRST cohomology, which turns
out to be automatically selected by the descent construction.

The above reasoning applies equally well to local gauge symmetries, associated to a
connection $A$ with curvature $F$, and to diffeomorphisms, associated to the
spin-connection $\omega$ with curvature $R$, and shows that gauge and gravitational
anomalies in a $2n$-dimensional theory are characterized by a gauge-invariant
$(2n+2)$-form, which suggests a $(2n+2)$-dimensional interpretation. A deeper analysis
shows indeed that the gauge and gravitational anomalies in $2n$ dimensions are related
to the gauge and gravitational contributions to the chiral anomaly in $2n+2$
dimensions\cite{AG-G2}. We shall see that the chiral anomaly in $2n+2$ dimensions is an
integer topological index given by the integral of some gauge-invariant form
$\Omega_{2n+2}(F,R)$ constructed out of the gauge and gravitational curvature
$2$-forms $F$ and $R$, ${\cal Z} = \int_{M_{2n+2}} \!\! \Omega_{2n+2}$. The gauge
and gravitational anomalies in $2n$ dimensions are then obtained through the
descent procedure, defined respectively with respect to gauge transformations
and diffeomorphisms, as ${\cal I} = 2 \pi i \int_{M_{2n}} \!\! \Omega_{2n}^{(1)}$.
A detailed derivation of the Stora--Zumino descent relations for the gauge case is
reported in Appendix A.

%%%%
In the following, we review in some detail the computation of the perturbative gauge
and gravitational contributions to the chiral anomaly for a Dirac fermion of spin 1/2 in an
arbitrary smooth Euclidean $2n$-dimensional manifold, following closely
section 11 of Ref.~\refcite{AGW}. We adopt Fujikawa's approach\cite{fuj}, in which the anomaly is
identified with the variation of the measure in the functional integral defining the
effective action. Using the relation between chiral and gauge/gravitational anomalies,
we can then easily deduce the latter. As we will see, this way of computing anomalies
is also easily generalized to study localized perturbative anomalies in
orbifold field theories. Fermions of spin 3/2 can be similarly analyzed, along the lines
of Ref.~\refcite{AGW}.
%%%%

\subsection{Chiral anomalies}

Let $\psi_A(x)$ be a massless Dirac fermion defined on a $2n$-dimensional manifold $M_{2n}$
with Euclidean signature $\delta_{\mu\nu}$ ($\mu,\nu=1,\ldots,2n$), and in an arbitrary
representation ${\cal R}$ of a gauge group $G$ ($A=1,\ldots,{\rm dim}\,{\cal R}$). The
minimal coupling of the fermion to the gauge and gravitational fields is described by the
Lagrangian
\be
{\cal L} = e\,\bar \psi(x)_A i \gamma^\mu D^A_{\mu\;B} \,\psi^B(x)\,,
\label{FerLag}
\ee
where $e$ is the determinant of the vielbein $e_\mu^a$ defining the tangent space and
$\gamma^\mu$ are the curved $\gamma$-matrices. The covariant derivative is given by
\be
D^A_{\mu\;B} = \Big(\partial_\mu + \frac 18 \omega_\mu^{ab}
\big[\gamma_a, \gamma_b\big] \Big) \delta^A_{\;B}
+ A^\alpha_\mu\, T^A_{\alpha\;B} \,,
\ee
in terms of the gauge connection $A_\mu^\alpha$ and the anti-Hermitian generators
$T^A_{\alpha\;B}$ of the group $G$ in the representation ${\cal R}$
($\alpha = 1,\ldots,{\rm dim}\,G$) for the gauge part, and the torsion-free
spin-connection $\omega_\mu^{ab}$ and the tangent space $\gamma$-matrices $\gamma_a$
satisfying the anticommutation relations $\{\gamma_a,\gamma_b\}=2 \delta_{ab}$
($a,b=1,\ldots,2n$) for the gravitational part.

The classical Lagrangian (\ref{FerLag}) is invariant under the global chiral
transformation $\psi \rightarrow e^{i \alpha \gamma_{2n+1}} \psi$, where
$\gamma_{2n+1}=i^n\prod_{a=1}^{2n} \gamma_a$ is the chirality matrix in $2n$
dimensions, normalized so that $\gamma_{2n+1}^2 = I$ for any $n$, and $\alpha$
is a constant parameter. This implies that the chiral current
$J^\mu_{2n+1}=\bar \psi_A \gamma_{2n+1}\gamma^\mu \psi^A$ is classically
conserved. At the quantum level, however, this conservation law can be violated
and turned into an anomalous Ward identity. To derive it, we consider the quantum
effective action $\Gamma$ defined by
\be
e^{-\Gamma(e,\omega,A)}=\int{\cal D}\psi{\cal D}\bar\psi\,
e^{-\int\! d^{2n}\!x \,{\cal L}}\,,
\label{effectiveA}
\ee
and study its behavior under an infinitesimal chiral transformation of the
fermions, with a space-time-dependent parameter $\alpha(x)$, given by\footnote{For
simplicity of the notation, we omit the gauge index $A$ in the following equations.
It will be reintroduced later on in this section.}
\be
\delta_\alpha \psi = i \alpha \gamma_{2n+1} \psi\,, \ \ \
\delta_\alpha \bar \psi = i \alpha \bar\psi\gamma_{2n+1}\,.
\label{chiraltran}
\ee
Since the external fields $e$, $\omega$ and $A$ are inert, the transformation
(\ref{chiraltran}) represents a redefinition of dummy integration variables, and
should not affect the effective action: $\delta_\alpha \Gamma = 0$. This statement carries
however a non-trivial piece of information, since neither the action nor the integration
measure is invariant under (\ref{chiraltran}). The variation of the classical action under
(\ref{chiraltran}) is non-vanishing only for non-constant $\alpha$, and has the form
$\delta_\alpha \int {\cal L} = \int\! J^\mu_{2n+1} \partial_\mu \alpha$. The variation
of the measure is instead always non-vanishing, because the transformation (\ref{chiraltran})
leads to a non-trivial Jacobian factor, which has the form
$\delta_\alpha [{\cal D}\psi{\cal D}\bar\psi] = \exp \{- 2 i \int \alpha {\cal A}\}$,
as we will see below. In total, the effective action therefore transforms as
\be
\delta_\alpha \Gamma = \int\!d^{2n}\!x e\, \alpha (x)
\Big[2i {\cal A}(x)-\langle  \partial_\mu J^\mu_{2n+1}(x) \rangle \Big]\,.
\ee
The condition $\delta_\alpha \Gamma = 0$ then implies the anomalous Ward identity:
\be
\langle \partial_\mu J^\mu_{2n+1} \rangle = 2i {\cal A}\,.
\label{chiral-ano}
\ee
In order to compute the anomaly ${\cal A}$, we need to define the integration measure more
precisely. This is best done by considering the eigenfunctions of the Dirac operator
$i\Dslash \equiv i e^\mu_a \gamma^a D_\mu$. Since the latter is Hermitian, the set of
its eigenfunctions $\psi_k(x)$ with eigenvalues $\lambda_k$, defined by
$i\Dslash \psi_n = \lambda_n \psi_n$, form an orthonormal and complete basis of
spinor modes:
\be
\int\! d^{2n}\!x \,e \,\psi_k^\dagger(x) \psi_l(x) = \delta_{k,l}\,,\;\;
\sum_k e\, \psi_k^\dagger(x) \psi_k(y) = \delta^{(2n)}(x-y)\,.
\ee
The fermion fields $\psi$ and $\bar\psi$, which are independent from each other
in Euclidean space, can be decomposed as
\be
\psi = \sum_k a_k \psi_k\,, \ \ \
\bar\psi = \sum_k \bar b_k \psi_k^\dagger \,,
\ee
so that the measure becomes
\be
{\cal D}\psi{\cal D}\bar\psi=\prod_{k,l} da_k d\bar b_l\,.
\label{mis}
\ee
Under the chiral transformation (\ref{chiraltran}), we have
\be
\delta_\alpha a_k = i \int d^{2n}\!x\,e\,
\sum_l \psi_k^\dagger \alpha \gamma_{2n+1} \psi_l a_l\,,\;\;
\delta_\alpha \bar b_k = i \int d^{2n}\!x\,e\,
\sum_l \bar b_l \psi_l^\dagger \alpha \gamma_{2n+1} \psi_k\,,
\ee
and the measure (\ref{mis}) transforms as
\be
\delta_\alpha \Big[{\cal D}\psi{\cal D}\bar\psi\Big] =
{\cal D}\psi{\cal D}\bar\psi \exp \bigg\{\-\!- 2i \sum_k\int\! d^{2n}\!x\,e\,
\psi_k^\dagger \alpha \gamma_{2n+1}\psi_k\bigg\}\,.
\label{varmeasure}
\ee
For simplicity we can take $\alpha$ to be constant. It is clear that this formal
expression is ill-defined as it stands, since it decomposes into a vanishing trace
over spinor indices (${\rm tr}\,\gamma_{2n+1} = 0)$ times an infinite sum over the modes
($\sum_k 1 = \infty$). A convenient way of regularizing this expression is to introduce a
gauge-invariant Gaussian cut-off. The integrated anomaly ${\cal Z} = \int {\cal A}$
can then be defined as
\bea
{\cal Z} \a=\a \lim_{\beta \rightarrow 0}
\sum_k \psi_k^\dagger \gamma_{2n+1}\psi_k e^{-\beta \lambda_n^2/2} \nn \\
\a=\a  \lim_{\beta \rightarrow 0} {\rm Tr}\,
\Big[\gamma_{2n+1} e^{-\beta (i\Dslashs)^2/2}\Big]\,,
\label{anomalyDef}
\eea
where the trace has to be taken over the mode and the spinor indices, as well as
over the gauge indices.

Our next task is to compute ${\cal Z}$, as defined in (\ref{anomalyDef}). We notice that,
by definition, this is equivalent to computing the high-temperature limit of the partition
function of a system with Hamiltonian $H=(i\Dslash)^2/2$ and density matrix
$\rho=\gamma_{2n+1}e^{-\beta H}$: ${\cal Z} = {\rm Tr}\,\rho$. Luckily, a system with
such properties can be obtained by dimensional reduction to $1$ dimension of a particular
$2$-dimensional supersymmetric $\sigma$-model (see Appendix B). The Lagrangian of this
supersymmetric quantum mechanics is
\bea
{\cal L} = \a\a \frac 12 g_{\mu\nu} \dot x^\mu \dot x^\nu
+ \frac i2 \psi_a \dot \psi^a +
\frac{i}4 \big[\psi_a,\psi_b\big] \omega_\mu^{ab} \dot x^\mu  \nn \\
\a\a +\, i c_A^\star \Big(\dot c^A + A^A_{\mu\,B}\,\dot x^\mu c^B \Big)
+ \frac 12 c_A^\star c^B \psi^a \psi^b F_{ab\;B}^{A}\,,
\label{sqm1}
\eea
where the dot stands for the time derivative, $x^\mu$ are bosonic fields that take
values on the manifold $M_{2n}$, $\psi^a$ are fermionic fields constructed from
the superpartners of $x^\mu$ as $\psi^a = e_\mu^a \psi^\mu$ and finally
$c_A^\star$ and $c^A$ are a set of complex fermionic fields transforming respectively
in the representation ${\cal R}$ and its complex conjugate $\bar {\cal R}$ of the
gauge group $G$. The spin-connection and the gauge connection are precisely those
appearing in (\ref{FerLag}), whereas we have $F_{ab\;B}^A =e_a^\mu e_b^\nu F_{\mu\nu\;B}^A$,
with $F_{\mu\nu\;B}^A$ as in (\ref{F-Def}).

The Lagrangian (\ref{sqm1}) can be canonically quantized. The (anti)commutation
relations between the fields $x^\mu$, $c^A$ and their canonical momenta are given by
\be
\big[\pi_\mu, x^\nu\big] = -i \delta_\mu^\nu \,,\;\;
\big\{c_A^\star,c^B\big\} = \delta_A^B\,,
\label{comm-rel-1}
\ee
where the momentum $\pi_\mu$ conjugate to $x^\nu$ is found to be
\be
\pi_{\mu} = g_{\mu\nu} \dot x^\nu + \frac i4 \big[\psi_a,\psi_b\big] \omega_\mu^{ab}
+ i c_A^\star A^A_{\mu\;B} c^B  \,.
\ee
The situation for the fields $\psi^a$ is slightly more involved, because they
coincide with their canonical conjugate momenta, $\pi_\psi^a=i \psi^a/2$, and
this leads to an inconsistent set of anticommutation relations. This problem is
easily solved by considering Dirac brackets with the second class
constraints $\pi_\psi^a - i \psi^a/2=0$. In this way, one finds the consistent
commutation relations
\be
\big\{\psi_a,\psi_b\big\}_D = \delta_{ab} \,,\;\;
\big\{\pi_\psi^a,\pi_\psi^b\big\}_D = -\frac 14 \delta^{ab}\,,\;\;
\big\{\pi_\psi^a,\psi_b\big\}_D = \frac{i}2 \delta^a_b\,.
\label{comm-rel-2}
\ee
The relations (\ref{comm-rel-1}) and (\ref{comm-rel-2}) can be realized on a
standard Hilbert space by identifying $\pi_\mu \rightarrow -i \partial_\mu$,
$\psi^a\rightarrow \gamma^a/\sqrt{2}$ and $c_A^\star T^A_{\alpha\;B} c^B
\rightarrow T_\alpha$. The supercharge (\ref{supercharge}) then becomes
\be
Q = \frac{i}{\sqrt{2}} e_a^\mu \gamma^a \bigg[
\Big(\partial_\mu + \frac 18 \omega_\mu^{ab} \big[\gamma_a, \gamma_b\big] \Big)
\delta^A_{\;B} +  A^\alpha_\mu\, T^A_{\alpha\;B}\bigg] =
\frac{1}{\sqrt{2}}\,i \Dslash \,,
\ee
where $i\Dslash$ is exactly the covariant derivative entering in (\ref{FerLag}).
The Hamiltonian of the above quantum mechanical model, being supersymmetric,
is simply given by $H=Q^2$ and thus $H=(i\Dslash)^2/2$.

The integrated anomaly ${\cal Z}$ can now be derived by computing directly
the partition function of the above supersymmetric model in a Hamiltonian
formulation. In order to obtain the correct result, however, one has to pay
attention when tracing over the fermionic operators $c_A^\star$ and $c^A$.
Indeed, considering them as creation and annihilation operators, the Hilbert
space of the system consists of the vacuum $|0\rangle$, $1$-particle states
$c_A^\star |0\rangle$, $2$-particle states $c_A^\star c_B^\star |0\rangle$,
etc. Among all these states, only the $1$-particle states correspond to the
representation ${\cal R}$ of the gauge group $G$, the vacuum being a singlet
of $G$ and multiparticle states leading to tensor products of the representation
${\cal R}$. In order to compute the anomaly for a single spinor in the representation
${\cal R}$, it is therefore necessary to restrict the partition function to
$1$-particle states. This is most conveniently done by using a hybrid formulation,
which is Hamiltonian with respect to the fields $c_A^\star$ and $c^A$, and Lagrangian
with respect to the remaining fields $x^\mu$ and $\psi^a$. Technically, this can be
done by introducing the Routhian ${\cal R} = {\cal L} - i c_A^\star \dot c^A$.
After Wick-rotating to Euclidean time $\tau \rightarrow -i \tau$, the partition function
can then be written as
\be
{\cal Z} = {\rm Tr}_{c,c^\star} \int_P \! {\cal D} x^\mu \int_P \! {\cal D} \psi^{a}
\exp \bigg\{\!-\! \int_0^\beta \! d\tau\, {\cal R} \Big(x^\mu(\tau),\psi^{a} (\tau),
c_A^ \star, c^A \Big) \bigg\} \,.
\label{pathspinor}
\ee
%%%%
The subscript $P$ on the functional integrals stands for periodic boundary
conditions along the closed time direction $\tau$, ${\rm Tr}_{c,c^\star}$
represents the trace over the $1$-particle states $c_A^\star |0\rangle$, and the Euclidean
Routhian ${\cal R}$ is given by
\bea
{\cal R} \a=\a
\frac 12 g_{\mu\nu} \dot x^\mu \dot x^\nu + \frac 12 \psi_a \dot \psi^a +
\frac{1}4 \big[\psi_a,\psi_b\big] \omega_\mu^{ab} \dot x^\mu \nn \\
\a\;\a +\, c_A^\star A^A_{\mu\;B}\dot x^\mu  c^B
- \frac{1}2 c_A^\star c^B \psi^a \psi^b F_{ab\;B}^{A}\,.
\label{Rsqm}
\eea
We notice that the only effect of the matrix $\gamma_{2n+1}$ in the density matrix
$\rho$ is to change the boundary conditions of the fermionic fields $\psi^a$ from
antiperiodic --- the natural boundary conditions for fermions --- to
periodic (see {\it e.g.} Ref.~\refcite{luisbonn}
for more details).
%%%%
Equation (\ref{pathspinor}) should then be understood as follows: after integrating
over the fields $x^\mu$ and $\psi^a$, one gets an effective Hamiltonian $\hat H(c,c^\star)$
for the operators $c_A^\star$ and $c^A$, from which one finally computes
${\rm Tr}_{c,c^\star} e^{-\beta \hat H(c,c^\star)}$. Although this procedure looks
quite complicated, we will see that it drastically simplifies in the high-temperature
limit we are interested in. Since $\gamma_{2n+1}$ anticommutes with all fermionic fields,
it can be identified with the fermion-number operator $(-1)^F$. The anomaly is then given
by the Witten index\cite{wit} of the above-described supersymmetric quantum mechanical
model,
\be
{\cal Z} = {\rm Tr} \Big[ (-1)^F e^{-\beta H} \Big]\,.
\ee
This quantity is an integer number and coincides with the index of the Dirac operator.
Thanks to this interpretation, the anomaly can be computed in a mathematical way by
means of index theorems.

The path-integral in (\ref{pathspinor}) is dominated by constant paths in the limit
$\beta \rightarrow 0$. Its computation is greatly simplified by using the
background-field method and expanding around constant bosonic and fermionic configurations
in normal coordinates\cite{agfm}, defined around any point $x_0$ in a such a way that
the spin-connection (or equivalently, the Christoffel symbol), as well as its
symmetric derivatives, vanish at $x_0$.
It is also convenient to rescale $\tau \rightarrow \beta \tau$
and define
\be
x^\mu(\tau) = x_0^\mu + \sqrt{\beta}\, e^\mu_a (x_0) \xi^a(\tau)\,, \ \ \
\psi^a(\tau) = \frac 1{\sqrt{\beta}}\psi_0^a + \lambda^a(\tau)\,.
\ee
In this way, it becomes clear that it is sufficient to keep only quadratic terms in the
fluctuations, which have a $\beta$-independent integrated Routhian, since higher-order
terms in the fluctuations come with growing powers of $\beta$. We can therefore use a
quadratic effective Routhian given by
\bea
{\cal R}^{\rm eff} \a=\a \frac 12 \Big[\dot \xi_a \dot \xi^a + \lambda_{a} \dot \lambda^{a}
+ R_{ab}(x_0,\psi_0) \xi^{a} \dot \xi^{b} \Big] -c_A^\star F^A_{\;B}(x_0,\psi_0) c^B\,,
\label{Reff}
\eea
where
\bea
R_{ab}(x_0,\psi_0) \a = \a \frac 12 R_{abcd}(x_0) \psi_0^c \psi_0^d \,,\nn \\
F^A_{\;B}(x_0,\psi_0) \a = \a \frac 12 F_{ab~B}^A(x_0)  \psi_0^a \psi_0^b \,.
\eea
Since the fermionic zero modes $\psi_0^a$ anticommute with each other,\footnote{The
anticommuting $\psi_0^a$'s are simply Grassmann variables in a path integral and
should not be confused with the operator-valued fields $\psi^a$ entering in (\ref{sqm1}),
which satisfy the anticommutation relations appearing in (\ref{comm-rel-2}).} they define
a basis of differential forms on $M_{2n}$, and the above quantities behave as curvature
$2$-forms.

{}From (\ref{Reff}) we see that the gauge and gravitational contributions to the chiral
anomaly are completely decoupled. The former is determined by the trace over the
$1$-particle states $c_A^\star |0\rangle$, and the latter by the determinants arising
from the Gaussian path integral over the bosonic and fermionic fluctuation fields:
\be
{\cal Z} = \int\! d^{2n}\!x_0 \int\! d^{2n}\!\psi_0\,
{\rm Tr}_{c,c^\star} \Big[e^{c_A^\star F^A_{\;B}c^B}\Big]\,
{\rm det}^{-1/2}_P\Big[\!-\!\partial_\tau^2 \delta_{ab} + R_{ab} \partial_\tau\Big]\,
{\rm det}_P^{1/2}\Big[\partial_\tau \delta_{ab}\Big]\,.
\ee
The trace yields simply
\be
{\rm Tr}_{c,c^\star} e^{c_A^\star F^A_{\;B} c^B} = {\rm tr}_{\cal R} e^{F}\,.
\label{F-anomaly}
\ee
The determinants can be computed by decomposing the fields on a complete basis of
periodic functions of $\tau$ on the circle with unit radius, and
using the standard $\zeta$-function regularization. It is also useful to bring the
curvature $2$-form $R_{ab}$ into a block-diagonal form (this can always be done
by an appropriate rotation of the fields $\xi^a$) of the type
\bea
R_{ab}= \pmatrix{
0 \a \lambda_1 \a \a \a \a \cr
-\lambda_1 \a 0 \a \a \a \a \cr
\a \a ... \a \a \a \cr
\a \a \a 0 \a \lambda_n \a \cr
\a \a \a -\lambda_n \a 0 \a \cr }\,,
\label{matrix}
\eea
where $\lambda_i$ are skew-eigenvalues $2$-forms, so that the bosonic determinant
decomposes into $n$ distinct determinants with trivial matrix structure. Proceeding
in this way we find:
\bea
{\rm det}^{-1/2}_P\Big[\!-\!\partial_\tau^2 \delta_{ab} + R_{ab} \partial_\tau\Big] \a=\a
(2 \pi)^{-n} \prod_{i=1}^{n} \frac {\lambda_i/2}{\sin (\lambda_i/2)} \,,
\label{deter1} \\
{\rm det}_P^{1/2} \Big[\partial_\tau \delta_{ab}\Big] \a=\a (-i)^n \,.
\label{deter2}
\eea
The final result for the anomaly is obtained by putting together (\ref{F-anomaly}),
(\ref{deter1}) and (\ref{deter2}), and integrating over the zero modes.
The Berezin integral over the fermionic zero modes vanishes unless all of them
appear in the integrand, in which case it yields
\be
\int d^{2n}\!\psi_0\, \psi_0^{a_1} \ldots \psi_0^{a_{2n}}
= (-1)^n\,\epsilon^{a_1 \ldots a_{2n}} \,.
\label{intferm}
\ee
This amounts to selecting the $2n$-form component from the expansion of the integrand in
powers of the $2$-forms $F$ and $R$. Since this is a homogeneous polynomial of degree $n$
in $F$ and $R$, the factor $(i/(2\pi))^n$ arising from the normalization of (\ref{deter1}),
(\ref{deter2}) and (\ref{intferm}) amounts to multiplying $F$ and $R$ by $i/(2\pi)$.
The final result for the integrated anomaly can therefore be rewritten more concisely as:
\be
{\cal Z} = \int_{M_{2n}}\!\!\, {\rm ch}_{\cal R}(F) \,\hat A(R) \,,
\label{chiral-anomaly}
\ee
where
\bea
{\rm ch}_{\cal R}(F) \a=\a {\rm tr}_{\cal R} e^{i F/(2\pi)} \,,
\label{chern} \\
\hat A(R) \a=\a \prod_{i=1}^{n}\frac {\lambda_i/(4\pi)}{\sinh [\lambda_i/(4\pi)]}\,.
\label{roof}
\eea
The quantities (\ref{chern}) and (\ref{roof}) are characteristic classes constructed
from the curvature $2$-forms $F=1/2 F_{\mu\nu}dx^\mu\wedge dx^ \nu$ and
$R=1/2 R_{\mu\nu}dx^\mu\wedge dx^ \nu$, respectively the Chern character of the gauge
bundle and the Dirac genus of the tangent bundle. They must be understood as power
series in their arguments, whose leading terms are
\bea
\!\!\!{\rm ch}(F) \a=\a r + \frac i{2\pi} {\rm tr}\, F - \frac 1{2(2\pi)^2} {\rm tr}\, F^2
- \frac i{6(2\pi)^3} {\rm tr}\, F^3 + \frac 1{24(2\pi)^4} {\rm tr}\, F^4 + \ldots \,,
\label{ch-dec} \\
\!\!\!\hat A(R) \a=\a 1 + \frac{1}{(4\pi)^2}\frac{1}{12} {\rm tr}\, R^2 +
\frac{1}{(4\pi)^4} \bigg[\frac{1}{288} ({\rm tr}\,R^2)^2 + \frac{1}{360}{\rm tr}\,
R^ 4 \bigg] + \ldots \,,
\eea
where $r$ is the dimension of the representation ${\cal R}$.
In (\ref{chiral-anomaly}), as well as in all the analogous expressions
that will follow, it is understood that only the $2n$-form component
of the integrand has to be considered.

The anomalous Ward identity (\ref{chiral-ano}) for the chiral symmetry
can then be formally written, in any even dimensional space, as
\be
\langle \partial_\mu J^\mu_{2n+1} \rangle = 2i \,{\rm ch}_{\cal R}(F) \hat A(R)\,.
\label{chiral-ano2}
\ee
In $4$ space-time dimensions, for instance, (\ref{chiral-ano2}) gives
\be
\langle \partial_\mu J^\mu_{5} \rangle =
-i\epsilon^{\mu\nu\rho\sigma} \bigg[\frac{1}{16\pi^2}
{\rm tr} \,F_{\mu\nu}F_{\rho\sigma} +\frac{r}{384 \pi^2} R_{\mu\nu}^{\;\;\;\;\alpha\beta}
R_{\rho\sigma\alpha\beta}\bigg] \,.
\label{chiral-ano4}
\ee

\subsection{Gauge and gravitational anomalies}

Let us now turn to gauge and gravitational anomalies. In this case we are interested
to see whether the effective action defined in (\ref{effectiveA}) is invariant under
gauge and general coordinate (or equivalently local Lorentz) transformations.
Differently from the chiral anomaly, gauge and gravitational anomalies can arise only
from chiral fermions, and not from Dirac fermions. They can be computed with the same
method as used above for chiral anomalies, again along the lines of section 11 of
Ref.~\refcite{AGW}. Computed in this way, however, the anomaly does not automatically
satisfy the Wess--Zumino consistency conditions, because the Gaussian cut-off that is
used to regularize the computation does not preserve Bose symmetry. The resulting anomaly
controls the Ward identity of a classically conserved current or energy-momentum tensor
that differs from the usual one by a functional of the gauge or gravitational fields.
The anomaly associated to the standard currents is called ``consistent'', because it
satisfies the Wess--Zumino consistency conditions. The anomaly associated to the modified
current is instead called ``covariant'', because it transforms covariantly under the
local symmetry. These two forms of the anomaly contain the same information, and there
is a well-defined procedure to switch from one to the other\cite{AGG}. To put the
covariant anomaly emerging from a computation \`a la Fujikawa into a consistent form,
it is sufficient to take the leading-order part of it, which contains $n+1$ fields, add the
symmetrization factor required for a Bose-symmetric result, and then use the Wess--Zumino
consistency condition to reconstruct the subleading part. As already anticipated at the
beginning of this section, this leads to a very simple and nice result: the gauge and
gravitational anomalies induced by a chiral fermion are obtained by taking the appropriate
Stora--Zumino descents of the chiral anomaly induced by a Dirac fermion. To give the reader
a flavor of how this comes about, we shall sketch below how gauge and gravitational
anomalies can be computed \`a la Fujikawa.

The anomaly is by definition a variation of the quantum effective action under
a local symmetry transformation. Again, the only potential source of such a variation
is a Jacobian in the integration measure, since the classical action is invariant.
The computation is therefore technically analogous to the one we performed for the
chiral anomaly, except that the transformation law is now different and acts with
opposite signs on $\psi$ and $\bar \psi$. Moreover, in this case, the full Dirac
operator contains a projector ${\cal P}_\eta = 1/2(1 + \eta \gamma_{2n+1})$ on definite
chirality $\eta = \pm 1$, $i \Dslash_\eta = i \Dslash {\cal P}_\eta$, and is not Hermitian.
For this reason, we have to use the eigenfunctions $\phi^\eta_n$ of the Hermitian
operator $(i \Dslash_\eta)^\dagger i \Dslash_\eta$ to expand $\psi$ and the
eigenfunctions $\varphi^{\eta}_n$ of $i \Dslash_\eta (i \Dslash_\eta)^\dagger$ to
expand $\bar \psi$.

Let us first consider the case of gauge anomalies. Under an infinitesimal gauge
transformation with parameter $v_\alpha$, or $v^A_{~B} = v^\alpha T^A_{\alpha\;B}$,
the fermion fields transform as
\be
\delta_v \psi^A = - v^A_{~B} \psi^B\,,\;\;
\delta_v \bar \psi^A = v^A_{~B} \bar\psi^B\,.
\label{chiraltrangau}
\ee
This induces a variation of the integration measure given by
\be
\delta_v\Big[{\cal D}\psi{\cal D}\bar\psi\Big] =
{\cal D}\psi{\cal D}\bar\psi \exp \bigg\{\sum_k\int\! d^{2n}\!x\,e\,
\Big(\phi_k^{\eta \dagger} v_\alpha T^\alpha \phi_k^\eta
- \varphi_k^{\eta \dagger} v_\alpha T^\alpha \varphi_k^\eta \Big)
\bigg\}\,.
\label{varmeasuregau}
\ee
As in the case of the chiral anomaly, this formal expression needs to be regularized,
and we can define the integrated gauge anomaly to be
\bea
{\cal I}^{\rm gauge}(v) \a=\a -\lim_{\beta \rightarrow 0}
\sum_k \Big(\phi_k^\dagger v_\alpha T^\alpha
e^{-\beta (i\Dslashs_\eta)^\dagger i\Dslashs_\eta/2} \phi_k
- \varphi_k^\dagger v_\alpha T^\alpha
e^{-\beta i\Dslashs_\eta (i\Dslashs_\eta)^\dagger/2} \varphi_k \Big)\nn \\
\a=\a -\eta \lim_{\beta \rightarrow 0} {\rm Tr}\,
\Big[\gamma_{2n+1} Q^{\rm gauge} e^{-\beta (i\Dslashs)^2/2}\Big]\,.
\label{anomalyDefgau}
\eea
The operator $Q^{\rm gauge}$ is defined in such a way to act as $v_\alpha T^\alpha$
on the Hilbert space. A concrete realization of it within the supersymmetric quantum
mechanics introduced in the previous subsection is
\be
Q^{\rm gauge} \rightarrow c^*_A v^A_{~B} c^B\,.
\label{Qgau}
\ee
The computation of (\ref{anomalyDefgau}) is similar to that of (\ref{anomalyDef}).
The only difference is the insertion of the operator (\ref{Qgau}) into the trace
(\ref{F-anomaly}). It is clear from (\ref{F-anomaly}) that making this insertion is
equivalent to substituting $F \rightarrow F + v$ in the trace without insertion and
taking the linear part in $v$ of the result.
For the leading part of the result with $n+1$ fields, taking the linear part in
$v$ differs from taking the Stora--Zumino descent with respect to the gauge symmetry
just by the above-discussed Bose-symmetrizaton factor.\footnote{Consider for example
a term with $k+1$ gauge fields. The starting point is then the expression
$1/(k+1)!\, {\rm tr}\,F^{k+1}$. Shifting $F \rightarrow F + v$ and taking
the linear part in $v$ yields, at leading order in the gauge field $A$,
$1/k!\, {\rm tr}\,[v (d A)^k]$. Taking the Stora--Zumino descent of the original
expression yields instead, again at leading order in the gauge field $A$,
$1/(k+1)!\, {\rm tr}\,[v (d A)^{k}]$. This differs from the former by
the factor $1/(k+1)$, which is required to Bose-symmetrize the leg
on which the variation is implemented with the other $k$ identical legs.}
The non-linear completion specified by the Wess--Zumino consistency condition
is then just the non-linear completion of the Stora--Zumino descent. Taking into
account the factor $i/(2\pi)$ entering the definition of the Chern character, the
consistent form of the integrated gauge anomaly is therefore
\be
{\cal I}^{\rm gauge}(v) = 2\pi i \eta \int_{M_{2n}} \!\!
\Big[{\rm ch}_{\cal R}(F) \Big]^{(1)} \hat A(R) \,.
\label{Igauge}
\ee

The case of gravitational anomalies is similar. Under an infinitesimal diffeomorphism
with parameter $\epsilon^\mu$, the fermion fields transform, modulo a local
Lorentz transformation, as
\be
\delta_\epsilon \psi = - \epsilon^\mu D_\mu \psi\,, \ \ \
\delta_\epsilon \bar \psi = \epsilon^\mu D_\mu \bar\psi\,.
\label{chiraltrangra}
\ee
This induces the following variation of the integration measure:
\be
\delta_\epsilon \Big[{\cal D}\psi{\cal D}\bar\psi\Big] =
{\cal D}\psi{\cal D}\bar\psi \exp \Big\{\sum_k\int\! d^{2n}\!x\,e\,
\Big(\phi_k^{\eta \dagger} \epsilon^\mu D_\mu \phi_k^\eta
- \varphi_k^{\eta \dagger} \epsilon^\mu D_\mu \varphi_k^\eta \Big)
\Big\}\,.
\label{varmeasuregra}
\ee
The regularized expression for the integrated gravitational anomaly is then:
\bea
{\cal I}^{\rm grav}(\epsilon) \a=\a -\lim_{\beta \rightarrow 0}
\sum_k \Big(\phi_k^\dagger e^{-\beta (i\Dslashs_\eta)^\dagger i\Dslashs_\eta/2}
\epsilon^\mu D_\mu \phi_k
- \varphi_k^\dagger e^{-\beta i\Dslashs_\eta (i\Dslashs_\eta)^\dagger/2}
\epsilon^\mu D_\mu \varphi_k
 \Big)\nn \\
\a=\a -\eta \lim_{\beta \rightarrow 0} {\rm Tr}\,
\Big[\gamma_{2n+1} Q^{\rm grav} e^{-\beta (i\Dslashs)^2/2}\Big]\,.
\label{anomalyDefgra}
\eea
The operator $Q^{\rm grav}$ must act as $\epsilon^\mu D_\mu$ on the Hilbert space. Since
$i \dot x^\mu \rightarrow D^\mu$ upon canonical quantization, we can identify
\be
Q^{\rm grav} \rightarrow i \epsilon_\mu \dot x^\mu \,.
\label{Qgrav}
\ee
It is now convenient to exponentiate the action of $Q^{\rm grav}$. This is realized in
the Euclidean path-integral representation by adding to the Routhian (\ref{Rsqm}) the term
$\epsilon_\mu \dot x^\mu$. After going to normal coordinates, and retaining the leading
term in $\beta$, one realizes that the insertion of the exponentiated form of $Q^{\rm grav}$
amounts to the shift $R_{ab} \rightarrow R_{ab} + D_a \epsilon_b - D_b \epsilon_a$
in the functional integral without insertion. The original expression (\ref{anomalyDefgra})
is obtained by keeping only the linear piece in $\epsilon$. After adding the appropriate
symmetrization factors and following the same procedure as for the gauge anomaly to switch
to a consistent form of the anomaly, this implements the Stora--Zumino descent with respect
to diffeomorphisms. The consistent form of the integrated gravitational anomaly is finally
found to be
\be
{\cal I}^{\rm grav}(\epsilon) = 2\pi i \eta \int_{M_{2n}} \!\! {\rm ch}_{\cal R}(F)
\Big[\hat A(R) \Big]^{(1)} \,.
\label{Igrav}
\ee

The above results show that the quantum effective action $\Gamma(e,\omega,A)$ is not
invariant under gauge transformations and diffeomorphisms, and correspondingly the
standard gauge current and energy-momentum tensor are not conserved at the quantum level:
\bea
\delta_v \Gamma(e,\omega,A) \a=\a \int v_\alpha \langle D_\mu J^{\mu \alpha} \rangle
= {\cal I}^{\rm gauge}(v)\,,
\label{ano--gauge} \\
\delta_\epsilon \Gamma(e,\omega,A) \a=\a \int \epsilon_\nu \langle D_\mu T^{\mu\nu} \rangle
= {\cal I}^{\rm grav}(\epsilon)\,.
\label{ano--grav}
\eea
It is worth noticing that local Lorentz transformations can also suffer from anomalies,
resulting in an energy-momentum tensor that is not symmetric at the quantum level.
However, these are not independent from anomalies in diffeomorphisms, and it is always
possible to switch to a situation in which one or the other of the anomalies vanish,
but not both\cite{AGG}. Correspondingly, the energy-momentum tensor can be chosen to
be symmetric or conserved at the quantum level, but not both simultaneously.

The Minkowski form of the anomaly is easily derived by Wick rotation.
We find that $dx^{2n} \rightarrow i dx^{0}$, $A_{2n} \rightarrow -iA_{0}$,
$\omega_{2n} \rightarrow -i \omega_{0}$. Moreover, if $\epsilon^{12\ldots 2n}=1$,
after Wick rotation $\epsilon^{01\ldots 2n-1}=-1$, and hence
${\cal I}_M = - {\cal I}_E$. Since $\Gamma_M=i \Gamma_E$, one obtains
then eqs.~(\ref{ano--gauge}) and (\ref{ano--grav}) with an additional factor of
$i$ in front of the ${\cal I}$'s.

\section{Anomaly-cancellation mechanisms}

Anomalies occur as a consequence of UV divergences that cannot be regularized
without manifestly breaking a symmetry, but they represent a computable IR
effect that does not depend on the UV completion of the theory and persists
at any energy scale. As mentioned in the introduction, there is an extremely
important distinction between anomalies in fundamental theories, where the
full quantum effective action can be reliably computed, and effective theories,
where this cannot be done. In a fundamental theory, anomalies must cancel and
this puts very restrictive constraints on the allowed spectra of particles.
Examples of such theories are renormalizable field theories such as the standard
model or its minimal supersymmetric extension, or string theory.
In an effective theory, valid only up to some energy scale, the situation is
radically different. The quantum effective action consists of two parts: the
first is non-local and generated by the quantum fluctuations of the fields
described by the effective theory, and can lead to anomalies. The second is
related to the physics above the scale of validity of the theory and is
parametrized by an infinite set of local non-renormalizable operators with
dimensionful coefficients suppressed by inverse powers of this scale. There
is no reason for this part of the effective action to share the classical
symmetries of the effective theory, and it can therefore have a non-vanishing
variation. The variation of the total effective action can therefore be made
to vanish through a cancellation between the two parts of it related to physics
below and above the validity scale.

There are several interesting mechanisms through which an anomaly can cancel. In
the following, we will consider two general classes of these that are available
in effective theories: the Green--Schwarz mechanism\cite{GS} and, when a
spontaneous symmetry breaking occurs, the possibility of adding Wess--Zumino
counterterms to the action (see {\em e.g.} Refs.~\refcite{AGGW,CHZ,Preskill});
the generalization of these mechanisms to orbifold field theory models will
be discussed in section 5. Another relevant anomaly-cancellation mechanism,
which we will not discuss in this review, is the anomaly inflow
of Refs.~\refcite{Callan:sa}. This applies to theories where chiral
fermions are dynamically localized on space-time topological defects, as a
consequence of some non-trivial bosonic background, and consists in the
compensation of the localized anomaly that is induced by the fermions
through an inflow of anomaly from the bulk of the space-time. This inflow of
anomaly is provided by certain Wess--Zumino couplings of tensor fields with
the gauge and gravitational fields, similarly to what happens in the Green--Schwarz
mechanism. In fact, the two mechanisms are not completely independent; see
Refs.~\refcite{Izquierdo:st} for an analysis of the relations between them.

\subsection{The Green--Schwarz mechanism}

The Green--Schwarz anomaly cancellation mechanism was first discovered by
Green and Schwarz in the context of string-derived effective supergravity
theories in $10$ dimensions\cite{GS}. It achieves in a non-trivial
and interesting way the cancellation of gauge and gravitational anomalies,
which is guaranteed in the full string theory by its finiteness, stemming
from general principles such as modular invariance\cite{mod} or tadpole
cancellation\cite{tad}. Thanks to this mechanism, it has been understood
that field theories with an anomalous spectrum of massless fermions can be
anomaly-free, and thus consistent, in certain particular circumstances.
The mechanism involves antisymmetric tensor fields, and the essential idea
is that the anomaly is canceled by the gauge variation of some counterterms,
constructed out of these tensor fields as well as the gauge and gravitational
connections and field strengths.

Before describing the Green--Schwarz mechanism and its generalization to any
space-time dimension, it is necessary to introduce the notion of ``reducible''
and ``irreducible'' forms of the anomaly. As shown in (\ref{Igauge}) and
(\ref{Igrav}), a generic gauge or gravitational anomaly can be written in
the form ${\cal I} = 2 \pi i \int_{M_{2n}} \!\! \Omega_{2n}^{(1)}$, where
$\Omega_{2n}^{(1)}$ is the Stora--Zumino descent of a closed and gauge-invariant
$(2n+2)$-form $\Omega_{2n+2}$, function of the curvature $2$-forms $F$ and
$R$.\footnote{Here and in the following we will refer to gauge symmetries in
a broad sense, including in particular local Lorentz symmetries, in order to
treat gauge and gravitational anomalies at once.}
The form $\Omega_{2n+2}(F,R)$ is said to be ``irreducible'' when it cannot be
decomposed as a sum of products of closed and gauge-invariant forms of lower
degree. Typical examples are ${\rm tr}\,R^{n+1}$ or ${\rm tr}\,F^{n+1}$ for
a representation that does not admit a decomposition to lower forms. It is instead
said to be ``reducible'' when $\Omega_{2n+2}(F,R)$ can be decomposed as
$\Omega_{2n+2}=\Omega_{2k}\Omega_{2n+2-2k}$ for some $k$. Examples of such a
type are ${\rm tr}\,F^k {\rm tr}\, F^{n+1-k}$, ${\rm tr}\,F^{k} {\rm tr}\, R^{n+1-k}$
or ${\rm tr}\,R^{k} {\rm tr}\, R^{n+1-k}$.

The original Green--Schwarz mechanism in $10$ dimensions requires the introduction
of a $2$-index antisymmetric tensor field, but we will describe here its generalization
to $2n$ dimensions, where a $2l$-index antisymmetric tensor field of the type
$C_{2l}^{\mu_1 \dots \mu_{2l}}$ is needed\cite{Witten:dg}. These fields generalize
the standard electromagnetic vector potential\footnote{The ``electric'' and ``magnetic''
sources of these fields in $2n$ dimensions are respectively $(2l-1)$- and
$(2n-2l-3)$-dimensional extended objects.} and are conveniently described in terms of
$2l$-forms $C_{2l}$, subject to the $U(1)$ gauge transformation $\delta C_{2l}=d\lambda_{2l-1}$,
with $\lambda_{2l-1}$ an arbitrary $(2l-1)$-form. The gauge-invariant field strengths
$H_{2l+1}^{\mu_1 \dots \mu_{2l+1}} = \partial^{\mu_1} C_{2l}^{\mu_2 \dots \mu_{2l+1}}
\pm \mbox{permutations}$, are correspondingly described by the $(2l+1)$-forms
$H_{2l+1} = d C_{2l}$.

As will become clear below, only reducible anomalies can be canceled through the
Green--Schwarz mechanism. We shall therefore consider a generic reducible anomaly of
the form
\be
{\cal I}= 2\pi i \int_{M_{2n}} \!\!\!\! \Omega_{2n}^{(1)}\,,\;\;{\rm with}\;\;
\Omega_{2n+2}=\Omega_{2k}\,\Omega_{2n+2-2k}\,.
\label{GSano}
\ee
Following the Stora--Zumino descent procedure reported in Appendix A, the Chern--Simons
form $\Omega_{2n+1}^{(0)}$ corresponding to $\Omega_{2n+2}$ is found to be
\bea
\Omega_{2n+1}^{(0)} \a=\a \frac{k}{n+1} \Omega_{2k-1}^{(0)} \Omega_{2n+2-2k}
+\frac{n+1-k}{n+1}\Omega_{2k}\Omega_{2n+1-2k}^{(0)} \nn \\
\a\;\a +\, \alpha\, d\Big(\Omega_{2k-1}^{(0)} \Omega_{2n+1-2k}^{(0)}\Big)\,,
\label{SZ-GS0}
\eea
where $\alpha$ is an arbitrary parameter taking into account the ambiguity in the
definition of $\Omega^{(0)}_{2n+1}$. The numerical factors in (\ref{SZ-GS0}) are
fixed by Bose statistics, which implies treating all field strengths symmetrically.
>From (\ref{SZ-GS0}) one derives
\be
\Omega_{2n}^{(1)} = \bigg(\frac{k}{n+1}-\alpha\bigg) \Omega_{2k-2}^{(1)} \Omega_{2n+2-2k}
+\bigg(\frac{n+1-k}{n+1}+\alpha\bigg) \Omega_{2k}\Omega_{2n-2k}^{(1)}\,.
\label{SZ-GS1}
\ee
The anomaly corresponding to (\ref{SZ-GS1}) can be canceled by introducing a
$(2k-2)$-index tensor field $C_{2k-2}$ with the following action:
\bea
S = \int_{M_{2n}} \bigg[ \a\a \!\!
\frac 12 \Big|d C_{2k-2} + \sqrt{2\pi}\,\xi\,\Omega_{2k-1}^{(0)}\Big|^2
+ i\frac{\sqrt{2\pi}}{\xi}\,C_{2k-2}\,\Omega_{2n-2k+2} \nn \\
\a\a \!\! -2\pi i \bigg(\frac{n+1-k}{n+1}+\alpha\bigg)
\Omega_{2k-1}^{(0)}\Omega_{2n+1-2k}^{(0)} \bigg]\,.
\label{action1}
\eea
The field $C_{2k}$ has dimension $n-1$, and the forms $\Omega_{2l}$ and
$\Omega_{2l-1}^{(0)}$ have dimensions $2l$ and $2l-1$. The arbitrary parameter
$\xi$ therefore has dimension $n-2k+1$. The action (\ref{action1}) is not invariant
under local symmetry transformations. The modified kinetic term of the field
$C_{2k-2}$ makes it clear that the appropriate definition of its field strength
$H_{2k-1}$ is
\be
H_{2k-1} = dC_{2k-2} + \sqrt{2\pi}\,\xi\, \Omega_{2k-1}^{(0)}\,.
\label{H1}
\ee
This field strength can be made gauge-invariant, provided that $C_{2k-2}$
transforms inhomogeneously under gauge transformations, in such a way as to
compensate the transformations of the Chern--Simons form $\Omega_{2k-1}^{(0)}$:
\be
\delta_\epsilon C_{2k-2} = - \sqrt{2\pi}\,\xi\, \Omega_{2k-2}^{(1)} \,.
\label{C2k-2tra}
\ee
In this way $\delta_\epsilon H_{2k-1}=0$ and the kinetic term is therefore
invariant. However, the Wess--Zumino coupling $C_{2k-2}\Omega_{2n-2k+2}$ and
the last counterterm in (\ref{action1}) transform non-trivially and lead to
a non-vanishing variation of $S$ that exactly compensates for the $1$-loop
anomaly (\ref{GSano}), independently of the value of $\xi$:
\be
\delta_\epsilon S = -2\pi i \,\int_{M_{2n}} \!\!\!\! \Omega_{2n}^{(1)}\,.
\label{GSanoA}
\ee
Although the form of the anomaly (\ref{GSano}) and of the last term in
(\ref{action1}) depend on the arbitrary parameter $\alpha$, the gauge
variation of $S$ due to the transformation (\ref{C2k-2tra}) is independent
of $\alpha$ and thus universal.

The above Green--Schwarz mechanism can be alternatively described in terms of a
``magnetic'' $(2n-2k)$-form potential $\tilde C_{2n-2k}$, dual to the ``electric''
$(2k-2)$-form potential $C_{2k-2}$. The relation between the two formulations
is defined in terms of the corresponding field strengths $H_{2k-1}$ and
$\tilde H_{2n-2k+1}$, and corresponds to the statement that these are Hodge-dual
to each other: $\tilde H_{2n-2k+1} = i {}^*H_{2k-1}$. In terms of the magnetic
potential $\tilde C_{2n-2k}$, the Green--Schwarz action reads
\bea
\tilde S = \int_{M_{2n}} \bigg[ \a\a \!\!
\frac 12 \Big|d \tilde C_{2n-2k} + \frac {\sqrt{2\pi}}{\xi}\,\Omega_{2n-2k+1}^{(0)}\Big|^2
+ i\sqrt{2\pi}\,\xi\,\tilde  C_{2n-2k}\,\Omega_{2k} \nn \\
\a\a \!\! -2\pi i \bigg(\alpha-\frac{k}{n+1}\bigg)
\Omega_{2k-1}^{(0)}\Omega_{2n+1-2k}^{(0)} \bigg]\,.
\label{action2}
\eea
The appropriate definition of the field strength is thus
\be
\tilde H_{2n-2k+1} = d \tilde C_{2n-2k}
+ \frac{\sqrt{2\pi}}{\xi}\, \Omega_{2n-2k+1}^{(0)} \,.
\label{H2}
\ee
As before, this is gauge-invariant, provided that the antisymmetric tensor field
transforms non-linearly:
\be
\delta_\epsilon \tilde C_{2n-2k} = - \frac {\sqrt{2\pi}}\xi\,\Omega_{2n-2k}^{(1)} \,.
\label{C2n-2ktra}
\ee
Using these expressions, it is straightforward to verify that the variation
of the action cancels the $1$-loop anomaly (\ref{GSano}), exactly as in
the electric formulation:
\be
\delta_\epsilon \tilde S = -2\pi i \int_{M_{2n}} \!\! \Omega_{2n}^{(1)} \,.
\label{GSanoB}
\ee
Actually, the above two formulations can be shown to be perfectly equivalent.
In the electric formulation, the equation of motion following from
(\ref{action1}) reads $d^*H_{2k-1} = -i(\sqrt{2\pi}/\xi)\,\Omega_{2n-2k+2}$,
whereas the Bianchi identity following from the definition (\ref{H1}) is
$d H_{2k-1} = \sqrt{2\pi}\,\xi\,\Omega_{2k}$. The magnetic formulation is
then defined through the substitution $H_{2k-1} \rightarrow i {}^* \tilde H_{2n-2k+1}$.
The old equation of motion is turned into the new Bianchi identity
$d \tilde H_{2n-2k+1} = (\sqrt{2\pi}/\xi)\,\Omega_{2n-2k+2}$, whose general solution
is given by (\ref{H2}), the old Bianchi identity is turned to the new equation
of motion $d {}^*\tilde H_{2n-2k+1} = -i\sqrt{2\pi}\,\xi\,\Omega_{2k}$, and
the old action (\ref{action1}) to the new action (\ref{action2}).

The Green--Schwarz mechanism described above, involving a single tensor field
$C_{2k-2}$ or its dual $\tilde C_{2n-2k}$, can cancel only reducible anomalies
of the form $\Omega_{2n+2}=\Omega_{2k}\Omega_{2n+2-2k}$, with $1\leq k \leq n$.
This is clear from eqs.~(\ref{action1}) or (\ref{action2}),
but also from the fact that the involved forms are physical propagating fields
only for $1\leq k \leq n$. Notice in particular that the cases $k=0$ or $k=n+1$,
corresponding to irreducible anomalies, would formally require $(-1)$-forms or
$2n$-forms, with field strengths dual to each other, which are clearly unphysical.
Indeed, the top
$2n$-form has no physical degrees of freedom, since it cannot have a sensible
field strength, and its equation of motion simply implies that the total charge
under it should vanish; its would-be dual $(-1)$-form is correspondingly not existing.
However, a straightforward generalization of the basic Green--Schwarz mechanism, involving
several physical tensor fields $C_{2k_i-2}^i$ or their duals $\tilde C_{2n-2k_i}^i$,
with $1\leq k_i \leq n$, can cancel anomalies that are not reducible but
can be decomposed into a sum of reducible ones, with
$\Omega_{2n+2}=\sum_i \Omega_{2k_i}\Omega_{2n+2-2k_i}$, each tensor field being
responsible for the cancellation of one of the terms in the anomaly\cite{GSstringa}.
We will see in section 5 that in the case of orbifold field theories many more
generalizations are possible, depending on whether the tensor fields live in the
bulk of space-time or are localized at the singularities of the internal space.

Notice finally that since the anomaly (\ref{GSano}) is a $1$-loop effect, either the
Wess--Zumino coupling or the Chern--Simons form modifying the kinetic term of the
antisymmetric tensor fields in the actions (\ref{action1}) or (\ref{action2}) must
arise at the $1$-loop level, depending on $n$ and $k$. One of these two terms can therefore
be thought of as being induced by the heavy states associated to the physics in the UV.
This was explicitly verified in string theory, where the microscopic theory is
known and computable. It is important to notice that the Green--Schwarz anomaly
cancellation mechanism cannot work in renormalizable $4$-dimensional theories, since
the actions (\ref{action1}) or (\ref{action2}) necessarily contain operators of
dimension greater than $4$.

\subsection{Spontaneous symmetry breaking and anomalies}

The analysis of gauge anomalies that we have made so far can also be extended to
spontaneously broken gauge theories. The term ``spontaneously broken'' is actually
improper for local symmetries, since it is well-known that these are never truly
broken\cite{Elitzur:im}, but at most non-linearly realized in a Higgs phase. In
the following, we will restrict for simplicity to the case where the Higgs phase
is parametrized by the vacuum expectation value of some fundamental charged scalar
fields --- through which one can construct gauge-invariant operators acting as order
parameters --- but most of the considerations that follow are actually of a more
general validity. In particular, they apply also to theories on extra compact
dimensions where, as we will see in the next section, other symmetry-breaking
mechanisms, such as Wilson lines and orbifold projections, can occur.

The fundamental issue we want to address is whether the restrictions imposed by
anomaly cancellation are alleviated in the presence of a spontaneous symmetry
breaking. We will see below that this is indeed the case. For a theory with gauge
group $G$ that is spontaneously broken to a subgroup $H$, the precise statement
is that the low-energy effective theory below the breaking scale can always be
made consistent through the addition of suitable local counterterms, provided
the unbroken symmetries in $H$ are free of anomalies and some other mild conditions
are satisfied. The required counterterms can be constructed in a systematic way in
terms of the would-be Goldstone bosons $\phi$ in the coset $G/H$, the gauge bosons
$A$ of the group $G$ and their field strength $F$; they are in general higher-dimensional
operators that make the effective theory non-renormalizable and limit its validity
range to the breaking scale. More in general, an anomaly in a spontaneously broken symmetry does
not lead to any true consistency constraint in an effective non-renormalizable theory,
since it is possible to cure it through counterterms involving a high energy scale,
and the study of such counterterms is therefore relevant also in the framework of the
higher-dimensional theories that are the subject of this review. For simplicity,
we will consider only gauge anomalies in flat space,
following\cite{AGGW,CHZ,Preskill,Fabbrichesi:2002am}.

All the essential features of the above situation are well illustrated with a very
simple and particular example. Consider a simple $U(1)$ gauge theory in $2n$
dimensions with a chiral fermion and a complex Higgs field $\phi$, both of unit
charge, and suppose that $\phi$ has a non-vanishing vacuum expectation value
$v/\sqrt{2}$ that completely breaks the $U(1)$ symmetry. The Higgs field $\phi$ is
most conveniently parametrized as $\phi = \rho e^{-\theta/v}$, with $\theta$
anti-Hermitian. The minimal kinetic Lagrangian of $\phi$ gives rise to the kinetic
term $1/2|d \theta - v A|^2$ for $\theta$, which transforms non-linearly as
$\delta \theta = v\,\lambda$ under a $U(1)$ gauge transformation with parameter
$\lambda$. The $U(1)$ symmetry is anomalous, and the $1$-loop effective action
$\Gamma$ transforms as
\be
\delta \Gamma = -\frac{i^{n}}{(n+1)!\,(2\pi)^n} \lambda \, F^n\,.
\label{U(1)}
\ee
This anomalous variation can however be compensated by adding the local counterterm
\be
\Gamma_{WZ} = \frac{i^{n}}{(n+1)!\,(2\pi)^n} \frac{\theta}{v} \, F^n\,.
\label{WZU(1)}
\ee
The total effective action $\Gamma +\Gamma_{WZ}$ is then gauge-invariant. Notice
that the counterterm involves the operator $\theta F^2$, which has dimension $5$
in the $4$-dimensional case with $n=2$ and makes, as anticipated, the theory
non-renormalizable\cite{GJ}.

It is not difficult to realize that the anomaly cancellation mechanism provided by
the counterterm (\ref{WZU(1)}) in the simple example above is nothing but a particular
case of the Green--Schwarz mechanism discussed in subsection 3.1, with $n=2$, $k=1$,
$\xi \sim v$ and $C_0 \sim \theta$, which in such a situation also implies
that the $U(1)$ symmetry is spontaneously broken. The Green--Schwarz perspective is
actually slightly more general. Indeed, as explained in subsection 3.1, two physically
different possibilities can arise for the shift in the kinetic term of the field $\theta$
and the Wess--Zumino term: either the first is already there from the beginning and the
latter is induced by integrating out the heavy modes of the UV, or vice versa. In the
first case, the situation is perfectly analogous to the one described above, with a
charged Higgs field taking a vacuum expectation value. The second
situation, instead, corresponds to an originally neutral field that becomes effectively
charged due to the heavy modes that have been integrated out. In other words, in the
first case the symmetry is spontaneously broken from the beginning and the anomaly is
canceled by adding a Wess--Zumino term, whereas in the second case the Wess--Zumino
term is present from the beginning and it is the symmetry breaking that is added.

The generalization to an arbitrary gauge symmetry $G$ spontaneously broken to a subgroup $H$
is not totally straightforward. We denote by $T_A$ the whole set of generators of the Lie
algebra associated to $G$, and with $T_i$ and $T_\alpha$ those associated to $H$ and $G/H$.
We then assume that the fermion content of the theory is such that anomalies arise
at most in the broken symmetries in $G/H$, but not in the unbroken ones belonging to $H$,
that is:
\be
\delta_\alpha \Gamma(A) \neq 0 \,,\;\;
\delta_i \Gamma(A) = 0 \,.
\label{anomalia}
\ee
As before, we want to construct a Wess--Zumino term $\Gamma_{WZ}$ involving the would-be
Goldstone bosons $\phi$ in $G/H$ and the gauge fields $A$ in $G$, in such a way that its
variation compensates (\ref{anomalia}) and $\Gamma+\Gamma_{WZ}$ is gauge-invariant. To
construct such a Wess--Zumino term, it is convenient to choose the unitary gauge in which
all the $\phi$'s are constant. This gauge choice is in general achieved through some gauge
transformation $U(x)$ that transforms the $G/H$-valued fields $\phi(x)$ to constants.
This gauge transformation $U$ is however clearly not unique: two $G$-valued maps $U$
and $U'$ correspond to the same $\phi$ if and only if they differ by a right transformation
$h$ belonging to $H$: $U'=U h$. It is then possible to use $U(x)$ as a dynamical variable
instead of $\phi(x)$, although this introduces an additional $H$-gauge degree of freedom
that should eventually decouple. This change of variables is very convenient, because $U$
transforms covariantly under a gauge transformation as $U\rightarrow U^g=g^{-1}U$, whereas
$\phi$ does not, and it allows to construct the required Wess--Zumino term for a
generic anomalous quantum effective action $\Gamma$ simply as
\be
\Gamma_{WZ}(U,A) = \Gamma(A^U)-\Gamma(A) \,.
\label{GammaWZdef}
\ee
In this way, the total effective action $\Gamma(A)+\Gamma_{WZ}(U,A)$ is automatically
gauge-invariant under all the transformations in $G$, since
\bea
\Gamma_{WZ} (U^g,A^g) - \Gamma_{WZ} (U,A) \a=\a \Big[\Gamma(A^{g U^g})-\Gamma(A^g)\Big]
- \Big[\Gamma(A^U) - \Gamma(A)\Big] \nn \\
\a=\a - \Big[\Gamma(A^g) - \Gamma(A) \Big] \,.
\eea
Notice that, as a consequence of the second equation in (\ref{anomalia}), the Wess--Zumino
term (\ref{GammaWZdef}) is, as it should, insensitive to the ambiguity in the definition of
$U$ modulo transformations $h$ of $H$: $\Gamma_{WZ}(Uh,A)=\Gamma_{WZ}(U,A)$. Parametrizing
the anomalous variation of the latter as in (\ref{G2}), and introducing again the space
$M^\prime_{2n+1}$ whose boundary is the $2n$-dimensional space-time $M_{2n}$, it can be
written more explicitly as
\be
\Gamma_{WZ}(U,A) = 2 \pi i \int_{M^\prime_{2n+1}} \!\!\!\!
\Big[\Omega_{2n+1}^{(0)}(A^U) - \Omega_{2n+1}^{(0)}(A) \Big]\,.
\label{GammaWZ}
\ee

The above construction can be generalized to the case in which the second
condition in eq.~(\ref{anomalia}) is replaced by the weaker condition that
the pure $H$ anomalies vanish:
\be
\delta_i \Gamma(A_H)=0 \,.
\label{weakercond}
\ee
Indeed, in such a situation it is possible\cite{AGGW,CHZ} to satisfy the second
relation in (\ref{anomalia}) by adding to the effective action a suitable local
functional $B_{2n}(A_H,A)$ of the gauge fields, provided that $G/H$ is a reductive
space, {\it i.e.} $[T_i, T_\alpha] = f_{i \alpha \beta} T^\beta$, which is in
particular true when $G$ or $H$ are compact groups.
In this more general setting,
all perturbative gauge anomalies can be canceled by adding to the action the
Wess--Zumino term $\Gamma^\prime_{WZ} (A,U) = \Gamma^\prime(A^U)-\Gamma^\prime(A)$,
which corresponds to the modified effective action $\Gamma^\prime(A) = \Gamma(A) +
B_{2n}(A_H,A)$. Once gauge invariance is restored in this way, it is possible to
choose the unitary gauge where $\phi$ decouples and the gauge bosons of $G/H$ are
massive. However, the presence of $\Gamma^\prime_{WZ}$ in a generic gauge suggests
that such a theory is nevertheless non-renormalizable. This has been proved explicitly
using the 't Hooft--Landau gauge in Ref.~\refcite{Preskill}.

A particularly nice physical interpretation of the Wess--Zumino terms can be given
in the case where the microscopic theory with gauge group $G$ contains
extra chiral fermions that make it completely anomaly free and become heavy
after the spontaneous symmetry breaking $G\rightarrow H$, thanks to large
Yukawa couplings, with a mass $M$ that is much larger than the breaking scale
$v$. The Wess--Zumino term that is needed to cancel the anomaly of the remaining
light modes is then automatically generated in the low-energy effective theory
when integrating out these heavy fermions\cite{hokerfari}. The fact that this
happens is in fact ensured by 't Hooft's anomaly matching condition\cite{tHooft}:
anomalies, being long-distance effects, have to match in going from the
fundamental to the effective theory. Possible topological obstructions or
global gauge anomalies will be briefly considered in section 6.

An important constraint on the above anomaly-cancellation mechanism for spontaneously
broken symmetries is that the addition of the Wess--Zumino term in the action
should not break other important symmetries that are possibly present in the theory.
This observation is particularly relevant when studying gauge theories with compact
extra dimensions, where most of the higher-dimensional local symmetries are
non-linearly realized on scalar fields arising from internal components of tensor
fields in higher dimensions. To illustrate the point, let us consider the
simplest example of a $U(1)$ gauge theory on $R^4\times S^1$. After Kaluza--Klein
decomposition, the $5$-dimensional gauge field $A_M$ yields an infinite number
of $4$-dimensional gauge bosons $A_{\mu, n}$ plus an infinite number of charged
scalar fields $A_{5,n}$, and similarly the $5$-dimensional gauge parameter
$\lambda$ yields an infinite number of $4$-dimensional gauge parameters $\lambda_n$.
The gauge transformation laws then become $\delta_n A_{\mu, n} = \partial_\mu \lambda_n$
and $\delta_n A_{5, n} = i v_n \lambda_n$, where $v_n=n/R$ in terms of the
radius $R$ of the internal dimension. This clearly shows that from the
$4$-dimensional point of view there is a linearly realized $U(1)_0$ gauge
symmetry associated to $A_{\mu, 0}$, with a neutral field $A_{5,0}$ from the
zero-mode sector, plus an infinite set of spontaneously broken $U(1)_n$
gauge symmetries associated to the $A_{\mu, n}$, with would-be Goldstone
bosons $\phi_n$ given by the $A_{5, n}$ and breaking scales $v_n$, from the
non-zero-mode sectors. In the notation of this section, the original and
final gauge groups are therefore given by $G = \times_n U(1)_n$ and $H=U(1)_0$.
In the case at hand, the circle compactification always leads to a vectorial
theory and the broken symmetries are never anomalous. However, in similar
but less trivial situations such as orbifold compactifications, these symmetries
might become anomalous and require the introduction of infinitely many
Wess--Zumino counterterms of the form $\phi_{n} F_m F_k$. The requirement that
these should satisfy the $5$-dimensional Poincar\'e symmetries implies that
these must correspond to the Kaluza--Klein decomposition of a $5$-dimensional
Chern--Simons term of the form $\epsilon^{MNPQR} A_M F_{NP} F_{QR}$. We will
see in section 5 that such a term can indeed play an important role in the
cancellation of anomalies in such theories.

\section{Localized anomalies on orbifolds}

In this section, we shall generalize the computation of section 2 to the case
of orbifold theories. We shall consider a general orbifold construction defined
by starting with a $(2n+m)$-dimensional space that is the product of a non-compact
$2n$-dimensional\footnote{We take $M_{2n}$ to be an even-dimensional space,
because the study of perturbative anomalies for odd-dimensional spaces is trivial.}
space-time $M_{2n}$ and a compact $m$-dimensional internal space $K_{m}$. The
orbifold projection is implemented by ``gauging'' some finite group $\Z_N$ of
geometrical symmetries of $K_{m}$, thereby turning the physical space to
${\cal M}_{2n+m} = M_{2n} \times K_m/\Z_N$. The coordinates $x^{M}$ on
${\cal M}_{2n+m}$, with $M=1,\dots,2n+m$, can be split into the space-time
coordinates $x^\mu$ on $M_{2n}$, with $\mu=1,\dots,2n$, and the internal
coordinates $x^i$ on $K_{m}/\Z_N$, with $i=1,\dots,m$. For convenience, we
shall often indicate these two sets of coordinates simply with $x$ and $z$.
If $\Z_N$ acts non-freely on $K_m$ and has fixed points, the orbifold $K_m/\Z_N$
fails in general to be a manifold. Nevertheless, orbifold theories are very
appealing, because they provide all the features of a compactification on a curved
and topologically non-trivial manifold, but are much more tractable if the original
compact manifold $K_m$ is simple enough. We will concentrate in the following on
the simplest and also most relevant case, where $K_m$ is a product of circles or
tori, which are themselves obtained by modding out the Euclidean space $R^m$ with a
group of translations $T$. In such a case, one can equivalently construct the orbifold
by starting directly from $R^m$ and projecting it with the larger group generated
by the combination of the orbifold transformations $\Z_N$ and the translations $T$,
which is the so-called space group, in contrast to the group of the sole $\Z_N$
orbifold transformations, which is called the point group. See Ref.~\refcite{orb}
for more details.

Field theories on orbifolds can be consistently studied at the quantum level,
as low-energy effective descriptions of a more fundamental theory such as a string
theory. Although such higher-dimensional quantum field theories are unavoidably
non-renormalizable, consistency at low energies requires that gauge and gravitational
anomalies be absent. As explained in the introduction, one must then study
the quantum realization of the full group $G$ of higher-dimensional local symmetries
and require it to be free of any anomaly. In other words, the quantum effective action
$\Gamma(A)$ should be invariant under local transformations in $G$ with arbitrary
parameter $v_\alpha(x,z)$ ($\alpha= 1,\ldots, {\rm dim}\,G$) depending on the space-time
coordinates $x$ as well as the internal coordinates $z$.\footnote{We are assuming
here that the gauge fields propagate on the whole ${\cal M}_{2n+m}$, so that the
gauge symmetry is local with respect to all the coordinates $(x,z)$.
However, this is not necessary, and we will comment below on the more general
situation in which gauge fields are localized on a subspace of the internal space.}
The general form of a possible anomalous variation of the effective action $\Gamma$
under this kind of transformations is given by
\be
{\cal I}(v) = \delta_{v} \Gamma(A)
= 2 \pi i \int\!d^{2n\!}x\,d^{m}\!z {\cal A}^{(1)}(x,z)\,.
\label{loc-ano}
\ee
A related issue is the quantum realization of the subgroup $H$ of symmetries surviving
in the dimensionally reduced theory, defined by discarding all the heavy Kaluza--Klein modes
from the beginning without integrating them out, that is truncating the theory by retaining
only the lower-dimensional zero modes. An anomalous variation of the corresponding effective
action $\tilde \Gamma$ under a symmetry transformation of $H$ with parameter $v_i(x)$
($i=1,\ldots,{\rm dim}\,H$) is then given by
\be
\tilde {\cal I}(\tilde v) = \delta_{\tilde v} \tilde \Gamma(\tilde A)
= 2 \pi i \int\!d^4x \, \tilde {\cal A}^{(1)}(x) \,.
\label{loc-ano4D}
\ee
Although the dimensionally reduced theory crucially differs from the more relevant low-energy
effective theory obtained by integrating out the heavy Kaluza--Klein modes,
the comparison between (\ref{loc-ano}) and (\ref{loc-ano4D}) proves to be very useful.
We call ${\cal A}(x,z)$, defined as in (\ref{loc-ano}), the localized anomaly and
$\tilde {\cal A}(x)$, defined as in (\ref{loc-ano4D}), the integrated anomaly. We will show
that $\tilde {\cal A}(x)$ is indeed related to the integral over the internal space of
${\cal A}(x,z)$. If the latter is zero, the former is guaranteed to vanish, but the
converse is in general not true, and this is the crucial new aspect to be studied in
theories with compact extra dimensions. A more precise statement can be given using
the general formalism presented in subsection 3.2. From the $2n$-dimensional point of
view, the original gauge group is represented by a infinite number --- one for
each Kaluza--Klein mode of the gauge fields --- of copies of the higher-dimensional
gauge group: $G = \times_n G_n$. The final gauge group after the orbifold projection is
instead a subgroup $H$ of the sole $G_0$ group associated to the zero modes.
Two different kinds of symmetry breaking can be identified. First, a spontaneous symmetry
breaking $G \rightarrow G_0$, involving the non-zero modes of the internal components
of the gauge fields as Higgs fields, and then a truncation of $G_0 \rightarrow H$.
Although the latter truncation is not spontaneous in the sense of subsection 3.2, the
results of that subsection still hold, since possible anomalies associated to the coset
$G_0/H$ identically vanish. The condition that (\ref{loc-ano4D}) vanishes corresponds
then precisely to the condition (\ref{weakercond}).
The arguments of subsection 3.2 then imply that when (\ref{weakercond}) holds --
possibly thanks to a Green--Schwarz mechanism as discussed in subsection 3.1 -- there
should always exist local counterterms allowing the anomalies (\ref{loc-ano}) to be
canceled. However, as already pointed out in the introduction and at the end of
subsection 3.2, the non-trivial issue is to understand whether these counterterms
are consistent with all the other symmetries of the theory. On the other hand, the
nature of the counterterms or Green--Schwarz mechanisms that are required to cancel
(\ref{loc-ano}) can be easily understood by explicitly computing (\ref{loc-ano}).
This is the motivation for studying localized anomalies on orbifolds.

We will present below a computation of anomalies in orbifold theories along the lines
of section 2, using Fujikawa's approach and mapping the problem to the evaluation of
traces of certain supersymmetric quantum mechanical models. The consistent form of the
anomaly can then be derived by using the Stora--Zumino descent relations, exactly as
for smooth manifolds. The only novelty arising from the singularities introduced by
the orbifold projection is that the localized anomaly can have not only the expected
contribution distributed over all the space ${\cal M}_{2n+m}$, but also other
contributions that are localized at the orbifold fixed points.

\subsection{Generalities about orbifold projections}

The orbifold projection is defined by a geometric action $g$ representing a symmetry
$z \rightarrow g z$ of the internal covering space $K_m$ and generating the finite group
$\Z_N$, which is also embedded into the gauge symmetries of the original theory. The
geometric part of the $\Z_N$ action on a bulk field $\Phi$ propagating on the whole space
${\cal M}_{2n+m}$ is fixed by the decomposition of its representation under the tangent
space group $SO(2n+m)$ of ${\cal M}_{2n+m}$ in terms of $SO(2n)\times SO(m)$, where
$SO(2n)$ and $SO(m)$ are the tangent space groups of $M_{2n}$ and $K_m$.\footnote{Recall
that the tangent-space group of a $D$-dimensional manifold is isomorphic to $SO(D)$ and
has nothing to do with the symmetry group of the manifold itself. The ``spin'' of
a field $\Phi$ is defined by its representation under $SO(D)$.} In the special case
in which $K_m$ is $1$-dimensional, the geometric action of $g$ on $\Phi$ corresponds
instead to a parity transformation. The projection acts on the gauge group $G$ through
an automorphism on its Lie algebra, {\em i.e.} through a transformation of the type
$T^A\rightarrow {\cal P}^A_{\;B} T^B$ that leaves the structure constants of the group
invariant. When the automorphism can be written as a group conjugation,
${\cal P}^A_{\;B} T^B = P T^A P^{-1}$, with $P\in G$, it is called inner automorphism;
otherwise it is called an outer automorphism (see Ref.~\refcite{Hebecker:2001jb} for
more details). For simplicity, we will restrict to inner automorphisms, where $P$
satisfies $P^N=I$. The transformation properties of the space-time components of the
gauge fields, being insensitive to the geometric action of $g$, are uniquely given
by the matrix $P$:
\be
A_\mu(g z) = P A_\mu (z)\, P^{-1} \,.
\label{bcgauge}
\ee
The orbifold boundary conditions (\ref{bcgauge}) break the gauge group $G$ to the
subgroup $H$ that commutes with $P$; more precisely, as discussed at the end of
subsection 3.2, the transformations belonging to $H$ admit a linear realization
that involves the zero modes of $A_\mu$, whereas those in $G/H$ are non-linearly
realized on the non-zero modes.

The general gauge symmetry breaking $G \rightarrow H$ is most efficiently
described\cite{Ibanez:1986tp,Ibanez:1987pj} by distinguishing the Cartan generators
$H_I$ of the Lie algebra ${\cal G}$ associated to $G$, with $I = 1,\dots,{\rm rank}\,G$,
from the remaining ones $E_A$, with $A=1,\dots,{\rm dim}\,G-{\rm rank}\,G$.
The structure of the algebra is then as follows:
\bea
\big[H_I,H_J\big] \a=\a 0 \,, \label{comm1} \\
\big[H_I,E_A\big] \a=\a \rho_I^A E_A \,, \label{comm2} \\
\big[E_A,E_B\big] \a\propto\a E_{A+B} \,.\label{comm3}
\eea
The commutation relation (\ref{comm2}) defines the root vector $\rho_I^A$ associated to each
$E_A$, and the commutation relation (\ref{comm3}) is vanishing whenever the right-hand side does
not exist. Assuming without loss of generality that $P$ is diagonal, its general form
involves only the Cartan generators $H_I$ and is parametrized by a vector $V_I$ as
\be
P = e^{2 \pi i V_{I} H_I}\,.
\label{formM}
\ee
The twist vector $V_I$ is constrained by the condition $P^N=I$, but is otherwise arbitrary.
The $4$-dimensional gauge field can accordingly be decomposed as $A_\mu=A_\mu^I H_I +A_\mu^A E_A$.
It follows from the commutations relations (\ref{comm1}) and (\ref{comm2}) that all the modes
$A_\mu^I$ are even and lead to a zero mode in (\ref{bcgauge}), whereas the modes $A_\mu^A$ have
boundary conditions twisted by the phase $e^{2\pi i V \cdot \rho_A}$. All the non-Cartan generators
for which $V\cdot \rho_A$ is an integer will lead to zero modes and thus to $4$-dimensional
unbroken symmetries. Thanks to (\ref{comm3}), all the elements of the group $G$ that admit
zero modes form a subgroup $H\subset G$, whose rank coincides with that of $G$.

The compact manifold $K_m$ is in general not simply connected and hence, in addition to the
orbifold boundary conditions (\ref{bcgauge}), one has also to specify the periodicity conditions
of space-time fields around its non-contractible cycles\cite{orb,Witten-Wen,Ibanez:1986tp}.
Denoting by $e_a$ the basis vectors of the non-contractible cycles $\gamma_a$ in $K_m$,
one can impose for $A_\mu$ a general boundary condition that is twisted through arbitrary
matrices $W_a$ of the gauge group $G$ in the fundamental representation:
\be
A_\mu(z + e_a) = W_a A_\mu(z)\, W_a^{-1}\,.
\label{tran}
\ee
The twist matrices $W_a$ can be interpreted\cite{hos} as Wilson lines along the cycles $\gamma_a$:
\be
W_a = {\cal P} \, \exp{\oint_{\gamma_a} A }\,,
\label{Wilson-a}
\ee
where ${\cal P}$ (not to be confused with the twist matrix $P$) denotes the usual path
ordering.\footnote{Recall that in our conventions the gauge fields are taken to be
anti-Hermitian and this explains the absence of a factor $i$ in the exponent of (\ref{Wilson-a}).}
Only a subset of the Wilson lines that are allowed on  $K_m$ give well-defined Wilson lines on
$K_m/\Z_N$. The precise consistency conditions for the latter depend on the explicit form of
$K_m$ and must be discussed case by case. In subsections 4.2 and 4.3, we will do this in some
detail for the simplest $1$- and $2$-dimensional orbifold constructions, in which the Wilson
lines (\ref{Wilson-a}) arise from constant connections and the path ordering is irrelevant.
A general feature distinguishing the solutions to the Wilson line consistency conditions is
that they may or may not depend on continuous parameters. The first are called continuous
and the other discrete Wilson lines.

Wilson lines represent an additional possibility for gauge symmetry breaking, since only the
gauge fields $A_\mu$ left unbroken by the projection $P$ and periodic around all the
cycles of the internal space admit $4$-dimensional massless modes.
The combined gauge symmetry breaking due to the boundary conditions (\ref{bcgauge}) and
(\ref{tran}) can be alternatively understood in terms of the symmetry breaking that is locally
effective at the various fixed points of the orbifold action.\footnote{The orbifold action
may also happen to leave a subspace of the internal space fixed, rather than a point, in which
case one gets fixed planes, rather than fixed points. In the following we shall restrict to
fixed points for simplicity.}  The crucial property allowing this reinterpretation is that a
generic fixed point $z_{i_k}$ is left fixed by the element $g^k$ only modulo a suitable translation
in the covering space. More precisely, bringing back the image $g^k z_{i_k}$ to the original
$z_{i_k}$ will require some integer numbers $n_{i_k a}$ of translations along the basis vectors
$e_a$ specifying the cycles $\gamma_a$ of $K_m$, so that:
\be
z_{i_k} = g^k z_{i_k} + \sum_a n_{i_k a} e_a \,.
\label{fixed-tran}
\ee
The numbers $n_{i_k a}$ depend on the particular fixed point $z_{i_k}$ and are in general
different for different fixed points of the same element $g^k$. Combining (\ref{bcgauge}),
(\ref{tran}) and (\ref{fixed-tran}), we deduce that at a given $g^k$ fixed point $z_{i_k}$,
with associated integers $n_{i_k a}$, the effective orbifold projection is implemented by a
matrix that is not just $P^k$ but rather
\be
P_{i_k} = \prod_a W_a^{n_{i_k a}} P^k \,.
\ee
More precisely, this means that only those components of the gauge field that commute with
$P_{i_k}$ can possibly have zero modes. The gauge group $G$ is therefore locally broken at
$z=z_{i_k}$ to the subgroup $H_{i_k}$ of $G$ commuting with $P_{i_k}$ at $z=z_{i_k}$.
The globally unbroken gauge group $H$ in $2n$ dimensions is then the intersection $H$ of
the gauge groups surviving at all the fixed points: $H=\cap_{i_k} H_{i_k}$.
Depending on whether the Wilson lines $W_a$ commute or not with the projection $P$,
rank-preserving or rank-reducing gauge symmetry breaking are possible.\footnote{More
generally, a rank-reducing Wilson line symmetry breaking occurs whenever the orbifold
space group is non-Abelian and can be implemented in the gauge group in a non-Abelian
way\cite{Ibanez:1987xa}.}
In the following, we will take into account the possible presence of Wilson lines,
and discuss in some detail their effect on anomalies (see also
Ref.~\refcite{GrootNibbelink:2003gd} for a recent similar analysis).

Charged matter fields can propagate either in the whole space-time ${\cal M}_{2n+m}$ and be
in representations of the full gauge group $G$ (bulk fields) or localized at one of the
fixed points $z_{i_k}$ and be in representations of the corresponding subgroup $H_{i_k}$
(boundary fields). Although we have assumed, in writing (\ref{bcgauge}), that the gauge
fields are bulk fields, this is not necessary; in fact they can be localized on some
subspace of the whole space-time, as happens in string theory in the presence of $D$-branes,
for example. On the contrary, gravity is always a bulk field, of course. As we will show
more concretely below, in such a general orbifold model the localized anomaly (\ref{loc-ano})
consists of a delocalized piece arising from possible anomalies already present on the
covering space $K_m$ and concerning the full gauge group $G$, plus additional pieces
localized at the fixed points $z_{i_k}$ and concerning the gauge subgroups $H_{i_k}$.

\subsection{Boundary fermions}

The contribution to the anomaly (\ref{loc-ano}) induced by a boundary fermion of
$2n$-dimensional chirality $\eta$, living at some fixed point $z_{i_k}$ and transforming
in a representation ${\cal R}_{i_k}$ of the locally unbroken group $H_{i_k}$ at
$z_{i_k}$, is easily derived. It is obviously localized at $z_{i_k}$ and given by the
standard expression in terms of the gauge and gravitational curvatures:\footnote{More
in general, the boundary fermion can couple also to gravitational fluctuations in the
transverse space $K_m$. This happens for instance for a set of fermions transforming
non-trivially under the transverse $SO(m)$ tangent space group. A typical example is
given by boundary fermions arising from a trivial dimensional reduction of a bulk
fermion, which transform in the spinor representation of $SO(m)$. This is the case
of the fermions living on $D$-branes in string theory\cite{Witten:1995im}. See
Ref.~\refcite{Dbrane} for the general form of the gauge and gravitational anomaly
induced by these fermions.}
\be
{\cal A}(x,z) = \eta\,{\rm ch}_{{\cal R}_{i_k}}\!(F) \hat A(R)\,
\delta^{(m)}(z-z_{i_k}) \,.
\label{anomalygauge-localized}
\ee
The case of a fermion that is localized only on a subspace of the internal
space is more similar to the case of a bulk field, because the orbifold projection
affects the result for the anomaly. The relevant formulae are perfectly
similar to those derived in the next subsections for bulk fields, and we will
therefore not discuss any further this intermediate case. An even more general
situation is obtained when the matter fermion and the gauge fields are localized
on different subspaces of the internal space with a non-empty intersection. In
this case too, the relevant formulae do not present significant new features
with respect to the two extreme cases of fields localized at a point of the internal
space or propagating on all of it.

\subsection{Bulk fermions on $S^1/\Z_2$}

The simplest example of orbifold construction, with $2n$ space-time dimensions
and $m=1$ internal dimension, is based on a simple circular covering space
$K_1 = S^1$, which is defined by modding out the real line with coordinate $z$
by the translation $T:z \rightarrow z + e$, with $e = 2 \pi R$ in terms of the radius
$R$. The orbifold projection is defined by identifying points on the circle $S^1$
that are related by a $\Z_2$ reflection $R:z\rightarrow -z$. There are two fixed
points, $z_1=0$ and $z_2 = \pi R$, and the physical part of the internal space is
the segment of length $\pi R$ that connects them. The $\Z_2$ projection is embedded
into the gauge group $G$ through an arbitrary matrix $P\in G$ satisfying $P^2 = I$.

Consider a bulk Dirac fermion field $\Psi$ in an arbitrary representation
${\cal R}$ of the gauge group $G$ in interaction with external gauge and
gravitational fields. The $(2n+1)$-dimensional Lagrangian before the $\Z_2$
projection is given by (\ref{FerLag}). The action of the reflection on
the fields depends on the orbifold matrix $P$ and an overall sign choice
$\eta = (-1)^{r_\eta}$ for the fermion, where $r_\eta=0,1$ can in general
depend on the representation ${\cal R}$. The orbifold boundary conditions
then read:
\bea
\Psi(-z) \a=\a \eta\,\gamma_{2n+1} P_{{\cal R}} \Psi(z)\,, \nn \\
A_{\mu}(-z) \a=\a P A_{\mu}(z) P^{-1}\,,\;\;
A_{z}(-z) = -\,P A_{z}(z) P^{-1}\,, \nn \\
g_{\mu\nu}(-z) \a=\a g_{\mu\nu}(z)\,,\;\;
g_{\mu z}(-z)= - g_{\mu z}(z)\,,\;\;
g_{zz}(-z)=g_{zz}(z)\,.
\label{Z2def}
\eea
In these expressions, $P=P_{\rm fund}$ and $P_{{\cal R}}$ are the twist matrices
in the fundamental defining representation and in the generic representation ${\cal R}$
respectively. Since $S^1$ is not simply connected, we also need to specify the
corresponding periodicity conditions. These are in general twisted by a
matrix $W$, and read
\bea
\Psi (z + 2\pi R) \a=\a W_{{\cal R}} \Psi(z)\,, \nn \\
A_{M} (z + 2\pi R) \a=\a W A_{M} W^{-1}(z) \,, \nn \\
g_{MN} (z + 2\pi R) \a=\a g_{MN} (z) \,,
\label{per-ab}
\eea
where $W=W_{\rm fund}$ and $W_{{\cal R}}$ represent the Wilson line in the
fundamental representation and in the generic representation ${\cal R}$.

The fixed points $z_1=0$ and $z_2=\pi R$ have $n_1=0$ and $n_2 = 1$ in (\ref{fixed-tran}),
and the corresponding effective projections are $P_1=P$ and $P_2=W P$. Denoting by $H_1$
and $H_2$ the associated gauge subgroups, the surviving gauge group in $2n$ dimensions
is $H=H_1\cap H_2$. The gauge twist can be interpreted as a Wilson line
$W = \exp\,\{\!\oint \langle A_z \rangle\} = \exp\{2 \pi R \langle A_z \rangle\}$,
constructed from a non-vanishing $\langle A_z \rangle$ that is constant and compatible
with the boundary conditions for $A_z$. This is possible as a consequence of the fact
that $\langle A_z \rangle$ does not necessarily commute with $P$ and is moreover defined
only up to the equivalence class $\langle A_z \rangle = \langle A_z \rangle + ip/R$, where
$p$ is any integer, dictated by periodic gauge transformations on $S^1$ (see {\em e.g.}
Ref.~\refcite{Hall:2001tn}). The allowed values for $W$ can be determined by noting
that the geometrical actions $T$ and $R$ of the translation and of the $\Z_2$ action
satisfy the relation $(RT)^2=I$. As a consequence, the generic boundary conditions
(\ref{Z2def}) and (\ref{per-ab}) on the fields are mutually consistent only if the
corresponding twist matrices $P$ and $W$ satisfy the relation\cite{orb,consistency}
\be
(P W)^2 = I \;.
\label{conscondz2}
\ee
Two possibilities can then arise, depending on whether the Wilson line originates
from an even or an odd component of $A_z$. The generators $T_m$ of the Lie algebra
of $G$ that correspond to components of $A_z$ that are even under both projections
effectively implemented at the two fixed points, and therefore lead to zero modes
for $A_z$, are specified, as a consequence of (\ref{conscondz2}) and (\ref{Z2def}),
by the following two conditions:
\be
\big\{P,T_m\big\} =  0\,,\;\;
\big\{WP,T_m\big\} = 0\,.
\label{cond-Ay}
\ee
Together, these also imply that $[W,T_m]=0$. These Wilson lines
are continuous, since the even fields $A_z^m$ from which they are constructed can take
an arbitrary constant vacuum expectation values.\footnote{Recall that no potential
for the scalar fields $A_z$ is allowed by gauge invariance on the orbifold $S^1/\Z_2$
and thus $A_z^m$ are moduli fields, at least at tree level.}
Due to the first of the conditions (\ref{cond-Ay}), $W$ cannot commute with $P$,
$[P,W]\neq 0$, so that continuous Wilson lines typically induce a spontaneous
rank-reducing gauge symmetry breaking.
On the other hand, the generators $T_{\hat m}$ that correspond
to components of $A_z$ that are odd under both local projections, and therefore
do not lead to zero modes for $A_z$, are specified by the conditions:
\be
\big[P,T_{\hat m}\big] = 0\,,\;\;
\big[WP,T_{\hat m}\big] = 0 \,.
\label{cond-Ay2}
\ee
As before, these imply that $[W,T_{\hat m}]=0$. In this case, the Wilson lines
constructed from the corresponding $A_z^{\hat m}$ are discrete. Indeed, only the
two specific values $\langle A_z^{\hat m}\rangle = 0$ and $i/(2R)$ satisfy the
odd orbifold boundary condition about each fixed point, thanks to the fact that a
shift by $i/R$ in $\langle A_z^{\hat m}\rangle$ is irrelevant, and are allowed.
In this case $W$ commutes with $P$, $[P,W]=0$, so that discrete Wilson lines
induce a rank-preserving gauge symmetry breaking. This can also be understood
from the fact that the orbifold projection acts with opposite signs on $A_\mu$ and
$A_z$; the gauge fields $A_\mu^{\hat m}$ are therefore even under both $P$ and
$WP$ and the generators $T^{\hat m}$ correspond to the unbroken gauge group $H$
in $2n$ dimensions. This is not a spontaneous symmetry breaking, but rather a
truncation and as such it is more similar to an orbifold projection (see below).
Finally, it can easily be verified that the remaining components
of $A_z$, which are even under one of the local projections and odd under the other,
can never give rise to consistent Wilson lines on $S^1/\Z_2$.

It is worth mentioning that the orbifold $S^1/\Z_2$ in the presence of a discrete
Wilson line $W$ can be equivalently described in terms of an other orbifold,
constructed from a circle $S^{1\prime}$ of radius $R^\prime = 2R$ that is the
double cover of the original $S^1$. Since $W^2=I$, all fields are periodic
around $S^{1\prime}$, and the projection $P^\prime = WP$ is now realized
through a new independent $\Z_2^\prime$ reflection that is orthogonal to the
original $\Z_2$ reflection and acts as inversion around the point $\pi R^\prime/2$
of the circle. The resulting space is thus an $S^{\prime 1}/(\Z_2\times \Z_2^\prime)$
orbifold.

Let us now come to the computation of the anomaly induced by the fermion $\Psi$
in the general case in which discrete Wilson lines are allowed. Since the
dimensionality of the whole space is odd, it is not possible to compute the
chiral anomaly, as explained in subsection 2.1, and then apply the Stora--Zumino
descent relation. Rather, one has to consider the local transformations directly,
as done in subsection 2.2, but with the difference that the fermion $\Psi$ is
in this case necessarily of the Dirac type. To begin with, it is convenient to
consider anomalies on the covering space $M_{2n}\times S^1$. Proceeding along
the lines of subsection 2.2, these are found to be
\be
{\cal I} = \lim_{\beta \rightarrow 0} \sum_k \Big(\phi_k^\dagger v_\alpha T^\alpha
e^{-\beta (i\Dslashs)^2/2}\phi_k - \varphi_k^\dagger v_\alpha T^\alpha
e^{-\beta (i\Dslashs)^2/2} \varphi_k \Big)\,.
\label{anomalyDiracS1}
\ee
Since the two sets of eigenfunctions $\phi_k$ and $\varphi_k$ are equivalent,
the anomaly vanishes trivially, reflecting the fact that it is possible to regularize
the theory in a gauge-invariant way.
On the orbifold $S^1/\Z_2$, we still obtain eq.~(\ref{anomalyDiracS1}), but the two
sets of eigenfunctions $\phi_k$ and $\varphi_k$ are no longer equivalent, because
they are now two different subsets of the unique set of eigenfunctions on $S^1$
that have definite and opposite $\Z_2$ parity, as required to expand the fields
$\Psi$ and $\bar\Psi$. From the first equation in (\ref{Z2def}), it follows indeed
that the generator $g$ acts as $g\phi_k(z)= \eta \gamma_{2n+1} P \phi_k(-z)$ and
$g\varphi_k(z)=- \eta \gamma_{2n+1} P \varphi_k(-z)$ on the two sets. In order to
compute the resulting anomaly, it is convenient to represent these two relevant
sets of eigenfunctions on $S^1/\Z_2$ as the subsets of those on $S^1$ that are
invariant under the orbifold projection. This is achieved by inserting into the
trace over the eigenfunctions of $S^1$ the $\Z_2$ projector ${\cal P}_{\Z_2}=(1+g)/2$.
The part proportional to $1$ in the projector cancels between the two terms in
eq.~(\ref{anomalyDiracS1}), whereas the part proportional to $g$ sums up
to give:
\be
{\cal I} = \lim_{\beta \rightarrow 0} {\rm Tr}\,
\Big[g\, Q\, e^{-\beta H}\Big]\,.
\label{anomalyz2}
\ee
As in subsection 2.2, $H=(i\Dslash)^2/2$ is the Hamiltonian associated to the Lagrangian
(\ref{sqm1}), whereas $Q$ is given by (\ref{Qgau}) and (\ref{Qgrav}) for gauge transformations
and diffeomorphisms, and implements again the Stora--Zumino descent on the trace computed
without its insertion, modulo a Bose-symmetrization factor. The operator $g$ implements
instead the reflection along the internal direction, and the trace includes a sum over
all the Dirac components of the space-time fermions, which are realized in the standard way
from the Clifford algebra associated to the fermions $\psi^M$, as in (\ref{comm-rel-2}).
The relevant Euclidean Routhian associated to (\ref{anomalyz2}) is
\bea
{\cal R} = \a\a \frac 12 g_{MN} \dot x^M \dot x^N + \frac 12 \psi_{\underline{M\!}\,}
\dot \psi^{\underline{M\!}\,} + \frac{1}4
\big[\psi_{\underline{N\!}\,},\psi_{\underline{P\!}\,}\big]\,
\omega_{M}^{\underline{N\!}\,\underline{P\!}\,} \dot x^M  \nn \\
\a\a +\, c_A^\star A^A_{M\,B}\dot x^M  c^B
- \frac{1}2 c_A^\star c^B \psi^{\underline{M\!}\,}\psi^{\underline{N\!}\,}
F^A_{\underline{M\!}\,\underline{N\!}\,\,B}\,.
\label{sqm1Z2}
\eea
The underlined letters denote flat indices. The periodicity conditions of the fields
$X^M$ and $\psi^{\underline M}$ in the functional-integral representation of
(\ref{anomalyz2}) are determined by the action of the $\Z_2$ twist $g$ on them.
Since $g$ acts as $+1$ for space-time indices and $-1$ for the internal index,
the fields $x^\mu$ and $\psi^{\underline \mu}$ are periodic, whereas $x^z$ and
$\psi^z$ are antiperiodic. The action of $g$ on the operators $c^*$ and $c$ is instead
given in terms of the matrix $P$ as $gc^* g^{-1} = c^* P$ and $g c g^{-1} = P^{-1} c$.

To evaluate (\ref{anomalyz2}), we proceed as in section 2.1. The antiperiodic fermion
field $\psi^z$ does not admit zero modes. On the contrary, the antiperiodic $x^z$ does
admit a zero mode that is quantized to assume the discrete values $z_{1,2}$ corresponding
to the two points left fixed by $g$. After going to normal coordinates, the effective quadratic
Routhian is found to be
\bea
{\cal R}^{\rm eff} = \a\a
\frac 12 \Big[\dot \xi_{\underline \mu}\dot \xi^{\underline \mu}
+ \lambda_{\underline \mu} \dot \lambda^{\underline \mu}
+ R_{\underline \mu\underline\nu}(x_0,z_i,\psi_0)
\xi^{\underline \mu} \dot \xi^{\underline \nu} \Big]
- c_A^\star F(x_0,z_i,\psi_0)^A_{\;\;B} c^B \nn \\
\a\a +\, \frac 12 \Big(\dot \xi_{\underline z}\dot \xi^{\underline z}
+ \lambda_{\underline z} \dot \lambda^{\underline z} \Big)
+ (-2z_i)\, c_A^\star A^A_{z\,B} c^B \,,
\label{ReffZ2}
\eea
where $A^A_{z\,B}$ denotes the background associated to a possible discrete
Wilson line and
\bea
R_{\underline \mu\underline\nu}(x_0,z_i,\psi_0) \a = \a \frac 12
R_{\underline \mu\underline\nu\underline \rho\underline\sigma}(x_0,z_i)
\psi_0^{\underline \rho} \psi_0^{\underline \sigma} \,,\nn \\
F(x_0,z_i,\psi_0) \a = \a \frac 12 F_{\underline \mu\underline\nu}(x_0,z_i)
\psi_0^{\underline \mu} \psi_0^{\underline \nu} \,.
\label{R-F}
\eea
The last term in (\ref{ReffZ2}) originates from the term proportional to $A_z$ in
(\ref{sqm1Z2}). For constant $A_z$, the latter is a total derivative with respect to
$\tau$, but its integral $\oint \dot x^z = x^z(\beta) - x^z(0)$ does not vanish,
since $x^z$ is antiperiodic and thus $x^z(\beta) = -x^z(0)$. Replacing $x^z(0)$
with one of its two zero modes $z_i$, we then find $-2 z_i$, as in (\ref{ReffZ2}).
Using (\ref{fixed-tran}), $-2z_i$ can be conveniently rewritten as $n_i 2 \pi R$,
where $n_i$ denotes the number of translations that are required to identify $z_i$
and its $\Z_2$ image $-z_i$. The trace over the $1$-particles states of the fermion
fields $c_A$ can be easily evaluated. It depends on the orbifold matrix $P$ as well
as on the discrete Wilson line $W$ and yields:
\be
{\rm Tr}_{c,c^\star} \Big[P e^{2\pi n_i R \,c_A^\star A^A_{z\,B} c^B}
e^{c_A^\star F^A_{\;\;B} c^B}\Big] = {\rm tr}_{\cal R} \Big[W^{n_i}P\, e^{F}\Big]\,.
\label{TrZ2}
\ee
The path integral over the fluctuations of the periodic fields is computed as
in section 2.1. The one over the fluctuations of the antiperiodic fields yields
instead:
\bea
{\rm det}_A^{-1/2}\,\Big[\!-\!\partial_\tau^2 \Big] \a=\a \frac 12 \,, \\
{\rm det}_A^{1/2}\,\Big[\partial_\tau \Big] \a=\a 1 \,.
\label{dettfer}
\eea
The result for the fermionic determinant (\ref{dettfer}) arises in a subtle way.
In fact, the determinant itself would give a factor $\sqrt{2}$ when evaluated
using the $\zeta$-function regularization, but an extra factor $1/\sqrt{2}$ must
be added in this case to obtain a correctly normalized fermionic measure, due to
the fact that the spacetime is odd-dimensional. This normalization factor can be
fixed by requiring that for a free theory the fermionic contribution to the
ordinary partition function, without any operator insertion, should be equal to
the dimension of the spinor representation in $2n+1$ dimensions, that is $2^n$.
In the path-integral representation of this quantity, all the
$2n+1$ fermions are antiperiodic and yield each a determinant given by $\sqrt{2}$,
leading to a total factor $2^{n+1/2}$. A normalization factor $1/\sqrt{2}$ must
then be introduced to properly normalize the measure. More in general, the
$\zeta$-function regularization automatically produces a correctly normalized
measure in even dimensions, but requires an extra $1/\sqrt{2}$ normalization
factor in odd dimensions.

Putting the above results together, the consistent form of the $(2n+1)$-dimensional
localized gauge and gravitational anomalies is finally found to be
\be
{\cal A}(x,z) =  \frac{\eta}2 \sum_{i=1,2}
{\rm ch}_{\cal R} (P_i, F)\hat A(R)\,\delta(z-z_i)\,,
\label{chiral-anomalyZ2}
\ee
where $P_i$ is the effective projection at each fixed point $z_i$, {\em i.e.}
$P_0=P$ and $P_1= W P$, and the corresponding twisted Chern class is given,
as in Ref.~\refcite{dm}, by
\be
{\rm ch}_{\cal R} (P_i, F) = {\rm tr}_{\cal R}\Big[P_i e^{i F/(2\pi)}\Big]\,.
\label{cherntwisted}
\ee
The anomaly (\ref{chiral-anomalyZ2}) is the sum of two contributions localized at
the orbifold fixed points $z_i$. It is interesting to observe that these contributions
can be interpreted as the anomaly arising from the chiral fermions that would
survive the projection implemented by $P_i$ at the point $z_i$. In fact, if taken on
its own, $P_i$ would break the gauge group $G$ to a subgroup $H_i$, with the representation
${\cal R}$ of $G$ decomposing into the representations ${\cal R}_{l_i}$ of $H_i$. Since
$P_i$ commutes with all the generators of $H_i$, it acts as $(-1)^{r_{l_i}}$ with integer
$r_{l_i}$ on the representations ${\cal R}_{l_i}$ arising from the decomposition of
${\cal R}$ under the $G\rightarrow H_i$ symmetry breaking. Each representation ${\cal R}_{l_i}$
is then counted in (\ref{chiral-anomalyZ2}) with the sign $\eta (-1)^{r_{l_i}}$, which
corresponds to the $2n$-dimensional chirality of the surviving fermion. This makes it
possible to rewrite (\ref{chiral-anomalyZ2}) in terms of the anomaly ${\cal A}_i(x,z)$
induced by the massless fields that would survive the orbifold projection associated to
the fixed point $z_i$, which depends on those gauge fields that would survive that same
projection. The result is simply
\be
{\cal A}(x,z) = \sum_{i=1,2} \frac 12 {\cal A}_i(x,z_i) \delta(z-z_i) \,.
\label{A0piR}
\ee

On the other hand, the $2n$-dimensional anomaly obtained by integrating
(\ref{chiral-anomalyZ2}) over the internal dimension is given by
\be
\tilde {\cal A}(x) =  {\rm ch}_{\cal R} ({\cal P}, F)\,\hat A(R) \,.
\label{chiral-anomalyZ2-int}
\ee
In this case ${\cal P} = P(1+W)/2$ and $F$ is the field-strength of the gauge
fields commuting  with $P$ and $W$ and defining the unbroken gauge group. Again,
since ${\cal P}$ commutes with all the generators of $H$, we have
${\rm ch}_{\cal R} ({\cal P}, F) = \sum_{l} \eta_l \,{\rm ch}_{{\cal R}_{l}}(F)$,
where $\eta_l$ are the $2n$-dimensional chiralities of the surviving fermions in
the representations ${\cal R}_l$ of $H$, arising from the decomposition of
${\cal R}$. The integrated anomaly (\ref{chiral-anomalyZ2-int}) then coincides with
the anomaly induced by the massless $2n$-dimensional fermion zero modes, as expected.

The above general results are best illustrated with the simplest case $G=U(1)$
with $P=1$, so that $H=U(1)$. The allowed discrete Wilson lines are then given
by $W=\pm 1$. The case $W=1$ corresponds to a model where the projections at
the two fixed points $z_1=0$ and $z_2 = \pi R$ act in the same way and
preserve a fermion zero mode of the same chirality $\eta$; this implies that
${\cal A}_1$ and ${\cal A}_2$ in (\ref{A0piR}) are equal. This result for the
basic $S^1/\Z_2$ model was derived in Refs.~\refcite{Horava:1995qa,anomaly1}.
The case $W=-1$ corresponds instead to a model where the projections at the
two fixed points $z_1=0$ and $z_2 = \pi R$ preserve fermion zero modes of
opposite chiralities $\eta$ and $-\eta$; ${\cal A}_1$ and ${\cal A}_2$ are
then equal in magnitude but opposite in sign. The integrated anomaly
(\ref{chiral-anomalyZ2-int}) vanishes since ${\cal P}=0$ in
(\ref{chiral-anomalyZ2-int}); indeed, no massless fermion survives the
projection. This result was derived in Ref.~\refcite{anomaly2} within the
equivalent formulation of this model as an $S^1/(\Z_2 \times \Z_2^\prime)$
orbifold, as described before. It has also
been shown in Ref.~\refcite{Hirayama:2003kk} that these results are valid not
only for flat orbifolds, but also for warped ones, with a non-trivial metric
profile: the warp factor always cancels in both the localized and the
integrated anomaly. Another interesting class of models is obtained when
$G=SU(N+K)$ and $P$ is such that $H=SU(N) \times SU(K) \times U(1)$. There
are then various possibilities for $W$, among which $W=I$. This situation has
been studied in Refs.~\refcite{Kim:2002ab,Lee:2002qn}. A detailed study of
anomalies on a $7$-dimensional space based on an $S^1/\Z_2$ orbifold has
been performed in Ref.~\refcite{Gherghetta:2002nq}.

\subsection{Bulk fermions on $T^2/\Z_N$}

The next orbifold construction we consider is based on a toroidal covering space:
$K_2=T^2$. A torus $T^2$ is described by three real parameters and is obtained by
identifying points in the complex plane that are related by the two translations
$T_i:z \rightarrow z + e_i$ along the basis vectors $e_1=2 \pi R$ and $e_2 = 2 \pi R U$,
where $R$ is a radial parameter and $U$ is the complex structure of the torus.
The orbifold $T^2/\Z_N$ is obtained by further identifying points related by the
rotation $R:z \rightarrow \tau z$, where $\tau = e^{2 \pi i/N}$. There exist
consistent constructions for $N=2,3,4,6$, for which $2$-dimensional lattices
with non-trivial discrete rotational symmetries exist. The case $N=2$ is consistent
for arbitrary $R$ and $U$ and corresponds to a rather straightforward generalization
of the $1$-dimensional case of subsection 4.3. The cases $N=3,4,6$ are instead
consistent only when the complex structure of the torus is equal to the orbifold
twist: $U = \tau$, or other equivalent discrete choices.
The fundamental domain of these $T^2/\Z_N$ orbifolds can be
chosen to be a polygon of surface $|e_1 \times e_2|/N$ connecting the different
fixed points. The number of such points that are left fixed by the $k$-th power $g^k$
of the generator $g$ of $\Z_N$ is given by
\be
N_k = \bigg[ 2 \sin \Big(\frac{\pi k}{N}\Big) \bigg]^2\,.
\label{Nkprime}
\ee
It is therefore necessary to distinguish sectors labeled by different
$k=0,1,\dots,N-1$. However, since $g^{N-k}$ = $(g^k)^{-1}$, the $k$ and $N-k$
sectors have the same fixed-point structure and can be treated together, so
that the physically distinct sectors are labeled by $k=0,1,\dots,[N/2]$,
where $[N/2]$ denotes the integer part of $N/2$. The $\Z_N$ projection is
embedded as usual in the gauge group through an arbitrary matrix $P$ of $G$
satisfying $P^N = I$.

We consider a $(2n+2)$-dimensional complex fermion $\Psi$ of chirality $\rho$
in an arbitrary representation ${\cal R}$ of the gauge group $G$ and in
interaction with external gauge and gravitational fields. The action of the
$\Z_N$ rotation on the spinor indices is specified by the $SO(2)\simeq U(1)$
representation under the internal tangent space group. For a fermion of
spin $1/2$, this is a phase $\tau^s$, where $s=\pm 1/2$ defines the two
$2n$-dimensional components with chirality $\pm 1$. The orbifold action
on the gauge degrees of freedom is implemented by the matrix $P$ and can
involve a phase $\eta$ of the form $\eta = \tau^{1/2+r_\eta}$, with
$r_\eta=0,1,\dots,N-1$.\footnote{Notice that this phase is
actually necessary to have a $\Z_N$ action with $g^N=1$ on all fields.}
One then has
\bea
\Psi(\tau z) \a=\a \eta\, \tau^{s} P_{{\cal R}} \Psi(z)\,, \nn \\
A_{\mu}(\tau z) \a=\a P \, A_{\mu}(z)\, P^{-1},\;\;
A_{z}(\tau z) = \tau^{-1} \,P \, A_{z}(z)\, P^{-1},\;\;
A_{\bar z}(\tau z) = \tau \,P \, A_{\bar z}(z)\, P^{-1}, \nn \\
g_{\mu\nu}(\tau z) \a=\a g_{\mu\nu}(z)\,,\;\;
g_{\mu z}(\tau z)= \tau^{-1} g_{\mu z}(z)\,,\;\;
g_{\mu \bar z}(\tau z)= \tau g_{\mu \bar z}(z)\,,\;\; \nn \\
g_{z z}(\tau z) \a=\a \tau^{-2} g_{z z}(z)\,,\;\;
g_{\bar z \bar z}(\tau z)= \tau^{2} g_{\bar z \bar z}(z)\,,\;\;
g_{z \bar z}(\tau z)= g_{z \bar z}(z)\,.
\label{ZNdef}
\eea
Similarly, the actions of translations around the two independent cycles of
$T^2$ are encoded in boundary conditions that are in general twisted by two
Wilson lines $W_1$ and $W_2$ of the gauge group $G$:
\bea
\Psi (z + e_a) \a=\a W_{a,{\cal R}} \Psi(z)\,, \nn \\
A_{M} (z + e_a) \a=\a W_{a} A_{M} (z) \,W_a^{-1}, \nn \\
g_{MN} (z + e_a) \a=\a g_{MN} (z) \,.
\label{per-ab-Zn}
\eea
The notation for the twist and Wilson line matrices is as in subsection 4.3.
The Wilson lines are specified by the possible constant connections that can
exist around the two independent cycles $\gamma_a$ specified by the basis
vectors $e_a$: $W_a= \exp\,\{e_a \langle A_{z} \rangle + {\rm c.c.}\}$. Exactly as in the
$S^1/\Z_2$ case, the constant background value $\langle A_{z} \rangle$ can
be compatible with the orbifold boundary conditions, thanks to the equivalence
relation $\langle A_{z} \rangle = \langle A_{z} \rangle + 2 \pi i p_a/e_a$ on
$T^2$, with $p_a$ two arbitrary integers. The consistency conditions constraining
the matrices $P$ and $W_a$ are in this case quite severe. Indeed, the geometric
actions of $R$ and $T_a$ satisfy the relations $(T_a R^q)^{N/q}=I$ for each
integer $q=1,\dots,N/2$ such that $\Z_{N/q}$ is a subgroup of $\Z_N$, {\em i.e.}
$N/q$ is integer, and $[T_1,T_2] = 0$. These imply the conditions
\be
(W_a P^q)^{N/q} = I \,,\;\; [W_1,W_2]=0 \,.
\label{cond}
\ee
There is an additional condition depending on how the basis vectors $e_a$
are mapped within each other by the rotation. For $N=2$, each $e_a$ is
reflected to $-e_a$, and thus $R T_a = T_a^{-1} R$, but this does not lead to
any new condition. For $N=3,4,6$, since $U=\tau$, one has $Re_1=e_2$, and
hence $R T_1 = T_2 R$. This leads to the condition
\bea
W_{1} P \a=\a P W_{2} \,,\;\;{\rm for }\; N=3,4,6 \,. \label{condZN}
\eea
As for the case analyzed in subsection 4.3, there can can be continuous Wilson
lines with $[W_a,P] \neq 0$, associated to a constant connection of a field with
a massless mode, or discrete ones, with $[W_a,P] = 0$, where the constant connection
corresponds to a discrete deformation of the model. We again focus on discrete Wilson
lines. Since these commute with $P$, (\ref{cond}) implies that they have to satisfy the
relation $W_a^{N/q}=I$ for each $q$. The above conditions leave the following possibilities
for discrete Wilson lines in the various models. For $N=2$, the two Wilson
lines $W_a$ are independent and satisfy $W_a^2=I$. For $N=3,4,6$, they are instead
identified by the condition (\ref{condZN}), $W_1=W_2=W$, and satisfy respectively
$W^3=I, W^2=I, W=I$. In other words, there can be two independent $\Z_2$ Wilson lines
in the $\Z_2$ model, a $\Z_3$ Wilson line in the $\Z_3$ model, a $\Z_2$ Wilson line
in the $\Z_4$ model, and no Wilson lines at all in the $\Z_6$ model. As before, the
presence of discrete Wilson lines induces a distinction between the projections
occurring at the various fixed points $z_{i_k}$ in a given sector $k$, with
$i_k=1,\ldots, N_k$, depending on the numbers $n_{i_k 1}$ and $n_{i_k 2}$ of
$T_1$ and $T_2$ translations that are needed to relate $z_{i_k}$ and its image
$R^k z_{i_k}$. As a consequence of the presence of the discrete Wilson lines,
the effective $\Z_N$ projection at each fixed point $z_{i_k}$\footnote{Notice that
some of the points $z_{i_k}$ coincide, since the same fixed point $z_0$ is generally
fixed under more elements of the orbifold action.}
will then involve the matrix $P_{i_k} = W_1^{n_{i_k 1}}W_2^{n_{i_k 1}} P^k$.
Again, it is interesting to notice that a $T^2/\Z_N$ orbifold model with a $\Z_N^\prime$
discrete Wilson line can be equivalently understood as a freely acting orbifold
of the type $T^2/(\Z_N \times \Z_N^\prime)$. In this case, a precise map between
the two constructions is more difficult to define, because of the non-trivial complex
structure of the $T^2$. See Refs.~\refcite{ZMZN,sst} for some explicit constructions of string-derived
$\Z_N \times \Z_N^\prime$ orbifolds.

The computation of gauge and gravitational anomalies proceeds as in the previous example.
Since in this case the total space-time dimensionality is even, it is again possible to relate
them to the chiral anomaly in two more dimensions, as explained in section 2. To be precise,
we embed the $(2n+2)$-dimensional space-time ${\cal M}_{2n+2} = M_{2n} \times T^{2}/\Z_N$
into an auxiliary $(2n+4)$-dimensional manifold $\widetilde{\cal M}_{2n+4} = S^2 \times
{\cal M}_{2n+2}$ and compute the chiral anomaly ${\cal Z}$ induced on $\widetilde{\cal M}_{2n+4}$
by a Dirac fermion in a representation ${\cal R}$ of $G$ and with phase $\eta$ in the
orbifold twist. The corresponding $(2n+2)$-dimensional gauge and gravitational anomaly
induced on ${\cal M}_{2n+2}$ by $\Psi$ is then deduced through the usual Stora--Zumino
descent procedure, and the quantity ${\cal A}$ entering in (\ref{loc-ano}) is defined
by the relation ${\cal Z} = \int_{\widetilde{\cal M}_{2n+4}} A$.

Proceeding as in section 2, one finds for the chiral anomaly a straightforward generalization
of eq.~(\ref{anomalyDef}):
\be
{\cal Z} = \lim_{\beta \rightarrow 0} \frac 1N \sum_{k=0}^{N-1} {\rm Tr}\,
\Big[\gamma_{2n+5} \, g^k \, e^{-\beta (i\Dslashs)^2/2}\Big] \,.
\label{anomalyDef-orb}
\ee
The trace in (\ref{anomalyDef-orb}) is taken over the eigenfunctions on $T^{2}$, which are
then projected on the eigenfunctions on $T^2/\Z_N$, thanks to the $\Z_N$ projection operator
${\cal P}_{\Z_N} = \sum_k g^k/N$, and $\gamma_{2n+5}$ is the chirality matrix in $2n+4$
dimensions. As a consequence, ${\cal Z}$ naturally splits into the sum of $N$ contributions
${\cal Z}_k$ with $k=0,1,\dots,N-1$: ${\cal Z} = 1/N \sum_k {\cal Z}_k$. Correspondingly,
the quantity ${\cal A}$ parametrizing the gauge and gravitational anomaly (\ref{loc-ano})
can be decomposed as a sum of $N$ contributions ${\cal A}_k$, defined by the relation
${\cal Z}_k = \int_{\widetilde{\cal M}_{2n+4}} A_k$, as
\be
{\cal A}(x,z) = \frac 1 N \sum_{k=0}^{N-1}{\cal A}_k(x,z) \,.
\label{tot-anoZN}
\ee
Since the space-time dimensionality is even, the delocalized contribution from the $k=0$ sector
and the localized contributions from the $k \neq 0$ sectors are in general both non-vanishing.
Once again, the anomaly (\ref{anomalyDef-orb}) is computed as the high-temperature limit of
a partition function. The relevant Routhian is given in (\ref{sqm1Z2}) with an obvious
reinterpretation of the notation. It is convenient to compute separately the contributions
${\cal A}_k$ in (\ref{anomalyDef-orb}), for which the periodicity conditions of the various
fields are specified by the action of $g^k$. The space-time components $x^\mu$ and $\psi^\mu$
are periodic, the internal components $x^z$ and $\psi^z$ are twisted by $\tau^{-k}$, and
$x^{\bar z}$ and $\psi^{\bar z}$ by $\tau^{k}$, and finally $g^k c^* g^{-k}=c^* P^k$ and
$g^k c g^{-k}=P^{-k} c$.

In the sector $k=0$, all the fields are periodic and insensitive to the orbifold projection.
Proceeding exactly as described in section 2, we thus get the standard form (\ref{chiral-anomaly})
of the chiral anomaly:
\be
{\cal A}_0(x,z) = \rho\,{\rm ch}_{\cal R}(F) \,\hat A(R) \,.
\label{chi-ano}
\ee
Notice that discrete or continuous Wilson lines cannot modify the result
(\ref{chi-ano}), because ${\cal A}_0$ must coincide with the standard
anomaly in a $(2n+2)$-dimensional non-compact space in the decompactification
limit and this leaves no room for additional corrections. From a technical
point of view, their absence is due to the fact that all the fields with internal
indices are periodic, so that the terms in (\ref{sqm1Z2}) that involve
the gauge connection vanish at quadratic order in the fluctuations. Notice
also that (\ref{chi-ano}) is obviously independent of the twist $\eta$ of $\Psi$.

In the sectors with $k\neq 0$, the fields with internal indices have periodicity conditions
that are affected by the orbifold as described above. The fermions $\psi^z$ do not admit
zero modes, whereas the bosons $x^z$ admit some in correspondence of any of the $N_k$
fixed points $z_{i_k}$ of the $g^k$ element. After expanding in normal coordinates, one
finds the following effective quadratic Routhian:
\bea
{\cal R}^{\rm eff}_k \a=\a
\frac 12 \Big[\dot \xi_{\underline \mu}\dot \xi^{\underline \mu}
+ \lambda_{\underline \mu} \dot \lambda^{\underline \mu}
+ R_{\underline \mu\underline\nu}(x_0,z_{i_k},\psi_0)
\xi^{\underline \mu} \dot \xi^{\underline \nu} \Big]
-c_A^\star F(x_0,z_{i_k},\psi_0)^A_{\;\;B} c^B \nn \\
\a\;\a + \Big[{\dot{\bar{\xi}}}_{\underline z}\dot\xi^{\underline z}
+ \frac 12 \big(\bar\lambda_{\underline z}\dot \lambda^{\underline z}
+ \lambda_{\underline z} \dot{\bar\lambda}{}^{\underline z}\big) \Big]
+ c_A^\star\Big[(\tau^{k} - 1) z_{i_k} A^A_{\bar z\,B} + {\rm c.c.}
\Big] c^B \,.
\label{ReffZN-nonab}
\eea
The curvatures $R_{\mu\nu}$ and $F$ are defined as in (\ref{R-F}), with an
obvious adaptation of the notation. The last terms in (\ref{ReffZN-nonab})
take into account possible discrete Wilson lines and originates from the
term proportional to $A_{\bar z}$ and its complex conjugate in the generalization
of (\ref{sqm1Z2}). As in the $S^1/\Z_2$ case, $(\tau^{k} - 1) z_{i_k}$
can be most conveniently written as $n_{i_k 1} e_1 + n_{i_k 2} e_2$.
The trace over the $1$-particle states of the fermion fields $c_A$ can then be
easily evaluated. One gets
\be
{\rm Tr}_{c,c^\star}
\Big[P^k e^{c_A^\star [(\tau^{k}-1) z_{i_k} A^A_{z\,B} + {\rm c.c.}] c^B}
e^{c_A^\star F^A_{\;B} c^B}\Big]
= {\rm tr}_{\cal R} \big[W_1^{n_{i_k 1}} W_2^{n_{i_k 2}} P^k e^{F}\Big]\,.
\ee
The path integral over the periodic fields is computed as in section 2.1, whereas
the one over the fluctuations of the twisted fields can be computed using
the $\zeta$-function regularization and yields, as in Ref.~\refcite{Gindex}:
\bea
\Big|{\rm det}_{\tau^{k}}^{-1}\,\Big[\!-\!\partial_\tau^2 \Big]\Big| \a=\a
\bigg[2\sin\Big(\frac{\pi k}N\Big)\bigg]^{-2} = \frac 1{N_k}\,, \\
\Big|{\rm det}_{\tau^{k}}\,\Big[\partial_\tau \Big]\Big| \a=\a
\bigg[2\sin\Big(\frac{\pi k}N\Big)\bigg] = \sqrt{N_k} \,.
\eea
Putting all the contributions together, we finally find
\be
{\cal A}_k(x,z) = \frac \rho{N_k} \sum_{i_k=1}^{N_k} \eta^k N_k^{1/2}
{\rm ch}_{\cal R} (P_{z_{i_k}}, F)\,\hat A(R) \,\delta^{(2)}(z-z_{i_k})\,.
\label{chiral-ZN-nonabG}
\ee

The total anomaly (\ref{tot-anoZN}) is, as expected, the sum of a contribution
from the $k=0$ sector that is distributed over the whole space-time, and contributions
from the sectors $k \neq 0$ that are localized at the $N_k$ fixed points $z_{i_k}$ of
the elements $g^k$ of the orbifold group. The factor $1/N_k$ has the role of averaging
the contributions coming from the $N_k$ different fixed points of $g^k$. The localized
anomaly (\ref{chiral-ZN-nonabG}) can be understood as the total anomaly coming from the
chiral fermions surviving each projection $P_{i_k}$. The contribution localized at
$z_{i_k}$ involves a Chern character twisted by the matrix $P_{i_k}$ that represents
the projection that is active at that point, which taken on its own would break the gauge
group $G$ to a subgroup $H_{i_k}$, with the representation ${\cal R}$ decomposing into the
representations ${\cal R}_{l_{i_k}}$ of $H_{i_k}$. This contribution comes along, as
expected, with the overall phase $\eta$ as well as the factor $N_k^{1/2}$, which can
be rewritten in terms of phases as
\be
N_k^{1/2} = -i \sum_{s=\pm 1/2} (2s) \tau^{ks} \,.
\ee
In this form, $N_k^{1/2}$ is recognized to count the $2n$-dimensional chiral components
$s=\pm 1/2$ of the $(2n+2)$-dimensional fermion with opposite overall signs $2s$, where
the phase $\tau^{ks}$ is nothing else than the $\Z_N$ action on the spinor indices, as
dictated by (\ref{ZNdef}). On the other hand, the matrix $P_{i_k}$ commutes with all
the generators of the surviving group $H_{i_k}$ and acts as a phase $\tau^{k r_{l_{i_k}}}$
with integer $r_{l_{i_k}}$ on ${\cal R}_{l_{i_k}}$. The twisted Chern character occurring
at the fixed point $z_{i_k}$ then decomposes as ${\rm ch}_{\cal R} (P_{i_k}, F) =
\sum_{l_{i_k}} \tau^{k r_{l_{i_k}}} {\rm ch}_{{\cal R}_{l_{i_k}}}(F)$.
Each representation ${\cal R}_{l_{i_k}}$ is thus counted with a total phase corresponding
to the sum of the phases picked up by the two chiral and antichiral components of each
fermion under the orbifold twist acting at the fixed point:
$\sum_{s=\pm 1/2} \eta^k (2s) \tau^{ks} \tau^{k r_{l_{i_k}}}$.
This makes it possible to rewrite (\ref{tot-anoZN}) in terms of the anomaly ${\cal A}_{0}$
induced by the massless fields of the unprojected theory and the anomalies ${\cal A}_{i}$
induced by the massless fields that would survive the orbifold projection associated to all
the distinct fixed points $z_i$, which depends on those gauge fields that would survive that
same projection. The result is simply
\be
{\cal A}(x,z) = {\cal A}_{0} + \sum_{i} \sigma_{i}\,{\cal A}_{i}(x,z_i)\,
\delta^{(2)}(z-z_i)\,.
\label{ano-sigmaz0}
\ee
The sum now runs over all the distinct fixed points $z_i$ of the orbifold. In the absence of
discrete Wilson lines, the coefficients $\sigma_{i}$ in (\ref{ano-sigmaz0}) can be written as
\be
\sigma_{i} = -\frac iN \sum_{k^\prime=1}^{N/q_i-1}\frac 1{\sqrt{N_{q_i k^\prime}}}
\tau^{(\frac 12 + r_i + r_\eta) q_i k^\prime} \,,
\ee
where $q_i$ is defined in such a way that $\Z_{N/q_i}$ is the maximal orbifold subgroup that
leaves the point $z_i$ fixed, $r_i$ characterizes the action of the gauge twist on each
representation and $r_\eta$ is the integer entering in the definition of $\eta$.
The numerical values of the coefficients $\sigma_{z_i}$ are reported in table 1 of
Ref.~\refcite{vonGersdorff:2003dt} for all possible values of $r$ and $r_\eta=0$, in our
notation. In the presence of Wilson lines, the value of $r$ differs between fixed points of
the same orbifold subgroup.

The $2n$-dimensional anomaly is obtained by integrating the $(2n+2)$-dimensional
anomaly computed above and dividing by a factor of $i$, because of our convention
on the chiral gamma matrices in various dimensions (see the beginning of subsection 2.1).
The result can be written in the form
\be
\tilde {\cal A}(x) =  \rho \,{\rm ch}_{\cal R} ({\cal P}, F) \,\hat A(R)\,.
\label{chiral-anomalyZN-int}
\ee
The twist matrix appearing in this expression is
${\cal P} = 1/N \sum_k \eta^k N_k^{-1/2} \sum_{i_k} P_{z_{i_k}}$,
where the sum over $k$ runs only on the twisted sectors $k \neq 0$.
It is straightforward to verify case by case that the above matrix
can in fact be rewritten as ${\cal P} = {\cal P}_R{\cal P}_T$,
where ${\cal P}_R = 1/N \sum_k \eta^k N_k^{1/2} P^k$ is the index associated
to the orbifold projection that counts each fermion component with a sign
related to its chirality, whereas $P_{W}$ is the projector associated
to the Wilson line twists that keeps only fields periodic on $T^2$.
The projector ${\cal P}_R$  does not need to contain the $k=0$ term because
$N_k$ would vanish in that sector, reflecting the fact that the corresponding
states occur in pairs of fermions with opposite chiralities. The projector $P_W$
depends on the model: it is given by $1/4(1+W_1)(1+W_2)$ for $N=2$, $1/3(1+W+W^2)$
for $N=3$, $1/2(1+W)$ for $N=4$, and $1$ for $N=6$. This rewriting makes
it obvious that only the fermionic zero modes surviving
both the orbifold projection and periodic on $T^2$ do contribute
to the integrated anomaly, as must be. See Ref.~\refcite{Asaka:2002my} for an
analysis of anomalies in six dimensions.

\subsection{Bulk fermions on higher-dimensional orbifolds}

There exist many orbifold constructions with $2n$ space-time dimensions and $m>2$
internal dimensions. We will not attempt at a general classification of the various
types of known constructions, but rather make a few comments on $\Z_N$ orbifold
models based on even-dimensional tori $K_{2l}=T^{2l}$, since these are particularly
simple and interesting in the context of both string theory and higher-dimensional
field theories. Orbifolds of the form $T^{2l}/\Z_N$ for $l>1$ can exist for more
values of $N$ than those available for $l=1$ and represent a straightforward
generalization of the $T^2/\Z_N$ case discussed in subsection 4.4. The action of
the twist on the $l$ complex internal coordinates $z_j$ is now defined by a vector
$v_j$ such that $z_j \rightarrow e^{2\pi i v_j/N} z_j$, with $j=1,\ldots,l$.
A new feature of these orbifolds with respect to the $T^2/\Z_N$ case is the possibility
that entire hyperplanes be left fixed by the orbifold action. These fixed planes
arise whenever $v_j k/N$ is an integer for some $k$ and $v_j$, since the orbifold
action is then trivial along the $j$-th torus. Denoting by $n_k$ the number of complex
dimensions in the sector $k$ such that $k v_j/N$ is an integer, the number of
fixed planes of dimension $2n_k$ is given by a product of expressions like (\ref{Nkprime})
for all the remaining $n-n_k$ internal $T^2$'s:
\be
N_k^\prime = \prod_{j^\prime=1}^{n-n_k}
\bigg[2 \sin \Big(\frac{\pi k v_{j^\prime}}{N}\Big) \bigg]^2\,.
\label{Nkprime2}
\ee
The orbifold action on the various $2n$-dimensional components of the $(2n+2l)$-dimensional
fermion is specified by a vector $w_j$ encoding the weights under the $l$ $U(1)$ subgroups
of the $SO(2l)$ internal tangent space group. The $(2n+2l)$-dimensional anomaly induced
by a $(2n+2l)$-dimensional fermion with chirality $\rho$ can be computed as in
subsection 4.4, and the result is:
\bea
{\cal A}_0(x,z_j) \a=\a
\rho\, {\rm ch}_{\cal R}(F) \,\hat A(R) \,, \nn \\
{\cal A}_k(x,z_j) \a=\a \frac {\rho}{N^\prime_k} \sum_{i_k=1}^{N^\prime_k} \eta^k N_k^{\prime 1/2}
{\rm ch}_{\cal R} (P_{i_k}, F)\,\hat A(R) \,\delta^{(2l-2n_k)}(z-z_{i_k})\,.
\label{ZN-nonabG-higher}
\eea
The matrices $P_{i_k}$ define once again the local projection at the fixed point/plane
$z_{i_k}$, constructed from the twist matrix $P$ and the allowed Wilson lines on the orbifold.
By manipulations similar to those performed in the last subsection, one can show that the
integrated anomaly arising from (\ref{ZN-nonabG-higher}) coincides with the one induced by the
massless $2n$-dimensional modes of the fermion.
See Refs.~\refcite{sst,GrootNibbelink:2003gb,Gmeiner:2002es} and
Ref.~\refcite{Faux:2000dv} for studies of localized gauge and gravitational anomalies in
orbifold theories derived from string theory and M-theory.

\subsection{Inequivalent regularizations}

An important remark is now in order. The anomalies (\ref{A0piR}), (\ref{ano-sigmaz0}) and
(\ref{ZN-nonabG-higher}) have been derived with a specific heat-kernel regularization scheme.
A crucial property of this scheme is that it is independent of the internal directions and
therefore preserves automatically any symmetry related to the latter. In particular, it
preserves the symmetry relating all the fixed points of a same kind in the absence of localized
matter or discrete Wilson lines that explicitly distinguish them, and the results derived so
far for the anomalies have this symmetry. Using instead a different regularization scheme
that does not manifestly preserve this kind of symmetry, leads in general to anomalies that
are distributed in a different way over the internal space.

The simplest example of a regulator that does not respect the symmetry among the fixed points
is a Pauli--Villars fermion with a piecewise constant mass term\cite{CS1,CS2}. Let us consider the
$S^1/\Z_2$ orbifold of subsection 4.3 with $G=U(1)$ and $W=-1$, for which the integrated anomaly
vanishes, since no light fermions survive. The heat-kernel regularization leads in this case to
a distribution over the two fixed points with equal magnitude and opposite signs, whereas the
Pauli--Villars regularization yields a strictly vanishing anomaly at each fixed point. The
reason is that, in addition to the same anomaly that is found within the heat-kernel regularization
scheme, also a Chern--Simons counterterm that exactly cancels it is generated. In this example,
which as already mentioned can be reinterpreted as an $S^1/(\Z_2\times\Z_2^\prime)$ orbifold, there
is formally no global symmetry that can distinguish between the two regularizations;
see Ref.~\refcite{CS1} for a discussion.
However, the discrete Wilson line
interpretation makes it clear that this model is naturally connected to the $S^1/\Z_2$ model, which
has a global $\Z_2^\prime$ symmetry interchanging the fixed point and privileging the heat-kernel
regularization.

The above example shows that, as usual, different regularizations produce results for the
anomaly that differ by the variation of local counterterms that can be added to the action.
We will study such terms in the next section, together with the more general question of when
and how a generic higher-dimensional anomaly can be canceled.

\section{Anomaly cancellation on orbifolds}

In this section we shall address the issue of anomaly cancellation in orbifold
theories with $2n$ space-time dimensions and $m$ internal dimensions. When $m$ is
even, there can be anomalies originating in the unprojected theory. These fall
into the class described in section 2 and the question of their cancellation
therefore follows the general discussion of section 3. In addition, the projected
theory has in general anomalies that are localized at the fixed points of the
orbifold action. As explained in section 4, these localized $(2n+m)$-dimensional
anomalies are determined by the content of chiral fermions that would survive
the projection that is effective at the given fixed point, in contrast to the
$2n$-dimensional anomaly defined by their integral, which is instead determined
by the content of chiral fermions surviving the full orbifold projection.
At first sight, the cancellation of $(2n+m)$-dimensional anomalies is a much
stronger constraint than the cancellation of $2n$-dimensional ones. However,
we will see that there are in this case also new possibilities for canceling
anomalies, besides the obvious extensions of the mechanisms exposed
in section 3.
The main issue concerning the cancellation of anomalies on orbifolds is then to
investigate the physical relevance of those anomalies that integrate to zero, as
well as the new cancellation mechanisms that are available for them.

A $(2n+m)$-dimensional anomaly that is localized at the fixed points in the internal
space can be canceled through various generalizations of the Green--Schwarz mechanism
described in section 3.1, depending on its form. The most general form of such a
localized anomaly is:
\be
{\cal I} = 2 \pi i \int_{{\cal M}_{2n+m}} \hspace{-10pt}
\Omega_{2n+m}^{(1)}\;,\;\;{\rm with}\;\;
\Omega_{2n+m+2} = \sum_i \delta_m^i\Omega_{2n+2}^{i} \,.
\label{locano}
\ee
The sum runs over the orbifold fixed points $z_i$. The $i$-th term involves the
$m$-form $\delta_m^i = \delta^{(m)}(z-z_i) dz^1 \wedge \dots \wedge dz^m$ supported
at $z_i$, and the $2n$-form $\Omega_{2n}^{i(1)}$ involving traces with respect to
the gauge group $H_i$ surviving at that point. Finally, the Stora--Zumino descent of
$\Omega_{2n+m}^{(1)}$ is defined to be $\sum_i \delta_m^i\Omega_{2n}^{i{(1)}}$.
Notice that the $m$-forms $\delta_m^i$ are trivially closed, $d \delta_m^i = 0$, but
not exact. Indeed, globally well-defined $(m-1)$-forms $\eta_{m-1}^{i}$ such that
$\delta_m^i = d \eta_{m-1}^{i}$ cannot exist, since the integral
$\int_{M_m} \!\! \delta_m^i = 1$ does not vanish. The $2n$-dimensional anomaly
corresponding to (\ref{locano}) is instead given by
\be
\tilde {\cal I} = 2 \pi i \int_{M_{2n}} \hspace{-7pt}
\omega_{2n}^{(1)}\;,\;\;{\rm with}\;\;
\omega_{2n+2} = \sum_i \omega_{2n+2}^{i} \,.
\label{globano}
\ee
It depends on the sum of the descents of the forms $\omega_{2n+2}^{i}$ obtained
from the original forms $\Omega_{2n+2}^{i}$ by retaining only the $2n$-dimensional
zero modes of the gauge and gravitational fields and re-expressing all the traces
with respect to the surviving gauge group $H$.

A $2n$-dimensional anomaly of the form (\ref{globano}) can be canceled through
a Green--Schwarz mechanism if it is reducible from the $2n$-dimensional point of view.
This is the case if the form $\omega_{2n+2}$ defining it can be factorized as
$\omega_{2n+2} = \omega_{2k} \omega_{2n+2-2k}$ for some $k$, the descents
$\omega_{2n+1}^{(0)}$ and $\omega_{2n}^{(1)}$ having the general form given by
eqs.~(\ref{SZ-GS0}) and (\ref{SZ-GS1}). On the other hand, we will see below that a
$(2n+m)$-dimensional anomaly of the form (\ref{locano}) can be canceled through a
Green--Schwarz mechanism if it is reducible from the $(2n+m)$-dimensional point of view.
A first case where this happens is of course when all the $\Omega_{2n+2}^i$ are
separately reducible from the $2n$-dimensional point of view, and so
\be
\Omega_{2n+m+2}^{\rm I} =
\sum_i \delta_m^i \Omega_{2k_i}^i \Omega_{2n+2-2k_i}^i \,.
\label{caseI}
\ee
A second interesting case, which is a particular case of (\ref{caseI}) but is
worth to be considered separately, arises when all the $\Omega_{2n+2}^i$ are
commonly reducible from the $2n$-dimensional point of view, with the integers
$k_i$ all equal to some common $k$, and the forms $\Omega_{2k}^i$ are all equal
to some $\Omega_{2k}$, so that
\be
\Omega_{2n+m+2}^{\rm II} =
\sum_i \delta_m^i \Omega_{2k} \Omega_{2n+2-2k}^i \,.
\label{caseII}
\ee
Finally, a third important case is achieved when all the $\Omega_{2n+2}^i$ are obtained by
simply restricting at the various fixed points $z_i$ a single form $\Omega_{2n+2}$ ---
defined in all of the space ${\cal M}_{2n+m}$ and involving traces with respect
to the original gauge group $G$ --- that is $\Omega_{2n+2}^i = c_i \Omega_{2n+2}$
at $z_i$. In such a situation, we can rewrite the integrand
$\sum_i \delta_i \Omega_{2n}^{i(0)}$ of (\ref{locano}) as $\delta_m \Omega_{2n}^{(0)}$,
where $\delta_m = \sum_i c_i \delta_m^i$. This is reducible from the $(2n+m)$-dimensional
point of view, provided that the closed form $\delta_m$ is also exact and can be
rewritten in terms of some form $\eta_{m-1}$ as $\delta_m = d \eta_{m-1}$, so that it
behaves in all respects as a standard $\Omega_m$. A necessary and in fact also
sufficient condition for this to happen is that $\int \delta_m = 0$, implying that
$\sum_i c_i = 0$ and that the corresponding $2n$-dimensional anomaly vanishes.
The general form of such an anomaly then is
\be
\Omega_{2n+m+2}^{\rm III} =
\sum_i c_i \delta_m^i \Omega_{2n+2} \,,\;\;{\rm with}\;\;
\sum_i c_i = 0 \,.
\label{caseIII}
\ee
We shall now describe separately the different types of Green--Schwarz mechanisms
that can cancel the above three types of anomalies. Note that the same mechanisms
can be easily generalized to the case in which the anomaly is localized at fixed planes,
rather than fixed points. For simplicity, we will however restrict to the case of
fixed points.

\subsection{Green--Schwarz mechanism with localized forms}

Whenever the $(2n+m)$-dimensional anomaly is of the type I of (\ref{caseI}),
it can be canceled through a straightforward generalization of the standard
Green--Schwarz mechanism described in section 3.1, which involves localized forms
$C_{2k_i-2}^i$ at all the points $z_i$ where the anomaly is non-vanishing.
The relevant action is a sum of actions of the same type as (\ref{action1}),
restricted at the fixed points $z_i$, and reads:
\bea
S = \sum_i \int_{{\cal M}_{2n+m}} \hspace{-15pt} \delta_m^i \bigg[\a\a \!\!
\frac 12 \Big|d C_{2k_i-2}^i + \sqrt{2\pi}\,\xi_i\, \Omega_{2k_i-1}^{i(0)}\Big|^2
+ i\frac {\sqrt{2\pi}}{\xi_i}\,C_{2k_i-2}^i\,\Omega_{2n-2k_i+2}^i \nn \\
\a\a \!\!-2\pi i \bigg(\frac{n+1-k_i}{n+1}+\alpha\bigg)
\Omega_{2k_i-1}^{i(0)}\Omega_{2n+1-2k_i}^{i(0)} \bigg]\,.
\label{action1loc}
\eea
The parameters $\xi_i$ are arbitrary and have dimensions equal to $n-2k_i+1$.
Going through the same steps as in section 3.1, it can be easily verified that the
variation of $S$ cancels the anomaly:
\be
\delta_\epsilon S = - 2\pi i\sum_i \int_{{\cal M}_{2n+m}} \hspace{-15pt}
\delta_m^i \Big[\Omega_{2k_i}^i \Omega_{2n-2k_i}^i\Big]^{(1)} \,.
\label{varloc}
\ee

At energies much below the compactification scale, the effective Green--Schwarz
action reduces to a collection of actions of the form (\ref{action1}) for the
various form fields and the zero modes of the gauge and gravitational fields.
The case of a globally vanishing anomaly does not present any special feature
with respect to the general case of a globally non-vanishing anomaly, apart from the
fact that this effective action is then gauge-invariant from the $2n$-dimensional
point of view. Notice in particular that for the case $k_i=1$, corresponding to a
mixed $U(1)$ anomaly, eq.~(\ref{action1loc}) implies that the $2n$-dimensional
$U(1)$ symmetry is necessarily realized non-linearly. In the unitary gauge, in
which the mixing between the fields $C^i_0$ and the gauge field involved in
the forms $\omega_1^{i(0)}$ vanishes, the gauge field has a mass term proportional to
$\sum_i |\omega_1^{i(0)}|^2$. The $U(1)$ symmetry is therefore always spontaneously
broken\cite{sst}, as soon as the anomaly is non-vanishing at one of the fixed points, and
independently of the value of the integrated anomaly.
\footnote{A similar situation occurs when the $U(1)$ anomaly is localized on a fixed plane,
and integrates to zero from the lower-dimensional point of view.
In this case, the Green--Schwarz mechanism that cancels this anomaly leads again to the
spontaneous breaking of the $U(1)$ factor. In such a situation,
the mass of the $U(1)$ photon is proportional
to the compactification scale\cite{Antoniadis:2002cs}.}
One combination of the $C^i_0$
fields is eaten by the gauge boson and the remaining independent linear combinations
remain as massless axions with non-vanishing couplings to the form $\omega_{2n}$.

As in section 3.1, there exists an equivalent dual description involving the magnetic
form $\tilde C_{2n-2k}^i$, which are dual to the electric forms $C_{2k-2}^i$ from the
point of view of the $2n$-dimensional space-time, with the action
\bea
\tilde S = \sum_i \int_{{\cal M}_{2n+m}} \hspace{-15pt} \delta_m^i\bigg[ \a\a \!\!
\frac 12 \Big|d \tilde C_{2n-2k_i}^i
+ \frac {\sqrt{2\pi}}{\xi_i}\,\Omega_{2n-2k_i+1}^{(0)}\Big|^2
+ i\sqrt{2\pi}\,\xi_i\,\tilde  C_{2n-2k_i}\,\Omega_{2k_i} \nn \\
\a\a \!\! -2\pi i \bigg(\alpha-\frac{k_i}{n+1}\bigg)
\Omega_{2k_i-1}^{(0)}\Omega_{2n+1-2k_i}^{(0)} \bigg]\,.
\label{action2loc}
\eea

\subsection{Green--Schwarz mechanism with bulk forms}

A $(2n+m)$-dimensional anomaly of the type II of (\ref{caseII}) can be
canceled also by a different generalization of the standard Green--Schwarz mechanism
described in section 3.1, involving a single form $C_{2k-2}$ propagating
in the bulk\cite{Horava:1995qa,sst,GrootNibbelink:2003gb,vonGersdorff:2003dt}.
The Green--Schwarz action must then be taken as follows:
\bea
S = \int_{{\cal M}_{2n+m}} \bigg[ \a\a \!\!
\frac 12 \Big|d C_{2k-2} + \sqrt{2\pi}\,\xi\, \Omega_{2k-1}^{(0)}\Big|^2
+ i \frac {\sqrt{2\pi}}{\xi}\,\sum_i \delta_m^i C_{2k-2}\,\Omega_{2n-2k+2}^i
\raisebox{20pt}{} \nn \\
\a\a \!\! -2\pi i \bigg(\frac{n+1-k}{n+1}+\alpha\bigg) \sum_i \delta_m^i
\Omega_{2k-1}^{(0)}\Omega_{2n+1-2k}^{i(0)} \bigg]\,.
\label{action1locbis}
\eea
In this case a single arbitrary parameter $\xi$ with dimension equal
to $n+m/2-2k+1$ appears, and the variation of the action cancels the anomaly independently
of its value:
\be
\delta_\epsilon S = - 2\pi i \sum_i \int_{{\cal M}_{2n+m}} \hspace{-15pt}
\delta_m^i \Big[\Omega_{2k} \Omega_{2n-2k}^{i}\Big]^{(1)}\,.
\label{varlocbis}
\ee

To gain a better understanding of the physical implications of the action
(\ref{action1locbis}), let us examine in this case too the Green--Schwarz
effective action valid at energies below the compactification scale. Since
the involved form is now a bulk field, the effective action for its light
zero modes must be derived by carefully integrating out its heavy modes.
The bulk form $C_{2k-2}$ leads to $2n$-dimensional zero modes that consist
of a $(2k-2)$-form $c_{2k-2}$ arising from the component $C_{\parallel 2k-2}$
of $C_{2k-2}$ with only space-time indices, plus possibly other forms with
lower degree, arising from the original bulk form when some of the indices
are taken to be internal, depending on the details of the orbifold projection.
Notice now that only the former will have a kinetic term that is non-trivially
affected by the $2n$-dimensional zero mode $\omega_{2k-1}^{(0)}$ of the
Chern--Simons form $\Omega_{2k-1}^{(0)}$, and the latter will therefore be
completely irrelevant for anomaly cancellation in the low-energy limit. The
effective Green--Schwarz action is thus obtained through a simple dimensional
reduction, without any non-trivial effect from heavy modes. Defining also
$\omega_{2n-2k+2} = \sum_i \omega_{2n-2k+2}^i$, and the volume $V$ of the
internal space, the $2n$-dimensional effective action obtained by dimensional
reduction is then found to be\footnote{We normalize the $2n$-dimensional
zero modes $c$ and $\omega$ of the $(2n+m)$-dimensional forms $C$ and $\Omega$
as $c = V^{-1/2} \int_{M_m} C$ and $\omega = V^{-1} \int_{M_m} \Omega$.}
\bea
S^{\rm eff} = \int_{M_{2n}} \bigg[ \a\a \!\!
\frac 12 \Big|d c_{2k-2} + \sqrt{2\pi}\,\xi^{\rm eff}\, \omega_{2k-1}^{(0)}\Big|^2
+ i\frac {\sqrt{2\pi}}{\xi^{\rm eff}} c_{2k-2}\,\omega_{2n-2k+2} \nn \\
\a\a -2\pi i \bigg(\frac{n+1-k}{n+1}+\alpha\bigg)
\omega_{2k-1}^{(0)}\omega_{2n+1-2k}^{(0)} \bigg]\,.
\label{action1locbiseff}
\eea
The new arbitrary parameter is given by $\xi^{\rm eff} = \xi V^{1/2}$ and
has dimension $n-2k+1$. Differently from the previous case, the fate of
the anomalous $U(1)$ when $k=1$ depends on whether the integrated form
$\omega_{2n-2k+2}$ vanishes or not. If $\omega_{2n-2k+2}\neq 0$, the axion
$c_0$ is eaten by the gauge field, which gets a mass, and the $U(1)$ symmetry
is non-linearly realized; if instead $\omega_{2n-2k+2}=0$, the corresponding
axion coupling in (\ref{action1locbiseff}) vanishes, $c_0$ remains neutral
under gauge transformations, and the $U(1)$ symmetry is linearly realized.

Once again, there exists an equivalent dual description in terms of a magnetic
form $\tilde C_{2n+m-2k}$, which is dual to the electric $C_{2k-2}$ form from
the $(2n+m)$-dimensional point of view. The relevant action is given by
\bea
\tilde S = \int_{{\cal M}_{2n+m}} \bigg[ \a\a \!\!
\frac 12 \Big|d \tilde C_{2n+m-2k} + \frac {\sqrt{2\pi}}{\xi} \sum_i
\Omega_{2n-2k+1}^{i(0)} \delta_m^i\Big|^2 + i \sqrt{2\pi}\,\xi\,\tilde  C_{2n+m-2k}\,\Omega_{2k}
\raisebox{18pt}{} \nn \\
\a\a \!\! -2\pi i \bigg(\alpha-\frac{k}{n+1}\bigg) \sum_i \delta_m^i
\Omega_{2k-1}^{(0)}\Omega_{2n+1-2k}^{i(0)} \bigg]\,.
\label{action2locbis}
\eea
In this case, the magnetic formulation involves singular interactions proportional
to $|\delta_m^i|^2$ and has to be treated with some care. However, since this formulation
is equivalent to the electric formulation (\ref{action1locbiseff}), where such singular
terms do not arise, physically relevant quantities are not expected to be singular.
In fact, the role of the singular terms in (\ref{action2locbis}) is to cancel the
singular behavior induced by those components of the form $\tilde C_{2n+m-2k}$
that are odd under the orbifold projection and have only massive Kaluza--Klein modes
from the $2n$-dimensional point of view. This point is best elucidated by studying as
before the $2n$-dimensional low-energy effective action $\tilde S^{\rm eff}$ below the
compactification scale.
The bulk form $\tilde C_{2n+m-2k}$ leads to $2n$-dimensional
zero modes that consist of a $(2n-2k)$-form $\tilde c_{2n-2k}$ arising from the component
$\tilde C_{\perp 2n+m-2k}$ of $\tilde C_{2n+m-2k}$ with $2n-2k$ space-time indices
and $m$ internal indices, plus possibly other forms of higher degree depending
on the orbifold projection. It is this $(2n-2k)$-form
$\tilde c_{2n-2k}$ that will have a kinetic term that is non-trivially affected
by the $2n$-dimensional zero modes $\omega_{2n-2k+1}^{i(0)}$ of the Chern--Simons
forms $\Omega_{2n-2k+1}^{i(0)}$, and the other zero modes will be irrelevant for
anomaly cancellation in the low-energy limit.
The bulk form $\tilde C_{2n+m-2k}$ leads also to $2n$-dimensional non-zero modes.
In particular, one finds a set of $m$ $(2n+m-1-2k)$-forms $\tilde c_{2n+m-1-2k}^{(j)}$,
$j=1,2,\dots,m$, arising from the components $\tilde C_{\parallel 2n+m-2k}$ of
$\tilde C_{2n+m-2k}$ with $2n-2k+1$ space-time indices and all the internal indices
but the $j$'s one, {\it i.e.} $m-1$ internal indices. Owing to the factors $\delta_m^i$
appearing in the modified kinetic term, these heavy modes $\tilde c_{2n+m-1-2k}^{(i)}$
have a non-trivial mixing with the light mode $\tilde c_{2n-2k}$ and the Chern--Simons
forms $\omega_{2n-2k+1}^{i(0)}$, and must be carefully  integrated out.
This can be done directly in the $(2n+m)$-dimensional theory and at the classical
level, by using the equations of motion of the forms $\tilde C_{\parallel 2n+m-2k}$,
where space-time derivatives are neglected with respect to internal derivatives, since
they would lead to higher-derivative effects suppressed by the compactification
scale. The relevant Lagrangian for the $\tilde C_{\parallel 2n+m-2k}$'s is obtained
from the modified kinetic term after decomposing $\tilde C_{2n+m-2k}$ into the components
$\tilde C_{\parallel 2n+m-2k}$ and $\tilde C_{\perp 2n+m-2k}$, and splitting similarly
the exterior derivative into a space-time part $d_\parallel$ and an internal part
$d_\perp$, so that $d \tilde C_{2n+m-2k} = d_\parallel \tilde C_{\perp 2n+m-2k}
+ d_\perp \tilde C_{\parallel 2n+m-2k}$. Retaining only the zero modes
$\omega_{2n-2k+1}^{i(0)}$ of the $\Omega_{2n-2k+1}^{i(0)}$, the approximate equations
of motion for $\tilde C_{\parallel 2n+m-2k}$ then read
\be
d_\perp {}^* \bigg(d_\perp \tilde C_{\parallel 2n+m-2k}
+ d_{\parallel} \tilde C_{\perp 2n+m-2k} + \frac {\sqrt{2\pi}}{\xi}\,
\sum_i \omega_{2n-2k+1}^{i(0)} \delta_m^i\bigg) = 0 \,.
\ee
In component form, these represent a set of $m$ independent equations. More precisely, the
$j$'s equation, which is associated to that component $\tilde C_{\parallel 2n+m-2k}$ that
has all the internal indices but the $j$'s, takes the form of a derivative $\partial_{j}$
with respect to $z^j$ of the bracket.

The general solution of all the $m$ equations is obtained by setting the content
of the bracket equal to a constant $(2n+m-2k+1)$-form $D_{2n+m-2k+1}$, with $m$ internal indices,
that does not depend on the internal coordinates. This
form can be fixed by considering the integral of the resulting equation
over all the $m$ internal directions. The integral of the first term vanishes,
that of the second yields $V^{1/2}d_{\parallel}(\tilde c_{\perp 2n-2k}V_m)$,
and that of the third yields $(\sqrt{2\pi}/\xi)\,\omega_{2n-2k+1}^{(0)}$,
where, as above, $\omega_{2n-2k+2} = \sum_i \omega_{2n-2k+2}^i$, and $V_m = dz^1 \wedge
\dots \wedge dz^m$ is the volume form for the internal space. This must be equated
to the integral of the constant form, which is $V D_{2n+m-2k+1}$. We then deduce that
$D_{2n+m-2k+1} = V^{-1/2} d_{\parallel} (\tilde c_{\perp 2n-2k}V_m) +
V^{-1}(\sqrt{2\pi}/\xi)\,\omega_{2n-2k+1}^{(0)}$, so that the solution of the equations
is
\bea
d_\perp \tilde C_{\parallel 2n+m-2k} \a=\a
- d_\parallel \Big(\tilde C_{\perp 2n+m-2k} - V^{-1/2} \tilde c_{\perp 2n-2k} V_m\Big) \nn \\
\a\;\a - \frac {\sqrt{2\pi}}{\xi} \Big(\sum_i \omega_{2n-2k+1}^{i(0)} \delta_m^i
- V^{-1}\omega_{2n-2k+1}^{(0)} V_m \Big) \raisebox{15pt}{}\,.
\label{dperpC}
\eea
Plugging (\ref{dperpC}) back into the Lagrangian and integrating over the
internal space, one finds an effective Green--Schwarz action that is
free of singularities, as expected:
\bea
\tilde S^{\rm eff} = \int_{M_{2n}} \bigg[ \a\a \!\!
\frac 12 \Big|d \tilde c_{2n-2k}
+ \frac {\sqrt{2\pi}}{\xi^{\rm eff}}\,\omega_{2n-2k+1}^{(0)}\Big|^2
+ i \sqrt{2\pi}\,\xi^{\rm eff}\,\tilde c_{2n-2k}\,\omega_{2k} \nn \\
\a\a \!\! - 2\pi i \bigg(\alpha-\frac{k}{n+1}\bigg)
\omega_{2k-1}^{(0)} \omega_{2n+1-2k}^{(0)} \bigg]\,.
\label{action2locbiseff}
\eea
As a consistency check, note that the effective actions
(\ref{action1locbiseff}) and (\ref{action2locbiseff}) are dual to each other
from the $2n$-dimensional point of view, with $c_{2k-2}$ and $\tilde c_{2n-2k}$
being electric and magnetic potentials for the same physical interaction.

The situation encountered above, namely the occurrence of singular contact
interactions that take care of the bad behavior induced by fields that vanish
at the fixed points, but have non-vanishing interactions localized at these points
through their internal derivatives, is perfectly analogous to the one encountered
for $5$-dimensional supersymmetric theories compactified on a $S^1/\Z_2$ orbifold.
The relevant odd bulk field, which couples through a derivative to the boundary fields
and leads to a bad behavior that is canceled by singular counterterms, is represented
by a real pseudoscalar field $\Sigma$ in the case of gauge interactions, and by a real
vector field $A_M$ --- the graviphoton --- in the case of gravitational interactions.
These fields are required by the bulk supersymmetry and have a kinetic term that
is exactly of the type encountered in eq.~(\ref{action2locbis}). In these cases, the
cancellation of singularities is guaranteed by the existence of an off-shell version
of the theory that is manifestly free of any bad behavior associated to the fixed
points. The cancellation of singularities has been explicitly verified at the tree
level, as above, as well as at the $1$-loop level\cite{mp}.

\subsection{Green--Schwarz mechanism with top bulk forms}

Let us finally come to the case of a globally vanishing $(2n+m)$-dimensional anomaly
of the type III of (\ref{caseIII}). This can be canceled with a new type of
Green--Schwarz mechanism that has no $2n$-dimensional analogue and involves a single
form $C_{2n}$ propagating in the bulk\cite{sst}. The relevant Green--Schwarz action is given by
\be
S = \int_{{\cal M}_{2n+m}} \bigg[
\frac 12 \Big|d C_{2n} + \sqrt{2\pi}\,\xi\,\Omega_{2n+1}^{(0)}\Big|^2
+ i\frac {\sqrt{2\pi}}{\xi}\,\delta_m C_{2n} \bigg]\,.
\label{action1loctop}
\ee
In this case, the arbitrary parameter $\xi$ has dimension $-n+m/2-1$.
The last term in (\ref{action1loctop}) is a source term for those components of
$C_{2n}$ that have only space-time indices. The equation of motion
of the latter therefore reads $d_\perp {}^*H_{2n+1} = -i(\sqrt{2\pi}/\xi)\,\delta_m$.
The integrability condition for such an equation requires that $\int \delta_m = 0$,
namely that $\delta_m$ be not only closed but also exact and can be written as
$\delta_m = d \eta_{m-1}$.\footnote{For $m>1$, $\eta_{m-1}$ is given by complicated
modular functions of the internal coordinates, breaking translational invariance. In the
case $m=2$, for example, the $1$-form $\eta_1$ can be written as ${}^*d_\perp \Delta_0$,
where ${}^*d_\perp$ is the $2$-dimensional dual of $d_\perp$ and $\Delta_0$ is the torus
Green function satisfying the Laplace equation $d_\perp {}^*d_\perp \Delta_0 = \delta_2$,
with a $\delta_2^i$-function source of magnitude $c_i$ at each fixed point $z_i$. A
similar situation was encountered in Ref.~\refcite{Lee:2003mc}.}
This implies that $\sum_i c_i = 0$, as assumed in (\ref{caseIII}). Stated in other words,
it is possible to have in this case a non-trivial action for the $2n$-dimensional top
form. Indeed, the usual requirement that the total charge of its sources should vanish is
substituted by the milder requirement that the integral of the charges over the internal
space should vanish. It is easy to verify in the usual way that the variation of
(\ref{action1loctop}) cancels the anomaly:
\be
\delta_\epsilon S = -2\pi i \int_{{\cal M}_{2n+m}} \hspace{-15pt}
\delta_m \Omega_{2n}^{(1)}\,.
\label{varloctop}
\ee
A globally vanishing anomaly can therefore be canceled through a Green--Schwarz
mechanism that is formally the generalization of the one working for reducible
anomalies to the case $k = n+1$ with $\alpha=0$. This implies that even localized
anomalies that are irreducible from the $2n$-dimensional point of view can be canceled,
provided that their integrated form vanishes.

The physical interpretation of the above-described mechanism depends on the
number $m$ of extra dimensions. For $m=1$, namely ${\cal M}_{2n+1}=M_{2n}\times
S^1/\Z_2$, the top form $C_{2n}$ does not have any propagating degree of
freedom.\footnote{Recall that the number of physical degrees of freedom of a
$k$-form in $D$ dimensions is given by $(D-2)(D-3)\ldots (D-k-1)/k!$.}
The form $C_{2n}$ is then an auxiliary field and can be integrated out at the classical
level by using its equation of motion $d C_{2n} =  - \sqrt{2\pi}\,\xi\,\Omega_{2n+1}^{(0)}
- i(\sqrt{2\pi}/\xi)\,{}^* \eta_{0}$ in the action (\ref{action1loctop}), after having
rewritten the term $\delta_1 C_{2n}$ as $- \eta_{0} d C_{2n}$ through an integration
by parts\cite{sst,vonGersdorff:2003dt}. The resulting action reads:
\bea
S_{CS} = \int_{{\cal M}_{2n+1}} \bigg[\frac {\pi}{\xi^2} \big|\eta_{0} \big|^2
+ 2 \pi i\, \eta_{0} \Omega_{2n+1}^{(0)} \bigg]\,.
\label{SCS1}
\eea
The first term produces just an irrelevant infinite constant, whereas the second
term is a Chern--Simons counterterm with an odd coefficient $\eta_{0}$ given by
a piecewise constant function of the internal coordinate that jumps at the
fixed points, which cancels the anomaly. This shows that the cancellation mechanism
involving the top form is in this case completely equivalent to a bulk Chern--Simons
counterterm letting the anomaly at one fixed point flow to the other fixed point
to cancel the opposite anomaly that occurs there.
Notice finally that the low-energy Green--Schwarz effective action below the compactification
scale is obtained by integrating (\ref{SCS1}) over the internal dimension,
with $\Omega_{2n+1}^{(0)}$ substituted with
its zero mode $\omega_{2n+1}^{(0)}$. The resulting action is trivially gauge-invariant,
since the Chern--Simons term integrates to zero due to the fact that $\eta_{0}$ does so.

For $m\ge 2$, the top form $C_{2n}$ does have propagating degrees of freedom.
This implies that the cancellation mechanism is not completely equivalent to a
simple Chern--Simons counterterm, since extra physical degrees of freedom remain
in the theory\cite{vonGersdorff:2003dt}.
This is particularly clear in the equivalent magnetic formulation,
which exists only for $m \ge 2$ and involves a magnetic form $\tilde C_{m-2}$
that is dual to the electric form $C_{2n}$ from the $(2n+m)$-dimensional point
of view. The dual action reads
\be
\tilde S = \int_{M_{2n+m}} \bigg[\!\!
\frac 12 \Big|d \tilde C_{m-2} + \frac {\sqrt{2\pi}}{\xi}\,\eta_{m-1} \Big|^2
+ i \sqrt{2\pi}\,\xi\,\tilde  C_{m-2}\,\Omega_{2n+2}
+ 2 \pi i\,\eta_{m-1} \Omega_{2n+1}^{(0)} \bigg]\,.
\label{action2loctop}
\ee
Since the kinetic term of $\tilde C_{m-2}$ does not involve any gauge field,
$\tilde C_{m-2}$ is perfectly gauge-invariant and the only variation of
(\ref{action2loctop}) comes from the last Chern--Simons term, which cancels
the anomaly, as in the electric formulation.

It is worth mentioning that the above way of realizing an interaction with a piece-wise
constant coupling through a non-dynamical top form as auxiliary field, is
extremely relevant in theories with local supersymmetry. Indeed, it has been shown
in Ref.~\refcite{bkv}, in the context of $5$-dimensional supergravity theories compactified
on $S^1/\Z_2$, that this is the only way to introduce such interactions without
spoiling local supersymmetry at the fixed points.

\subsection{Anomaly cancellation in string-derived orbifold models}

All the considerations of section 4 on localized and integrated anomalies,
as well as the various generalized Green--Schwarz mechanisms described above,
can be applied to any orbifold field theory. In general, anomaly cancellation
must be imposed for consistency and represents a severe constraint on model
building. In the particular case of string-derived models, on the other hand,
the cancellation is guaranteed by the fact that the underlying microscopic
string theory is a finite theory and therefore free of any anomaly. It can
then be explicitly verified that the cancellation mechanisms required to cancel the
quantum anomalies of the light modes is automatically generated by integrating
out the heavy modes. General techniques allowing a precise analysis of the
cancellation mechanism that takes place from a low-energy field-theoretical
perspective have been developed both for heterotic\cite{canchet} and unoriented
strings\cite{cancunor}. These techniques have recently been generalized and
applied to study anomaly cancellation in orbifold models from a higher-dimensional
perspective, {\em i.e.} locally in the internal space\cite{sst,Gmeiner:2002es}.

It is also interesting to note that the auxiliary supersymmetric quantum mechanical
system, which we have so extensively used in the anomaly computations, naturally
arises when evaluating anomalies in string theory. In particular, it can be
shown\cite{canchet,cancunor} that the leading anomalous $1$-loop $(n+1)$-point
functions that need to be computed in $2n$ dimensions, are nicely encoded in the
odd-spin-structure partition function of a $2$-dimensional world-sheet $\sigma$-model
in the presence of gauge and gravitational backgrounds.
Correspondingly, the contribution of the zero-energy states, which are the only ones
that are relevant to anomalies, is encoded in the supersymmetric quantum mechanical
system arising from the dimensional reduction to $1$ dimension of this $2$-dimensional
$\sigma$-model, which coincides with the one we used.\footnote{Most likely, this
supersymmetric quantum mechanics should arise also in the direct evaluation of field
theoretical anomalous amplitudes along the lines of Ref.~\refcite{Strassler:1992zr}.}

\section{Non-perturbative anomalies}

The perturbative gauge and gravitational anomalies that we have considered so far
concern local symmetry transformations connected to the identity, and
can therefore be infinitesimal. In general, however, there can be additional
non-perturbative gauge and gravitational anomalies concerning
symmetry transformations topologically non-trivial and disconnected
from the identity, that hence exist only in a finite form and cannot be infinitesimal.
The latter can occur both for gauge symmetries and for diffeomorphisms
(or local Lorentz transformations). They are also called global anomalies and
were first discovered by Witten\cite{Wittensu2} in an $SU(2)$ model in $4$ dimensions.
Differently from perturbative anomalies, the non-perturbative ones cannot be directly
detected through perturbative Feynman diagram computations, and this explains their name.
A general discussion of gauge and gravitational global anomalies lies beyond the
aim of this review. In the following, we will recall some basic features
of the former in flat space, and advise the reader interested in the latter to
see Ref.~\refcite{W-grav}.

Let us begin by examining which gauge groups $G$ can lead to global gauge anomalies
in a $2n$-dimensional flat Euclidean space-time $R^{2n}$. We consider
gauge transformations $g(x)$ that reduce to the identity at infinity, so that
they represent maps from $S^{2n}$ (the $2n$-dimensional sphere)
into the gauge group $G$. Such gauge transformations are
classified by the $2n$-th homotopy group of $G$, denoted by $\pi_{2n}(G)$.
If the latter is trivial, all the gauge transformations are connected to the identity
and no global anomalies can arise. On the contrary, if it is not, there exist classes
of topologically non-trivial gauge transformations that can potentially be anomalous.
Denoting by $A^g= g^{-1} A g + g^{-1} dg$ the gauge-transformed connection obtained
from $A$ through such a non-trivial gauge transformation $g$, a global anomaly
can occur if the effective action --- which is defined modulo physically irrelevant
multiples of $2 \pi i$ as in (\ref{effectiveA}) --- changes under the finite transformation
$g$ by an amount $\Gamma(A^g) - \Gamma(A)$ that is not a multiple of $2 \pi i$:
\be
\Gamma(A^g) - \Gamma(A) \neq 2\pi i n \,.
\label{Global-ano}
\ee

If the situation (\ref{Global-ano}) occurs, the quantum effective action and all the
correlation functions of gauge-invariant operators it describes are not well-defined,
and the theory is thus inconsistent\cite{Wittensu2}.\footnote{To be precise,
(\ref{Global-ano}) leads to an inconsistency only if $A$ and $A^g$ are connected in
field space without passing infinite action barriers. Otherwise, it is possible to
define a sensible quantum effective action by restricting the functional integral
to topologically trivial gauge configurations only.} As perturbative gauge
anomalies, also non-perturbative gauge anomalies can be induced only by Weyl fermions
in even-dimensional space-times and through the imaginary part of the Euclidean
effective action, since Dirac fermions always allow for a manifestly gauge-invariant
regularization. Computing the contribution of a Weyl fermion to the transformation
(\ref{Global-ano}) for a generic gauge group $G$ is, however, a complicated mathematical
problem. It can be addressed essentially in two different ways: either, as was
originally done in Ref.~\refcite{Wittensu2}, by studying the evolution of the
spectrum of the Dirac operator when the gauge connection is varied from $A$ to $A^g$,
or by embedding the homotopically non-trivial gauge transformation $g$ into a larger
group $\hat G\supset G$, where the global anomaly (\ref{Global-ano}) is
reinterpreted\cite{WittenGlobal,en} as a perturbative gauge anomaly for the group
$\hat G$. We shall adopt here this second point of view, following the general
treatment of Ref.~\refcite{en}.

It should be clear that asking whether a theory is afflicted by global anomalies or not
is a meaningful question only when all perturbative anomalies cancel, the former being
defined in terms of homotopy classes and hence modulo local gauge transformations.
Consider then a gauge theory with group $G$ in $2n$ dimensions, with a generic
spectrum of Weyl fermions
that is free of perturbative anomalies and that we symbolically denote by $[\psi]$.
Following Refs.~\refcite{WittenGlobal,en}, we will map the computation of the
non-perturbative anomaly induced by $[\psi]$ in such a theory to the computation
of perturbative anomalies in another theory, with gauge group $\hat G$ containing $G$,
such that $\pi_{2n}(\hat G)$ is trivial even if $\pi_{2n}(G)$ is not, and a spectrum
$[\hat \psi]$ of chiral fermions that reduces to the original spectrum $[\psi]$ under
the decomposition $\hat G \rightarrow G$, modulo extra singlets. This auxiliary
theory is by construction free of non-perturbative anomalies, but must have
perturbative ones, since these are the only quantities to which the
non-perturbative anomalies of the original theory can map to. These perturbative
anomalies of the auxiliary theory that encode the non-perturbative anomalies of the
original theory can again be computed using Fujikawa's approach and looking at the
integration measure ${\cal D} [\hat \psi] {\cal D} [\hat{\bar \psi}]$ of the functional
integral representation of the effective action. The variation of this measure under
a finite gauge transformation $\hat g \in \hat G$ is given by
\be
{\cal D}[\hat \psi^{\hat g}]{\cal D}[\hat{\bar \psi}{}^{\hat g}] =
e^{- \Gamma_{WZ} (\hat g,\hat A,\hat F)}\,
{\cal D}[\hat \psi]{\cal D}[\hat{\bar \psi}]\,,
\ee
where $[\hat \psi^{\hat g}]$ denotes the $\hat g$-gauge transformed of $[\hat \psi]$,
$\hat A$ and $\hat F$ are the connection and field strength of $\hat G$ and
$\Gamma_{WZ}(\hat g,\hat A,\hat F)$ is the Wess--Zumino action obtained by
integrating (\ref{ano--gauge}). The explicit form of $\Gamma_{WZ}$ can be obtained,
as in (\ref{GammaWZ}), by extending the fields to a $(2n+1)$-dimensional ball
$B_{2n+1}$ whose boundary is the space-time sphere $S^{2n}$:
\be
\Gamma_{WZ}(\hat g,\hat A) = 2 \pi i \int_{B_{2n+1}} \!\!
\Big[\Omega_{2n+1}^{(0)}(\hat A^{\hat g}) - \Omega_{2n+1}^{(0)}(\hat A) \Big]\,.
\label{WZu2}
\ee
For simplicity of notation, the Chern--Simons form appearing in (\ref{WZu2})
corresponds actually to a sum of the basic Chern--Simons forms, defined as in
Appendix A, in the same representations of $\hat G$ of the associated
chiral fermions $[\hat \psi]$.
Let us now consider the subclass of gauge configurations $A$ and $F$ that are
in $G$ and the subclass of gauge transformations $\hat g$ that reduce to some
transformation $g$ of $G$ on $\partial B_{2n+1}=S^{2n}$. Notice that
this includes all the transformations of $G$, and in particular the
homotopically non-trivial ones we are interested in. The Wess--Zumino
action (\ref{WZu2}) for these configurations therefore encodes also
the variation of the effective action $\Gamma(A)$, obtained by integrating
out our original fermion spectrum $[\psi]$ under a homotopically non-trivial
gauge transformation $g$ of $G$:
\be
\Gamma(A^g) - \Gamma(A) = \Gamma_{WZ}(\hat g, \hat A) \,.
\label{Global-ano2}
\ee
Equation (\ref{Global-ano2}) allows a mapping of the global anomaly (\ref{Global-ano})
for $G$ to the perturbative anomaly under gauge transformations of $\hat G$
that reduce to $g\in G$ on $\partial B_{2n+1}$. The evaluation of the
right-hand side of (\ref{Global-ano2}) is however still a non-trivial
mathematical problem, in general.

The simplest non-trivial case where global anomalies can arise in $4$ dimensions
is\cite{Wittensu2} for $G=SU(2)$, since $\pi_4[SU(2)]={\bf Z}_2$. The simplest
anomaly-free spectrum is in this case $[\psi]=\psi_2$, where $\psi_2$ is a Weyl
doublet of $SU(2)$. We can now apply the procedure described above by taking
$\hat G=SU(3)$ (since $\pi_4[SU(3)]=0$) and $[\hat \psi]=\psi_3$,
where $\psi_3$ is a Weyl triplet of
$SU(3)$. The right-hand side of (\ref{Global-ano2}) can then be evaluated
using homotopy considerations based on exact sequences (see Ref.~\refcite{en})
or by an explicit construction of the gauge transformations $g$ and $\hat g$
as in Ref.~\refcite{w}. When $g$ belongs to the trivial element of $\pi_4[SU(2)]$,
it of course vanishes, whereas for $g$ belonging to the non-trivial element of
$\pi_4[SU(2)]$ one finds
\be
\Gamma(A^g) - \Gamma(A) = i \pi \,.
\ee
This shows that an $SU(2)$ theory with one or any odd number of Weyl doublets
is non-perturbatively inconsistent. Generalizations to other groups or dimensions
can be found in Refs.~\refcite{en,kiritsis}. Global gauge anomalies have also been
studied for theories that have an anomalous spectrum of fermions and are free
of perturbative anomalies thanks to a Green--Schwarz mechanism, in
Ref.~\refcite{Bershadsky:1997sb}. Finally, global gauge and gravitational
anomalies in string-derived theories have been studied in Ref.~\refcite{W-grav}
(see also Ref.~\refcite{W-gloGrav}).\footnote{For global gravitational anomalies,
there exist no generalization of the procedure outlined above for the gauge case,
and one has to rely on a more direct analysis of the spectrum of the Dirac operator.}

Global anomalies in spontaneously broken symmetries can be canceled with the
help of a generalization of the mechanism reviewed in section 3.2 for perturbative
anomalies\cite{Fabbrichesi:2002am}. The main issue in doing this is the
possible topological obstruction in extending locally defined quantities
to globally defined ones. The question is whether the modified Wess--Zumino
term $\Gamma_{WZ}^\prime(A,U)$, which must be added in order to cancel the
perturbative anomalies, is globally well defined and whether the new action
is invariant under homotopically non-trivial gauge transformations. This has
been studied in Ref.~\refcite{Fabbrichesi:2002am}, where it was shown that
$\Gamma_{WZ}^\prime(A,U)$ can be globally defined and cancels the possible
global gauge anomalies of the theory, provided the following two conditions
are met. First, the spectrum of fermions must be such that there are no
global anomalies for the unbroken group $H$. Second, there has to exist a
group $\hat G\supset G$, with $\pi_{2n}(\hat G)=0$, such that the fermion
spectrum of the theory can be extended to the group $\hat G$ without
generating further anomalies for $G$.\footnote{The latter condition
arises in this fashion when global anomalies are studied as in
Ref.~\refcite{en}.}

The cancellation of global anomalies provides additional constraints on the
possible chiral fermion spectrum in a theory. For instance, it has been
argued in Ref.~\refcite{Dobrescu:2001ae} that the number of generations of ordinary
quarks and leptons could be understood by requiring global anomaly cancellation
of the standard model gauge group, if all standard model fields propagate in
$6$-dimensions.

Also higher-dimensional theories defined on orbifolds should be free of any
global anomaly. For this kind of theories, one might actually have more
constraints than in theories defined on flat spaces, owing to the non-trivial
topology of the internal space. But very little is known about this.

\subsection{Parity anomalies in odd dimensions}

In an odd number of space-time dimensions, the Dirac operator is always
Hermitian and there cannot be any perturbative or non-perturbative
gauge or gravitational anomaly. This is best understood by observing that
a gauge-invariant Pauli--Villars regularization is always available in an
odd number of dimensions.\footnote{The only exception are theories in
$8n+1$ dimensions with an odd number of Majorana fermions, where it is
not possible to construct mass terms for all of them\cite{AGW}.}
However, a mass term in $2n+1$ dimensions violates the parity symmetry
defined by the inversion of a single space-time direction, and a
Pauli--Villars regulator therefore leads in general to a parity-violating
effective action. Whenever this is the case, there is a parity
anomaly\cite{AGW,Redlich,Alvarez-Gaume:1984nf}, whose form is closely
related to chiral anomalies in $2n+2$ dimensions and gauge anomalies
in $2n$ dimensions\cite{Niemi,Alvarez-Gaume:1984nf} (see Ref.~\refcite{Forte}
for an explicit construction). More precisely, a parity anomaly implies
that it is not possible to retain at the quantum level both
the parity and the gauge symmetry. Ignoring subtle global issues (see
Ref.~\refcite{AGerice} for a precise discussion), it is possible to switch
from a gauge-invariant but parity-violating effective action to a
gauge-violating but parity-invariant effective action, through a suitable
local Chern--Simons counterterm. The latter is parity odd and invariant under
small gauge transformations but in general not under large ones.
In other words, a parity anomaly can be traded for a global gauge anomaly and
vice versa\cite{Redlich}.

Parity symmetries are particularly relevant to odd-dimensional orbifolds,
since they may be part of the orbifold action. Whenever this is the case, a
parity anomaly on the covering space might lead to some inconsistency, since
in the orbifold theory the parity symmetry becomes a discrete ``gauge'' symmetry.
However, although it is quite obvious that requiring that there appear no parity
anomaly in the covering space is a sufficient condition (see {\em e.g.}
Ref.~\refcite{GrootNibbelink:2002qp} for a discussion), it is not clear whether
this condition is also necessary. In order to verify this, one would have to
look for a possible anomalous violation of the selection rule imposed by
the orbifold projection directly in the orbifold theory.

\section*{Acknowledgments}

We would like to thank G.~Thompson and R.~Rattazzi for useful discussions,
and especially L.~Alvarez--Gaum\'e, for many clarifying explanations and
important remarks on the manuscript. This research work was partly
supported by the European Community through a Marie Curie fellowship and
the RTN network contracts HPRN-CT-2000-00131 and HPRN-CT-2000-00148.

\appendix

\section{Descent relations and Chern--Simons forms}

In this appendix, we provide the explicit construction of the forms $\Omega_{2n+2}$,
$\Omega_{2n+1}^{(0)}$ and $\Omega_{2n}^{(1)}$ entering the Stora--Zumino descent
relations, as explained at the beginning of section 2. For simplicity we consider here
only the gauge case, but similar considerations apply for the gravitational case as well.
We closely follow section 3.C of Ref.~\refcite{AGG}, adopting the same, convenient,
differential-form notation with the same conventions. The gauge field $A$ is a $1$-form
and its field-strength is the $2$-form $F = d A + A^2$, which satisfies the Bianchi
identity $D F = d F + [A,F] = 0$. As explained in the main text, the relevant starting
point is a gauge-invariant $(2n+2)$-form $\Omega_{2n+2}(F)$. This is exact and
therefore defines a Chern--Simons $(2n+1)$-form through the local decomposition
$\Omega_{2n+2}(F) = d \Omega_{2n+1}(A,F)$. Finally, the gauge variation of the latter
defines a $2n$-form $\Omega_{2n}(A,F)$ through the transformation law
$\delta_v \Omega_{2n+1}(A,F) = d \Omega_{2n}(v,A,F)$.

To begin with, let us introduce the relevant formalism. We shall need to define
a family of gauge transformations $g(x,\theta)$ depending on the coordinates $x^\mu$
and on some parameters $\theta^\alpha$.  These gauge transformations change $A$ and $F$
into
\bea
\bar A(x,\theta) \a=\a g^{-1}(x,\theta) \big(A(x)+d\big) g(x,\theta)\,, \\
\bar F(x,\theta) \a=\a g^{-1}(x,\theta) F(x) g(x,\theta)\,.
\eea
We can then define, besides the usual exterior derivative $d=dx^\mu \partial_\mu$ with
respect to the coordinates $x^\mu$, an additional exterior derivative
$\hat d = d\theta^\alpha \partial_\alpha$ with respect to the parameters $\theta^\alpha$.
These two operators anticommute and are both nilpotent, $d^2 = {\hat d}^2 = 0$. This
implies that their sum $\Delta = d + \hat d$ is also nilpotent: $\Delta^2=0$. The two
operators $d$ and $\hat d$ naturally define two transformations parameters $v$ and
$\hat v$ through the expressions:
\bea
v(x,\theta) \a=\a g^{-1}(x,\theta) d g(x,\theta)\,, \\
\hat v(x,\theta) \a=\a g^{-1}(x,\theta) \hat d g(x,\theta)\,.
\eea
It is straightforward to verify that
\be
\hat d \hat v = - \hat v^2 \,,\;\;
\hat d \bar A = - \bar D \, \hat v\,,\;\;
\hat d \bar F = -[\hat v, \bar F]\,.
\ee
These equations show that $\hat d$ generates an infinitesimal gauge transformation
with parameter $\hat v$ on the gauge field $\bar A$ and its field-strength $\bar F$.
Interestingly enough, these can also be interpreted as BRST transformations, the ghost
fields being identified with $\hat v$. At this point, it is possible to define yet
another connection ${\cal A}$ and field-strength $\cal F$ as
\bea
{\cal A} = \a\a g^{-1} \big(A + \Delta \big) g = \bar A + \hat v\,,
\label{calAA} \\
{\cal F} = \a\a \Delta {\cal A} + {\cal A}^2 = g^{-1} F g = \bar F\,.
\label{calAF}
\eea
The crucial point is now that ${\cal A}$ and ${\cal F}$ are defined with respect to
$\Delta$ exactly in the same way as $\bar A$ and $\bar F$ are defined with respect
to $d$. Therefore, the corresponding Chern--Simons decompositions must have the same
form:
\bea
Q_{2n+2}({\cal F}) \a=\a \Delta Q_{2n+1}({\cal A},{\cal F}) \,,\\
Q_{2n+2}(\bar F) \a=\a d Q_{2n+1}(\bar A, \bar F)\,.
\eea
On the other hand, (\ref{calAF}) implies that the left-hand sides of these two equations
are identical. Equating the right-hand sides and using (\ref{calAA}) yields:
\be
(d+\hat d)Q_{2n+1}(\bar A+\hat v,\bar F) = d Q_{2n+1}(\bar A, \bar F)\,.
\label{SZ-1}
\ee
In order to extract the information carried by this equation, it is convenient to
expand $\Omega_{2n+1}(\bar A+\hat v,\bar F)$ in powers of $\hat v$ as
\be
Q_{2n+1}(\bar A+\hat v,\bar F) = Q_{2n+1}^{(0)}(\bar A,\bar F) +
Q_{2n}^{(1)}(\hat v,\bar A,\bar F) + \ldots + Q_{0}^{(2n+1)}(\hat v,\bar A,\bar F) \,,
\label{SZ-exp}
\ee
where the superscripts denote the powers of $\hat v$ and the subscript the
dimension of the form. Substituting this expansion in (\ref{SZ-1}) and equating
terms with the same power of $\hat v$, we finally find the Stora--Zumino descent
relations\cite{SZ}:
\bea
\a\a \hat d Q_{2n+1}^{(0)}+ d Q_{2n}^{(1)}=0\,,\nn \\
\a\a \hat d Q_{2n}^{(1)}+ d Q_{2n-1}^{(2)}=0\,, \nn \\
\a\a \dots \nn \\
\a\a \hat d Q_{1}^{(2n)}+ d Q_{0}^{(2n+1)}=0\,,\nn \\
\a\a \hat d Q_{0}^{(2n+1)} = 0 \,.
\label{SZ-des-rel}
\eea
The operator $\hat d$ makes it possible to understand in a simple way why the Stora--Zumino
descent represents the most general non-trivial solution of the Wess--Zumino consistency
condition (\ref{WZ2}). For the case of gauge anomalies we are considering here,
${\cal I}(v) = \int {\rm tr} (v\, a(A))$, the latter reads
\be
\delta_{v_1} \int {\rm tr} (v_2\, a(A)) - \delta_{v_2} \int {\rm tr} (v_1\, a(A)) -
\int {\rm tr} ([v_1,v_2]\, a(A)) = 0\,.
\label{WZgau}
\ee
The two transformations with parameters ${v_1}$ and ${v_2}$ can be incorporated into a
family of transformations parametrized by $\theta^1$ and $\theta^2$, with parameter
$\hat v = v_\alpha d\theta^\alpha$. In this way, $v_\alpha = g^{-1} \partial_\alpha g$.
At $\theta^\alpha = 0$, $g(x,0)=1$ and therefore $\bar A(x,0) = A(x)$ and
$\bar F(x,0) = F(x)$. At that point, $\hat d$ generates ordinary gauge transformations
on $A$ and $F$, with $\hat d = d \theta^\alpha \delta_{v_\alpha}$.
The condition (\ref{WZgau}) can then be multiplied by $d\theta^1 d \theta^2$ and rewritten
as
\be
\int {\rm tr} (\hat v\, \hat d\, a(A)) + \int {\rm tr} (\hat v^2\, a(A)) = 0\,.
\label{WZgau2}
\ee
Since $\hat d \hat v = - \hat v^2$, this can be rewritten simply as:
\be
\hat d \int {\rm tr} (\hat v\, a(A)) = 0\,.
\label{WZgau3}
\ee
The Wess--Zumino consistency condition is therefore the statement that the anomaly is
$\hat d$-closed, $\hat d \,{\cal I}(\hat v) = 0$. It is clear that the trivial
$\hat d$-exact solutions ${\cal I}(\hat v) = \hat d \int f(A)$ in terms of a local
functional $f(A)$ of the gauge field correspond to the gauge variation of local
counterterms that can be added to the theory, and the non-trivial anomalies emerging
from the non-local part of the effective action are therefore encoded in the cohomology
of $\hat d$. From the second relation appearing in the Stora--Zumino descent relations
(\ref{SZ-des-rel}), we see that the general non-trivial element of this cohomology is
of the form ${\cal I}(\hat v) = \int Q_{2n}^{(1)}$. With the above definitions, we have
$Q_{2n+2} = d Q_{2n+1}^{(0)}$ and $\delta_v Q_{2n+1}^{(0)}= - d Q_{2n}^{(1)}$. We can
therefore identify $\Omega_{2n+2} \leftrightarrow Q_{2n+2}$,
$\Omega_{2n+1}^{(0)} \leftrightarrow Q_{2n+1}^{(0)}$ and
$\Omega_{2n}^{(1)} \leftrightarrow - Q_{2n}^{(1)}$.

Let us now compute the explicit expressions of the forms that are relevant to gauge
anomalies. The starting point is the $(2n+2)$-form characterizing the chiral anomaly
in $2n+2$ dimensions:
\be
Q_{2n+2}(F) = {\rm tr}\,F^{n+1}\,.
\label{Om}
\ee
This is a closed form, as it should, since $dF = dA A - A dA = -[A,F]$ and therefore
$d({\rm tr}\, F^{n+1})= -{\rm tr}\, [A,F^{n+1}]=0$. To derive its first descendent,
$Q_{2n+1}^{(0)}(A,F)$, let us consider a continuous family of connections $A_t$, with
field strengths $F_t = dA_t + A_t^2$, which linearly interpolate between $0$ and $A$
for $t \in [0,1]$:
\be
A_t = t A \,,\;\;
F_t = t F + (t^2-t) A^2 \,.
\ee
It is easy to verify that
\be
\partial_t F_t = d A + \big[A_t, A\big] = D_t A\,,
\ee
and thus, using the Bianchi identity $D_t F_t = 0$,
\bea
\partial_t\, {\rm tr}\, F_t^{n+1} = \a\a (n+1)\, {\rm tr}\,
\Big[\partial_t F_t\,F_t^{n}\Big]
= (n+1)\, {\rm tr}\, \Big[D_t A\, F_t^{n} \Big]\nn \\
= \a\a (n+1)\, {\rm tr}\, \Big[D_t (A\, F_t^{n}) \Big] =
(n+1)\, d \,{\rm tr}\,\Big[A\, F_t^{n}\Big] \,.
\eea
Integrating this result over $t$ finally yields
\be
{\rm tr}\, F^{n+1} = (n+1)\, d \int_0^1\!\!dt\, {\rm tr}\,\Big[A\, F_t^{n}\Big]\,.
\label{chara}
\ee
The Chern--Simons form associated to (\ref{Om}) is therefore given, modulo
exact forms, by the expression:
\be
Q_{2n+1}^{(0)}(A,F)= (n+1) \int_0^1\!\!dt\, {\rm tr}\,\Big[A\,F_t^{n}\Big] \,.
\label{Om0}
\ee
To compute the second descendent, $Q_{2n}^{(1)}(v,A,F)$, we use the result (\ref{Om0})
to determine the left-hand side of (\ref{SZ-exp}). Defining for convenience
$\tilde{\bar F_t} = t \bar F + (t^2-t) (\bar A+\hat v)^2$, this reads
\be
\!\!Q_{2n+1}(\bar A+\hat v ,\bar F)
= (n+1) \int_0^1\!\!dt\, {\rm tr}\,\Big[(\bar A + \hat v) \tilde{\bar F}_t^{n}\Big] \,.
\ee
The quantity we are after is then found by expanding in powers
of $\hat v$ and retaining the linear order. Defining the symmetrized trace of generic
matrix-valued forms $\omega_i = \omega_i^{\alpha_i} t^{\alpha_i}$ as
\be
{\rm str}\,(\omega_1\ldots\omega_p) = \frac{1}{p!} \omega_1^{\alpha_1}\wedge \ldots
\wedge \omega_p^{\alpha_p}\sum_{\rm perms.} {\rm tr}\,(t^{\alpha_1} \ldots t^{\alpha_p})\,,
\ee
we find
\bea
Q_{2n}^{(1)}(\hat v,\bar A,\bar F) \a=\a (n+1)\! \int_0^1\!\!dt\, {\rm str}
\Big[\hat v \bar F_t^n + n (t^2-t) \bar A \big[\bar A,\hat v\big] \bar F_t^{n-1}\Big] \nn \\
\a=\a (n+1)\! \int_0^1\!\!dt\, {\rm str}\,\Big[\hat v \Big(\bar F_t^n + n (t-1)
\big( t \big[\bar A,\bar A\big] \bar F_t^{n-1} \!
- \bar A \big[\bar A_t, \bar F_t^{n-1}\big]\big) \Big)\Big] \nn \\
\a=\a (n+1)\! \int_0^1\!\!dt\, {\rm str}\,\Big[\hat v \Big(\bar F_t^n + n (t-1) \big(
\big(\partial_t \bar F_t - d\bar A\big) \bar F_t^{n-1} \!
+\bar A d \bar F_t^{n-1}\big)\Big)\Big] \nn \\
\a=\a (n+1)\! \int_0^1\!\!\!\!dt\, {\rm str}\Big[\hat v \Big(\bar F_t^n + (t-1)
\partial_t \bar F_t^n + n (1-t) d\big(\bar A\bar F_t^{n-1}\big) \Big)\Big].
\label{SZ-5}
\eea
In the third equality we have used the identities $D_t \bar F_t^{n-1} = d \bar F_t^{n-1}
+ [\bar A_t,\bar F_t^{n-1}]=0$ and $\partial_t \bar F_t = d\bar A + t [\bar A,\bar A]$.
After integrating by parts, the first and second term in the last line of (\ref{SZ-5})
cancel. Setting $\theta=0$ finally gives:
\be
Q_{2n}^{(1)}(v,A,F) = n(n+1)\int_0^1\! dt \, (1-t)\, {\rm str}\,
\Big[vd(A F_t^{n-1})\Big]\,.
\label{SZ-(1)}
\ee

In order to give a concrete example, let us derive the contribution of a Weyl fermion
with chirality $\eta=\pm 1$ to the non-Abelian gauge anomaly in $4$ dimensions. In
this case, we have to start from $\Omega_6 = {\rm ch} (F)|_{6} = -i/(48 \pi^3) {\rm tr} F^3$.
Taking into account the sign flip required by our definitions, equation (\ref{SZ-(1)})
leads then to
\be
\Omega_{4}^{(1)}(v,A,F) = \frac i{48 \pi^3} {\rm str}\, \Big[ v d \Big( A F -  \frac 12 A^3 \Big)\Big]\,.
\ee
Using (\ref{Igauge}) and taking into account the coefficient arising from the expansion
(\ref{ch-dec}), we obtain the following anomalous variation of the effective action:
\be
\delta_v \Gamma(A) = - \frac{\eta}{24\pi^ 2} \int d^4x\,{\rm str}\,
\Big[ v d\Big( A F -  \frac 12 A^3\Big) \Big]\,.
\label{ano-quattrod}
\ee

The generalization of the above relations to the gravitational case is straightforward,
particularly when dealing with local Lorentz transformations. In this case, we
must simply substitute $A$ and $F$ with the spin connection $\omega$ and the curvature $R$,
and consider infinitesimal $SO(2n)$ local rotations.

\section{${\cal N}=1/2$ $\sigma$-model in $2$ dimensions}

In this appendix, we show that the supersymmetric quantum mechanical system (\ref{sqm1})
can be constructed as a $1$-dimensional truncation of the supersymmetric $2$-dimensional
$\sigma$-model introduced in Ref.~\refcite{Hull:jv} to describe the heterotic string.
The construction is based on the so-called ${\cal N}=1/2$ superspace, namely a superspace
involving a single Grassmann variable $\theta$, associated to a Majorana--Weyl fermionic
supercharge $Q$, which has one real component in $2$ dimensions. The relevant chiral
superfields are\cite{Hull:jv}
\bea
\Phi^\mu \a = \a x^\mu + \theta \psi^\mu\,, \nn \\
C^A \a = \a c^A + \theta F^A\,, \nn \\
C_A^\star \a = \a c_{A}^\star + \theta F_{A}^\star\,,
\eea
where $\psi^\mu$ are chiral (left-handed) world-sheet spinors and space-time vectors,
with $\mu=1,\ldots,2n$,
$c_A^\star$ and $c^{A}$ are complex antichiral (right-handed) world-sheet spinors,
transforming respectively in the representation ${\cal R}$ of the gauge group $G$
and its conjugate $\bar {\cal R}$, with $A=1,\ldots, {\rm dim}\,{\cal R}$, and
finally $F_A^\star$ and $F^{A}$ are complex auxiliary fields. The superspace covariant
derivative and the supercharge are defined as
\be
D = \frac{\partial}{\partial\theta} + i \theta \partial\,,\;\;
Q = i \frac {\partial}{\partial\theta} + \theta\partial \,,
\ee
where $\partial = \partial/\partial z = \partial_\tau + \partial_\sigma$
and $\bar\partial = \partial/\partial \bar z = \partial_\tau - \partial_\sigma$
are the holomorphic and anti-holomorphic derivatives with respect to the complex
coordinates $z,\bar z$. The Lagrangian is defined as
\be
{\cal L}(\Phi,C,C^*) = \int\! d\theta \bigg[\!-\!\frac i2 g_{\mu\nu}(\Phi)
D\Phi^\mu \bar \partial \Phi^\nu - C_A^\star
\Big( D C^A + A^A_{\mu\;B}(\Phi) D\Phi^\mu C^B \Big)\bigg]\,,
\ee
where $g_{\mu\nu}$ is the metric of the $2n$-dimensional target space $M_{2n}$ and
$A_{\mu\;B}^A = A_\mu^\alpha\, T_{\alpha\;B}^A$ is the connection of the gauge
bundle, with the matrices $T_\alpha$ satisfying the commutation relations
$[T_\alpha,T_\beta]=f^\gamma_{\;\;\alpha \beta} T_\gamma$. Integrating over the
Grassmann variable $\theta$ and solving for the auxiliary fields $F_A^*$ and $F^A$,
the component expression of the Lagrangian is found to be:
\bea
{\cal L} \a=\a \frac 12 g_{\mu\nu} \partial x^\mu \bar\partial x^\nu +
\frac i2 g_{\mu\nu} \psi^\mu \Big(\bar\partial \psi^\nu +
\Gamma^\nu_{\rho\sigma} \bar\partial x^\rho \psi^\sigma\Big) \nn \\
\a\;\a +\,i c_A^\star \Big(\partial c^A + A^A_{\mu\;B}\,\partial x^\mu c^B \Big)
+ \frac{1}2 c_A^\star c^B \psi^\mu \psi^\nu F_{\mu\nu\;B}^{A}\,,
\label{2d-2}
\eea
where
\be
F_{\mu\nu\;B}^{A} = \partial_\mu A_{\nu\;B}^A - \partial_\nu A_{\mu\;B}^A
+ \big[A_\mu,A_\nu\big]^A_{\;B}\,.
\label{F-Def}
\ee
The on-shell supersymmetry transformations of the various fields are given by
\bea
\delta x^\mu \a = \a i\epsilon \psi^\mu, \ \ \hspace{34pt}
\delta \psi^\mu = - \epsilon \partial x^\mu \, \nn \\
\delta c^A \a = \a i \epsilon c^B A_{\mu\;B}^A \psi^\mu, \ \
\delta c_A^\star =  -i \epsilon c_B^\star A_{\mu\;A}^B \psi^\mu\,.
\eea
The trivial dimensional reduction of the Lagrangian (\ref{2d-2}) to 1 dimension
is obtained by discarding the $\sigma$-dependence of the fields. Introducing
the new field $\psi^a = e_\mu^a \psi^\mu$, the Lagrangian reduces to (\ref{sqm1}),
the supersymmetry transformations become
\bea
\delta x^\mu \a = \a i\epsilon e_a^\mu \psi^a, \ \ \hspace{38pt}
\delta \psi^a = - \epsilon \Big(e_\mu^a \dot x^\mu +\frac{i}2 \big[\psi_b,\psi_c\big] e^{\mu b}
\omega_\mu^{ac} \Big) \, \nn \\
\delta c^A \a = \a i \epsilon c^B A_{\mu\;B}^A e^\mu_a \psi^a, \ \
\delta c_A^\star =  -i \epsilon c_B^\star A_{\mu\;A}^B e^\mu_a \psi^a\,,
\eea
and the supercharge simplifies to
\be
Q = - e^\mu_a \psi^a \dot x_\mu\,.
\label{supercharge}
\ee

\end{document}